\documentclass[12pt]{article}
\usepackage{amsfonts,amsthm,amsmath,amssymb,upgreek,bm}
\usepackage[hidelinks]{hyperref}
\usepackage[paper=letterpaper,margin=.85in]{geometry}
\usepackage{graphicx}
\usepackage{units}
\usepackage{float}
\usepackage[export]{adjustbox}
\usepackage{bbold}
\usepackage{xcolor}
\usepackage[shortlabels]{enumitem}
\usepackage{dsfont}
\usepackage[mathscr]{euscript}
\usepackage[normalem]{ulem}
\usepackage{physics}
\usepackage{tikz-cd}

\newcommand{\be}{\begin{equation}}
\newcommand{\ee}{\end{equation}}
\renewcommand{\tilde}{\widetilde}
\renewcommand{\i}{\mathrm{i}}
\renewcommand{\d}{\mathrm{d}}

\newtheorem{conjecture}{Conjecture}

\numberwithin{equation}{section}

\def\Tr{\text{Tr}}

\usepackage{diagbox} 
\usepackage{multirow} 
\usepackage{xcolor}
\definecolor{dgreen}{rgb}{0.0, 0.5, 0.0}
\definecolor{dred}{rgb}{0.5, 0.0 , 0.0}
\usepackage{slashed}
 
\usepackage[compress]{cite}

\usepackage{braket}

	\usepackage{tikz}
	\usepackage{makecell}

\def\ie{\begin{equation}\begin{aligned}}
\def\fe{\end{aligned}\end{equation}}
\newcommand{\cH}{\mathcal H}

\newtheorem{defn}{Definition}

\begin{document}
\thispagestyle{empty}

\vspace*{2.5cm}
\begin{center}

{\bf {\LARGE Fortuity in SYK Models}}

\begin{center}

\vspace{1cm}

 {\bf Chi-Ming Chang$^{1,2}$, Yiming Chen$^3$, Bik Soon Sia$^{1,4}$, Zhenbin Yang$^{5}$}\\
  \bigskip \rm

\bigskip 
${}^1$Yau Mathematical Sciences Center (YMSC), Tsinghua University, Beijing, China

${}^2$Beijing Institute of Mathematical Sciences and Applications (BIMSA), 
Beijing, China

${}^3$Stanford Institute for Theoretical Physics, Stanford University, Stanford, CA, USA

${}^4$Department of Mathematics, Tsinghua University, Beijing, China

${}^5$Institute for Advanced Study, Tsinghua University, Beijing, China

\rm
  \end{center}

\vspace{2cm}
{\bf Abstract}
\end{center}
\begin{quotation}
\noindent

We study the fortuity phenomenon in supersymmetric Sachdev-Ye-Kitaev (SYK) models. For generic choices of couplings, all the BPS states in the $\mathcal{N}=2$ SUSY SYK model are fortuitous. The SYK models reveal an intimate connection between fortuity and the Schwarzian description of supersymmetric black holes, reflected in a sharp feature of $R$-charge concentration - microscopically, all the fortuitous states are concentrated in particular charge sectors. We propose that both $R$-charge concentration and the random matrix behavior near the BPS states are key properties of a generic $q$-local supercharge and formulate these as a supercharge chaos conjecture. We expect supercharge chaos to hold universally for supercharges in holographic CFTs near their fortuitous states, potentially providing a microscopic interpretation for the charge constraints of supersymmetric black holes.

We also construct SYK models that contain both fortuitous states and monotonous states and contrast their properties, providing further evidence that monotonous states are less chaotic than fortuitous states.

\end{quotation}

\setcounter{page}{0}
\setcounter{tocdepth}{2}
\setcounter{footnote}{0}
\newpage

\setcounter{page}{2}

\tableofcontents

\section{Introduction}\label{sec:introduction}

Research in holography has uncovered many remarkable quantum properties of black holes, such as thermalization, hydrodynamics, maximal chaos, Page curve, and random matrix behavior \cite{Horowitz:1999jd,Kovtun:2003wp,Sekino:2008he,Shenker:2013pqa,Maldacena:2015waa,Hayden:2007cs,Penington:2019npb,Almheiri:2019psf,Almheiri:2019hni,Penington:2019kki,Almheiri:2019qdq,Cotler:2016fpe,Saad:2018bqo,Saad:2019lba,stanford2019jt}. 
Recently, a new property called ``fortuity" was discovered, particularly for supersymmetric (BPS) black holes \cite{Chang:2024zqi}. Fortuity distinguishes typical microstates of black holes from those of horizonless geometries, through a relation between BPS states in the dual theories of different numbers of degrees of freedom, quantified by $N$. The microstates associated with horizonless geometries can be extrapolated to infinite $N$, where they disassemble into non-interacting gravitons, while preserving supersymmetry. In contrast, such a limit does not exist for typical BPS black hole microstates since the states will be lifted as we take $N$ large.

The relation is, more precisely, based on a one-to-one correspondence between BPS states and supercharge $Q$-cohomology classes \cite{Witten:1982df,Kinney:2005ej,Grant:2008sk}, along with the observation that in holographic conformal field theories (CFTs), the Hilbert space of the finite $N$ theory (as a vector space) can be realized as a quotient of the infinite $N$ Hilbert space by certain equivalence relations. These equivalence relations are, for example, the stringy exclusion principle in the D1-D5 CFTs \cite{Maldacena:1998bw} and trace relations in large $N$ gauge theories. In this setup, the $Q$-cohomology classes are classified into two categories: {\it monotonous} cohomology classes, whose representatives remain $Q$-closed in the infinite $N$ theory, and {\it fortuitous} cohomology classes, where the $Q$-closedness of their representatives relies on the equivalence relations. Consequently, monotonous cohomology classes in small $N$ theories are simply given by imposing equivalence relations on their large $N$ counterparts, whereas such a relation does not exist for fortuitous cohomology classes.

Supercharge cohomology has been most extensively studied in the ${\cal N}=4$ super-Yang-Mills (SYM) theories. All monotonous cohomology classes were found in \cite{Chang:2013fba}, the first fortuitous cohomology class was discovered in \cite{Chang:2022mjp,Choi:2022caq}, and several infinite towers of fortuitous cohomology classes in the SU(2) and SU(3) theories were found in \cite{Choi:2023znd,Choi:2023vdm}. 
Using a criterion of BPS chaos proposed by LMRS \cite{Lin:2022rzw,Lin:2022zxd}, explicit computation suggests that monotonous states exhibit only weak chaos, while fortuitous states are conjectured to be strongly chaotic \cite{Chen:2024oqv}. Furthermore, the study of the dressing of fortuitous cohomology classes by monotonous cohomology classes reveals a partial ``no hair" phenomenon for quantum black holes \cite{Choi:2023znd,Choi:2023vdm}. However, most of these studies, particularly those involving fortuitous states, have so far been limited to small $N$ (or based on results from small $N$) due to the intractability of ${\cal N}=4$ SYM at large $N$. Developing a toy model that allows us to extend the study of fortuity to larger $N$ would be a significant progress.

A promising candidate for such a toy model is the Sachdev-Ye-Kitaev (SYK) model \cite{Sachdev:1992fk,Kitaev:2015aa}, which is solvable in the large $N$ limit \cite{Kitaev:2015aa,Polchinski:2016xgd,Maldacena:2016hyu,Gross:2017aos} and captures many key features of black holes. It exhibits maximal chaos and, at low temperatures, contains a Schwarzian sector that governs the near-horizon excitations of extremal black holes \cite{Kitaev:2015aa,Maldacena:2016hyu}. These properties make the SYK model (more precisely its supersymmetric cousins) an ideal testing ground for exploring the fortuity of black holes.

To build intuition, we start by studying the supercharge $Q$ cohomology in the standard ${\cal N}=2$ supersymmetric SYK model introduced in \cite{Fu:2016vas} (see Section~\ref{sec:sykfortuity}). This model consists of $N$ complex fermions with a $q$-local supercharge $Q$ constructed from these fermions, with coupling constants that are independent, identically distributed Gaussian random variables. It was found in \cite{Fu:2016vas} that all the $Q$-cohomology classes/BPS states are sharply concentrated around a fermion number $p\sim \frac N2$, and interestingly, there is no fluctuation across the ensemble either in the distribution of BPS states or their number. Furthermore, by refining the cohomology problem using a \({\mathbb Z}_q\) symmetry that commutes with the supercharge, it can be shown that for a cochain complex with a nonzero index, all BPS states are sharply concentrated on a single space in the complex. This phenomenon will be referred to as {\it $R$-charge concentration}, as schematically illustrated below:
\ie\label{eqn:cochain_complex}
    \begin{tikzcd}
    \cdots \arrow[r,"Q"]  & H^{n_c-1} \arrow[r,"Q"] & \color{red}{H^{n_c}} \arrow[r, "Q"] & H^{n_c+1} \arrow[r,"Q"]& \cdots\,,
    \end{tikzcd}
\fe
where all the BPS states are concentrated at the space $H^{n_c}$. The degree $n$ is related to the $R$-charge by $n=\lfloor R/q\rfloor$, when the supercharge $Q$ carries $q$ units of $R$-charge. The rigorous classification of cohomology class into monotonous and fortuitous will be given in Section~\ref{sec:defns}; however, we can already see that all concentrated cohomology classes are fortuitous, as their fermion numbers scale with $N$ in the large $N$ limit, precluding them from being monotonous cohomology classes.

$R$-charge concentration is a ``smoking-gun" signature of the super-Schwarzian theory \cite{Stanford:2017thb}. 
Super-Schwarzian has been proposed as a universal mode that dominates the low energy dynamics of near-BPS black holes \cite{Heydeman:2020hhw,Boruch:2022tno,Heydeman:2024ezi}. 
Therefore, from the gravity side, we expect $R$-charge concentration due to the universal AdS$_2$ region of BPS black holes and the dominance of the Schwarzian mode. Our emphasis in this paper will instead be on the understanding of this universality on the boundary side, where we attribute $R$-charge concentration to the genericity of the supercharge $Q$. In Section \ref{sec:Qchaos}, we further relate the genericity of supercharge with the $\mathcal{N}=2$ SUSY random matrix ensemble by Turiaci and Witten \cite{Turiaci:2023wrh}. We summarize the connection between these ideas into Conjecture \ref{conj:Qchaos} (Supercharge Chaos).

Besides generic supercharges that exhibit $R$-charge concentration, the flexibility of SYK models also allows us to explore less generic supercharges where the concentration breaks down. If we prune the random couplings $C_{ijk}$ randomly, it turns out that the concentration is highly robust - one needs to make the couplings $C_{ijk}$ highly sparse in order for it to break down. We explore this numerically in Section \ref{sec:sparse}. On the other hand, one can orchestrate the couplings $C_{ijk}$ to enjoy some additional structure, such that apart from the fortuitous states that concentrate, one also gets additional monotonous states. In Section \ref{sec:twoflavor}, by generalizing a two-flavor model of Heydeman, Turiaci, Zhao \cite{Heydeman:2022lse}, which generically only contains fortuitous states,\footnote{There are some exceptions to this statement as we will discuss in Section \ref{sec:twoflavorreview}.} we find new families of models that also contain monotonous states, which are analogous to the BPS graviton operators in $\mathcal{N}=4$ SYM. In these toy models, we can study the fine-grained properties of the fortuitous and monotonous states in detail. We find that the fortuitous states exhibit much stronger chaos compared to the monotonous states, in accordance with the conjecture in \cite{Chen:2024oqv}.

A nice way of visualizing the distinctions between fortuitous and monotonous states is to treat $N$ as a continuous variable and trace the energies of the states as functions of $N$. Using our toy models, we flesh out this picture of following $N$. Compared to the previous discussion in the $\mathcal{N} = 4$ SYM case \cite{Budzik:2023vtr}, where a single fortuitous state is followed to enter the BPS subspace,\footnote{We will often use the term ``fortuitous state" loosely, to refer to not only the actual BPS state but also its lift in theories with larger $N$ where it is not BPS.} here we get to witness an exponential (in $N$) number of states entering the BPS subspace at the same value of $N$. Furthermore, we will be able to track the detailed properties of individual states during this process, demonstrating a ``chaos invasion" picture proposed in \cite{Chen:2024oqv}.

The rest of the paper is organized as follows. 

In \textbf{Section \ref{sec:sykfortuity}}, to help the readers quickly understand the main idea, we give a brief overview of the notion of fortuity in the $\mathcal{N}=2$ SUSY SYK model. 

In \textbf{Section \ref{sec:Qchaos}}, we generalize the lesson from SYK models and propose the universal properties of a generic supercharge as Conjecture \ref{conj:Qchaos} (Supercharge Chaos). 

In \textbf{Section \ref{sec:defns}}, we give a more detailed and mathematically precise exposition of fortuitous and monotonous BPS states in the context of SYK models. Readers who are more interested in the physical results can choose to skip this section at first read. 

In \textbf{Section \ref{sec:twoflavor}}, we construct families of models that also contain monotonous states, analyze their large $N$ equations of motion, and numerically compare the chaos of the fortuitous states versus the monotonous states.

In \textbf{Section \ref{sec:followN}}, we study the problem of following $N$ in SYK models and illustrate the ``chaos invasion" picture.

In \textbf{Section \ref{sec:concentration}}, we illustrate the idea of $R$-charge concentration by studying a sparse version of $\mathcal{N}=2$ SUSY SYK model. We make some comments on the $\mathcal{N}=4$ SYM theory and the non-linear charge constraint for the SUSY black holes in AdS$_5$.

In \textbf{Section \ref{sec:discussion}}, we conclude by discussing some general lessons and open problems.

\subsection{Fortuity in the $\mathcal{N}=2$ SUSY SYK model}\label{sec:sykfortuity}

Let us consider the ${\cal N}=2$ supersymmetric SYK model, first introduced in \cite{Fu:2016vas} (see also \cite{Sannomiya:2016mnj}). The supercharge of the theory is constructed out of $N$ complex fermions $\psi_i$ for $i=1,\,2,\,\cdots,\,N$ as
\ie\label{eqn:SYK_supercharge}
Q=\sum_{i_1,...,i_q=1}^N C_{i_1i_2\cdots i_q}\psi_{i_1}\psi_{i_2}\cdots\psi_{i_q}\,.
\fe
Together with its Hermitian conjugate $Q^\dagger$ and the Hamiltonian $H$, the supercharge $Q$ satisfies the ${\cal N}=2$ supersymmetry algebra
\ie\label{eqn:N=2SUSY}
\{Q,Q^\dagger\}=H\,,\quad Q^2=0=Q^{\dagger2}\,.
\fe
The coupling constants $C_{i_1i_2\cdots i_q}$ are usually taken to be independent Gaussian random complex variables. 
However, in this paper, we do not consider performing ensemble average of the random variables unless otherwise noted. Note that since all the fermions anticommute, $C_{i_1i_2\cdots i_q}$ can be chosen to be a totally anti-symmetric tensor, namely a $q$-form.

We study the $Q$-cohomology classes, which are in one-to-one correspondence with the BPS states. The $Q$-cohomology is graded by the fermion number\footnote{In the paper, we use the notation that $\psi$ denotes fermion creation operator and $\bar{\psi}$ $(=\psi^\dagger)$ denotes annihilation operator. We will \emph{not} use the notation $\psi^\dagger$. } 
\ie
{\bf N}_\psi = \sum_{i=1}^N \psi_i\bar{\psi}_i\,.
\fe
We denote the Hilbert space of fermion number $p$ by ${\cal H}^p$. The fermion number also serves as the $R$-charge in this model, $R = {\bf N}_\psi$, satisfying
\begin{equation}\label{eqn:R-charge_N=2SUSY}
   [R,Q] = \hat{q} Q, \quad [R,Q^\dagger] = -\hat{q} Q^\dagger,\quad [R,H] = 0,\quad (-1)^{F} = e^{\pi i R}\,.
\end{equation}
We use $\hat{q}$ when we specifically refer to the $R$-charge of the supercharge $Q$, which can in general be different from $q$, i.e. the number of operators in $Q$, but for $\mathcal{N}=2$ SYK, $\hat{q} = q$.
The Hilbert space of the model can be constructed by starting with the vacuum state $\ket\Omega$ defined by $\bar{\psi}_i \ket\Omega=0$ for all $i$ and acting on it the fermion creation operators $\psi_i$. An arbitrary state $\ket{\alpha_p}$ with fermion number $p$ can be viewed as a differential form $\alpha_{i_1\cdots i_p}$ as
\ie
\ket{\alpha_p} = \frac1{p!}\sum_{i_1,\cdots,i_p=1}^N\alpha_{i_1\cdots i_p}\psi_{i_1}\cdots\psi_{i_p}\ket\Omega\,.
\fe
The Hilbert space ${\cal H}^p$ of fermion number $p$ is the space of constant holomorphic $p$-forms in ${\mathbb C}^N$, i.e. ${\cal H}^p\cong \Lambda^p({\mathbb C}^N)$, and we have
\ie
D_p \equiv \dim {\cal H}^p=\binom{N}{p}\,.
\fe

\begin{figure}[t]
\begin{center}
\includegraphics[width=0.3\textwidth]{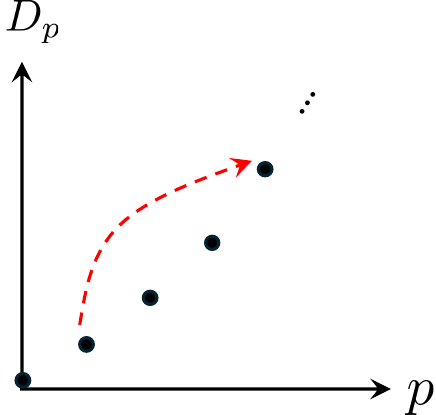}
\caption{If we fix $p$ and take $N\rightarrow\infty$, then generically the map from $\alpha_p$ to $C\wedge \alpha_p$ is full rank apart from the obvious kernel being the $Q$-exact states. In the drawing, we illustrate the case of $q=3$ and the red arrow denotes the $Q$-map from $\mathcal{H}^1$ to $\mathcal{H}^4$.}
\label{fig:nosoln}
\end{center}
\vspace{-1em}
\end{figure}

The supercharge \( Q \) maps a \( p \)-form \( \alpha_p \) to a \( (p+q) \)-form \( C \wedge \alpha_p \), as follows:  
\[
Q_p:~ \alpha_p \, \mapsto \, C \wedge \alpha_p\,,
\]  
where the subscript \( p \) indicates that \( Q_p \) acts specifically on \( p \)-forms.\footnote{In this formulation, the case $q=1$ coincides with the Koszul cohomology. Therefore, the $Q$-cohomology can be regarded as a natural generalization of Koszul cohomology to higher-degree forms.}
To search for $Q$-cohomology classes, one likes to find solutions to $C \wedge \alpha_p = 0$ that cannot be written as $\alpha_p = C \wedge \beta_{p-q}$. The $Q$-cohomology classes are in one-to-one correspondence with BPS states.
When $p\ll N$, with a generic $C$, we expect the map to be full rank apart from the obvious kernel being the $Q$-exact states. In other words, the only solution to the equation $C\wedge \alpha_p=0$ is $\alpha_p=C\wedge\beta_{p-q}$ for some $(p-q)$-form $\beta_{p-q}$. Therefore, there are no $Q$-cohomology classes for fixed $p$ and $N\rightarrow \infty$, as schematically shown in Figure~\ref{fig:nosoln}. This implies there are no monotonous cohomology classes.

\begin{figure}[t]
\begin{center}
\includegraphics[width=0.9\textwidth]{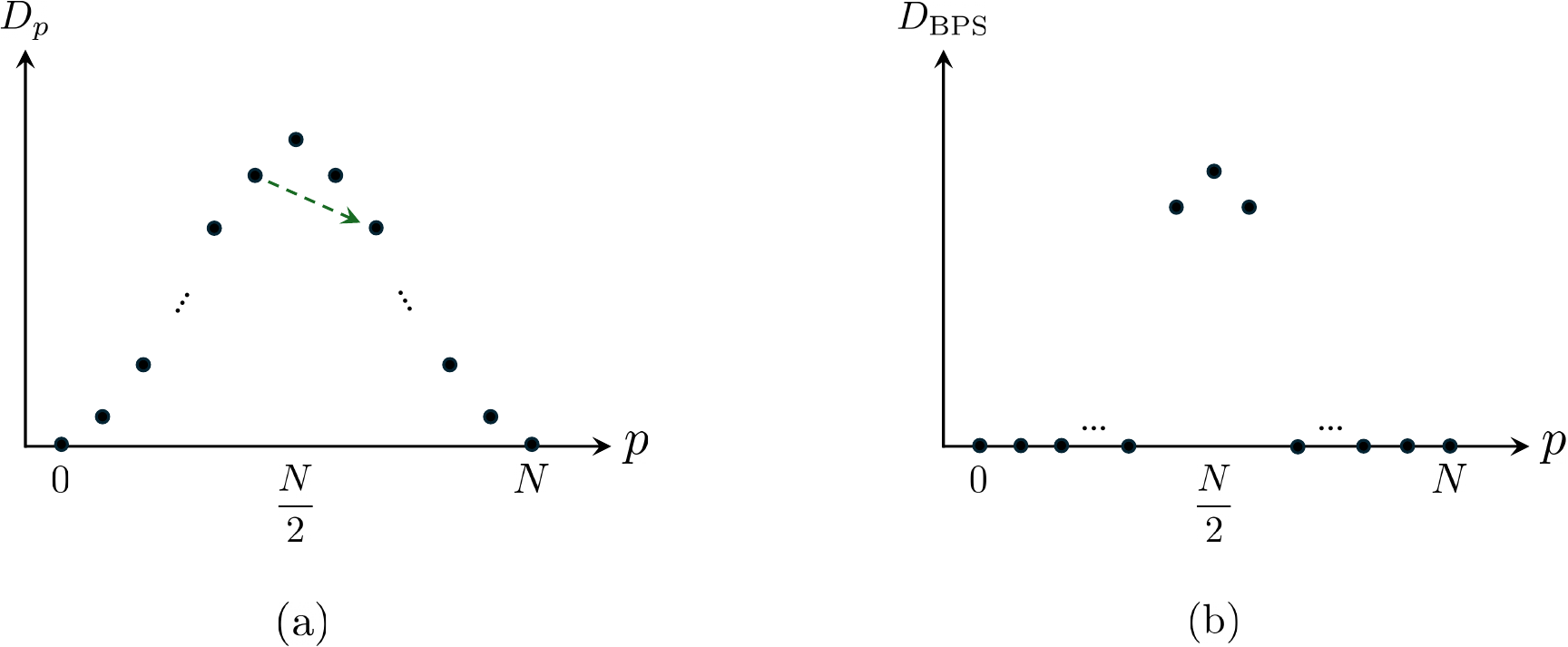}
\caption{(a) When $p \geq \frac{N}{2} - 1$, $Q \ket{\alpha_p} = 0$ has a large number of solutions since we are mapping from a larger space to a smaller one. (b) The BPS states are concentrated in the sectors $p = \frac{N}{2} , \frac{N}{2} \pm 1$. Here we consider $N$ even and $q = 3$.}
\label{fig:fortuitous}
\end{center}
\vspace{-1em}
\end{figure}

Now, consider finite $N$, even though the heuristic argument about the genericity of the map is only expected to hold when $p\ll N$, 
numerical study of explicit examples of $C$ drawn from a Gaussian ensemble suggests that we continue to have no $Q$-cohomology class/BPS state when $p< \frac{N- q}2$ \cite{Fu:2016vas}. Next, by the charge conjugation symmetry (exchanging $\psi_i$ and $\bar{\psi}_i$), there is also no $Q^\dagger$-cohomology class/BPS state when $p>\frac{N+q}2$. Hence, the fermion number $p$ of the BPS states are concentrated in a narrow window $\frac{N-q}2\leq p \leq \frac{N+q}2$. Schematic plots for the above properties are shown in Figure~\ref{fig:fortuitous}. 
This immediately suggests that all these BPS states are fortuitous, since if one fixes $p$ and takes $N$ to be large, these states will inevitably fall out of the range $\frac{N-q}2\leq p\leq \frac{N+q}2$ and will have to become non-BPS.

In \cite{Kanazawa:2017dpd,Berkooz:2020xne}, with the same assumption as above, the authors systematically accounted for the number of $Q$-exact (and $Q^\dagger$-exact) states in each charge sector and determined the number of BPS states. The final outcome indeed agrees with the (refined)-index of the model \cite{Fu:2016vas}.

Since $Q$ commutes with a $\mathbb{Z}_q$ charge $Q_f$, defined as $\textbf{N}_\psi$ modulo $q$, an irreducible cochain complex in the model contains only the states with a fixed $Q_f$. 
Along such a complex, only one space contains BPS states.\footnote{For some values of $N$ and $q$ (such as for odd $N$ when $q=3$), the cochain complex containing charge sectors $p = \frac{N \pm q}{2}$ could have an order one number of BPS states in both $p = \frac{N - q}{2}$ and $p = \frac{N + q}{2}$ spaces \cite{Fu:2016vas}. This case is special in that the index of the cochain complex is equal to zero. Such cases are excluded in our Conjecture \ref{conj:Qchaos} by demanding that the index is macroscopic.} Therefore, $\mathcal{N}=2$ SUSY SYK model is a simple example that satisfies the $R$-charge concentration phenomenon and further properties of supercharge chaos as we will discuss in Conjecture \ref{conj:Qchaos}.

Given an explicit example of $C$ found in numerics that exhibits $R$-charge concentration, we can prove that generic instances of $C$ should always satisfy $R$-charge concentration. 
Consider a basis \(\{e_1, e_2, \ldots, e_N\}\) for \({\mathbb R}^N\). In this basis, the map \( Q_p \) is represented by a \( D_p \times D_{p+q} \) matrix \( M_p(C) \). There is no \( Q \)-cohomology class with degree $p< \frac{N-q}{2}$ if and only if the rank of \( M_p(C) \) equal to \( \widetilde{D}_p \equiv \sum_{n=0}^{\lfloor p/q\rfloor } (-1)^n D_{p-n q} \). Suppose, by contradiction, that the rank of some of the matrix $M_p(C)$ is smaller than $\widetilde D_p$, then all \( \widetilde{D}_p \times \widetilde{D}_p \) minors of \( M_p(C) \) must vanish. This condition imposes a system of polynomial equations on the components of the \( q \)-form \( C \). Define the space  
\ie\label{eqn:polyconstaint}
S=\bigcup_{p<\frac{N-q}{2}} S_p\,,\quad S_p = \{C \in \Lambda^q({\mathbb R}^N) ~|~ \text{all \( \widetilde{D}_p \times \widetilde{D}_p \) minors of \( M_p(C) \) vanish}\}.
\fe
For \( p < \frac{N-q}{2}  \), numerical results yield a \( q \)-form \( C \) that gives rank-\( \widetilde{D}_p \) matrices \( M_p(C) \), i.e. a \( q \)-form \( C \notin S\)\,, namely for each $p$ at least some of the polynomial equations are nontrivial.
Therefore, \( S \) is a measure zero subset of $\Lambda^q({\mathbb R}^N)$, and generic $q$-forms $C$ must satisfy
\ie
\rank M_p(C)\ge \widetilde D_p\quad{\rm for}\quad p<\frac{N-q}{2} \,.
\fe
It is not hard to see that these bounds are saturated. Hence, there is no $Q$-cohomology class with degree $p< \frac{N-q}{2}$. By the charge conjugation symmetry and the isomorphisms between the $Q$-cohomology, the space of BPS states, and the $Q^\dagger$ cohomology,
we know that there is also no $Q$-cohomology class with degree $p>\frac{N+q}2$.

Finally, a consistency check for the $R$-charge concentration in this model comes from the large $N$ analysis of the Schwinger-Dyson equations \cite{Fu:2016vas,Heydeman:2022lse}. The equations admit conformal solutions for a large region of the $R$-charge, but the superconformal solutions exist only at $R\approx N/2$. We emphasize that even though the large $N$ analysis can only determine the concentrated $R$-charge $R_c$ approximately, microscopically the concentration is sharp.

\subsection{Supercharge chaos}
\label{sec:Qchaos}

A notable feature of the supercharge in the SYK model is its inherent \emph{genericity}: the number of independent parameters scales with 
$N^q$, aligning with the scaling behavior of generic $q$-local operators. For example, in fermion systems, three-local supercharges satisfying the ${\cal N}=2$ supersymmetric \eqref{eqn:N=2SUSY} and with an $R$-charge as defined in \eqref{eqn:R-charge_N=2SUSY} can only take the forms given in \eqref{eqn:SYK_supercharge} for $q=3$, or
\ie\label{eqn:BRST_Q}
Q'=\sum_{i,j,k=1}^N f^k_{ij}\psi^i\psi^j\bar\psi_k+\sum_{i=1}^Nc_i\psi^i\,.
\fe
Note that $Q'$ has $q=3$ but its $R$-charge $\hat{q} = 1$. The solution space of the nilpotency condition $Q'^2=0$ has dimensions that grow much slower than $N^q$.\footnote{If the space of supercharges comprises multiple connected components, only the supercharges within the component of maximum dimension are considered generic.}   Hence, the supercharge \eqref{eqn:SYK_supercharge} of the $q=3$, ${\cal N}=2$ SYK model is the most generic three-local supercharge. We expect similar argument to hold for larger $q>3$. This observation suggests that \( R \)-charge concentration is an intrinsic property of generic \( q \)-local supercharges.

From the bulk perspective, this genericity is tied to the universal structure of the near-horizon AdS\(_2\) region in BPS black holes and the domination of the Schwarzian mode, as well as to the inherent genericity of black hole systems themselves, being the geometries carrying most entropy. With the enhancement to \( \mathcal{N}=2 \) supersymmetry, the super-Schwarzian effective theory indeed predicts the \( R \)-charge concentration phenomenon in a precise manner \cite{Stanford:2017thb,Mertens:2017mtv}.

We emphasize that the super-Schwarzian description of BPS black hole \cite{Heydeman:2020hhw,Boruch:2022tno} is very robust. For instance, one could ask about corrections to the Schwarzian in the presence of low-tension strings, which is relevant for both $\mathcal{N}=4$ SYM at weak coupling and the putative bulk dual of the SYK model. Nevertheless, for the BPS states, the low-energy enhancement of the soft mode in JT gravity ensures that these additional stringy corrections do not interfere with the dominant Schwarzian effect.\footnote{For instance, the correction of the Lyapunov exponent goes to zero at extremality \cite{Maldacena:2016hyu}.} This suggests that the Schwarzian dominance is a universal feature, robust even in the limit of large \( \alpha' \) corrections,\footnote{An exception exists in rare cases where light particles with conformal dimensions between 1 and \( \frac{3}{2} \) appear \cite{Maldacena:2016upp}.} implying the validity of this description beyond the semiclassical gravity approximation and into the weakly coupled regime.
Another possible deformation is to deform the dilaton potential by adding supersymmetric defects. Again, since the dilaton approaches a constant near extremality, such deformations will only shift the extremal entropy $S_0$ and do not modify the distribution of BPS states, as observed by \cite{Turiaci:2023jfa}.

The universal bulk picture suggests that the Schwarzian sector exists universally in the dual boundary theories. In the non-supersymmetric setting, it has been understood that a Schwarzian sector universally exists in large \( c \) two-dimensional CFTs at large angular momentum, even when it's non-holographic \cite{Ghosh:2019rcj}. 
In the supersymmetric case, the bulk picture points toward the universal presence of the \( \mathcal{N}=2 \) supersymmetric Schwarzian theory in generic BPS sectors. 
The natural question arises: what boundary universality class accounts for the appearance of Schwarzian? Here, the most compelling candidate appears to be random matrix universality.

The precise notion of such a random matrix theory has been provided recently by Turiaci and Witten \cite{Turiaci:2023jfa}. They proposed a class of random matrix ensembles with $\mathcal{N}=2$ supersymmetry as an exact dual to \( \mathcal{N}=2 \) JT supergravity, thus extending the established JT/random matrix duality \cite{Saad:2019lba}. 
Microscopically, it is natural to expect that a generic $q$-local complex supercharge will be well approximated by the Turiaci-Witten random matrix ensemble. 

In holographic theories with precise bulk dual, despite the fact that the supercharge is usually non-generic due to extra symmetries and locality constraints, we would still expect that in proximity to its fortuitous BPS states, the supercharge should exhibit the same universal properties as a generic supercharge, 
in the same spirit of the usual random matrix universality.\footnote{We thank Douglas Stanford for insightful discussions on this topic.} This motivates us to formulate sharply the expected features of a generic supercharge near its BPS states. In analogy to the usual Hamiltonian chaos, we term the collection of these features as ``supercharge chaos", described in the following conjecture:  
\begin{conjecture}[Supercharge Chaos]\label{conj:Qchaos}

Let $Q$ be a generic \( q \)-local complex supercharge acting on a finite \( L \)-dimensional Hilbert space, satisfying the \({\cal N}=2\) supersymmetry algebra \eqref{eqn:N=2SUSY} with an $R$-symmetry \eqref{eqn:R-charge_N=2SUSY}. In an irreducible cochain complex of differential $Q$ with a macroscopic index of order \( L^\nu \) (where \( 0 < \nu \leq 1 \)) in the large \( L \) limit, the BPS states are sharply concentrated on a single space with a specific \( R \)-charge. Near the BPS states, the supercharge $Q$ is approximated by the random matrix ensemble of Turiaci-Witten, which is dual to the $\mathcal{N}=2$ JT gravity.

\end{conjecture} 

\noindent By an irreducible cochain complex, we refer to the complex that cannot be further refined by flavor symmetries that commute with the supercharge $Q$. By a generic supercharge, we refer to those that form a dense open subset of all supercharges satisfying \eqref{eqn:N=2SUSY} and \eqref{eqn:R-charge_N=2SUSY}. If the space of supercharges is equipped with a measure, the non-generic supercharges constitute a measure-zero subset. In other words, a generic supercharge retains its genericity under arbitrary deformations of $q$-local operators. 
By being approximated by the Turiaci-Witten random matrix ensemble, 
we mean that the spectrum of long multiplets in adjacent spaces $(r,r+1)$ that have order one energies above the BPS bound should be in the Altland-Zirnbauer class with parameters $(\alpha, \beta) = (1 + 2n_{\textrm{BPS}}, 2)$, where $n_{\textrm{BPS}}$ is the number of BPS states in the space $r$ or $r+1$ \cite{Turiaci:2023jfa}.  Furthermore, the BPS states should resemble a random subspace with respect to simple bases in the Hilbert space, and display the notion of strong chaos \cite{Chen:2024oqv}. A direct corollary of this conjecture is that the index is equal to the number of BPS states (up to a sign) as there will be no cancellations.

Given that $R$-charge concentration holds, one can conveniently determine the value of the $R$-charge $R_c$ at which concentration occurs through the supersymmetric index, which contains a sum over rapidly oscillating terms $e^{i\pi R}$. More precisely, what one can usually extract from the index is the large $N$ scaling of $R_c$, i.e. $r_c =R_c /N^{\#}$ where $\#$ depends on the specific model. This is done by a large $N$ saddle point approximation for the index\footnote{This is a special (infinite temperature) case of a more general computation $I = \textrm{Tr}[e^{-\beta H} e^{\beta \mu R}]$ with $\beta \mu = \pi \i$. The large $N$ saddle point approximation gives a one-parameter family of $r_* (\beta)$. In the $\beta\rightarrow \infty$ limit we have $r_* \rightarrow r_c$.}
\begin{equation}\label{index}
    I = \textrm{Tr}\,(-1)^R = \sum_{R} (-1)^R D_R \approx \int \d r \, e^{\i\pi  N^{\#}r + N^{\#} S(r)} \approx e^{\i\pi N^{\#} r_* + N^{\#} S(r_*)}, \quad \i\pi + S'(r_*) = 0\,.
\end{equation}
Matching the expectation from $R$-charge concentration $I = D_{\textrm{BPS},R_c} e^{\i\pi r_c N^{\#}}$, one finds
\begin{equation}\label{rc}
    r_c = \textrm{Im} \left[ \i r_* + \frac{1}{\pi} S(r_*)\right]\,.
\end{equation}
The calculation (\ref{index}) is not new, what is new is the interpretation of (\ref{rc}). Instead of interpreting $R_c$ as a macroscopic charge - a rough value around which potentially many sectors provide dominating contribution to the index - we are saying that microscopically, there is a \emph{single} charge sector $R=R_c$ (a single space) that contains BPS states along a cochain complex. 
In Section \ref{sec:N=4}, we will discuss the possible implication on the previous observation of the nonlinear charge constraint from the index calculation \cite{Cabo-Bizet:2018ehj,Choi:2018hmj,Benini:2018ywd}.

As an example, let us demonstrate this method by (re)determining $r_c=R_c/N$ of the ${\cal N}=2$ SYK model from the Witten index. 
Since the Witten index of the model vanishes, we can consider the refined index where we turn on a quantized chemical potential $\mu = \frac{2\pi k}{q}, \, k = 0, ..., q-1$ for the $\mathbb{Z}_q$ charge $Q_f$.  We have
\begin{equation}
   I(\mu) = \textrm{Tr} \left[(-1)^R e^{\i \mu R}\right] \approx \int_0^1 \d r\, e^{N \left[\i \pi r + \i \mu r + (r-1)\log (1-r) - r \log r\right]}\,.
\end{equation}
In the large $N$ limit, the saddle point of the integral is located at $r_* = 1 + 1/(e^{\i \mu}-1)$. At the saddle point, we find
\begin{equation}
   \textrm{Im}\left(\log  I(\mu) \right) \approx N \left(\i  \pi \frac{1}{2} + \i \mu \frac{1}{2}\right),
\end{equation}
therefore suggests that the BPS states are concentrated at $r_c =\frac{1}{2}$, regardless of the chemical potential $\mu$. We can get the index of a cochain complex with fixed $Q_f$ by taking simple linear combinations of these refined indices.\footnote{Consider the case with $q=3$ and even $N$, the Witten index for each cochain complex can be easily obtained through a discrete Fourier transformation over the refined Witten index $I_{Q_f}={1\over 3}\sum_{k=0}^3 I(2\pi k/3) e^{-\i 2\pi k Q_f/3}$. The exact answers are given by
$ I_{0}=2\times 3^{N/2-1}\cos{N\pi\over 6}, I_1=-2 \times 3^{N/2-1}\sin{(N+1)\pi\over 6},I_2=2\times 3^{N/2-1}\sin{(N-1)\pi\over 6}.$
 These expressions, when viewed as analytic functions in $N$, might naively lead one to conclude the location of the $R$-charge scales with $N/6$, which is incorrect. This mistake is due to that as $N$ increases, $Q_f$ of the cochain complex also shifts. For example, if we focus on the cochain complex that contains the maximum number of BPS states, its $Q_f$ is equal to $N/2$ mod $3$. Taking this shift into account, we find the index oscillates as $(-1)^{N\over 2}$, agreeing with the correct $R$-charge scaling of the location of BPS states.   } This implies that within a cochain complex with fixed $Q_f$, the BPS states are concentrated in a \emph{single} space with $R \approx N/2$ in the large $N$ limit, consistent with the observations in Section~\ref{sec:sykfortuity}.

The fact that the concentrated charge $R_c$ scales as $N^\#$ is consistent with the fact that these states are fortuitous. To illustrate this, we can start with $N=N_*$ and consider the BPS states concentrated with charge $R=R_c(N_*)$. If we keep $R = R_c(N_*)$ fixed and start increasing $N$, we will soon have $R \neq R_c (N)$ when $N-N_* \sim \mathcal{O}(1)$. Therefore, in a theory with slightly larger $N$, the space with the same $R$ charge no longer contains BPS states, thus the previous states must disappear by being lifted into non-BPS, and therefore are fortuitous.  The same reasoning can also be applied to decreasing $N$ from $N_*$. There the BPS states disappear by becoming null states due to equivalence relations.

Notice that, based on our Conjecture \ref{conj:Qchaos}, a generic supercharge should not have monotonous states that usually do not concentrate. This means, for instance, the supercharge of $\mathcal{N}=4$ SYM is not fully generic. In this context, the existence of monotonous states also led to other SUSY black hole solutions dressed by rotating gravitons or dual giant gravitons far outside the horizon, which can be in the same cochain complex as the ``pure" black hole solution \cite{Minwallatalk,Choi:2025lck}. This would naively suggest that $R$-charge concentration, in the most literal sense, fails in $\mathcal{N}=4$ SYM. However, it is likely that the pure black hole solutions exhibit partial no-hair properties \cite{Choi:2023vdm}.  This could imply that locally near the sector containing the pure black hole solution, we still have $R$-charge concentration for fortuitous operators, and only when we are far enough from the pure black hole in the cochain complex do we start to encounter solutions such as the SUSY grey galaxies.\footnote{Here we are excluding ``revolving black holes", which are simply pure black hole solutions that are revolving in AdS space. We thank Shiraz Minwalla for discussion on this.} 
We elaborate further on this scenario in Section \ref{sec:N=4}.

We emphasize also that due to the dominance of Schwarzian even in the presence of large stringy corrections, the question of whether the supercharge in a holographic CFT  behaves as a generic supercharge or not does not require taking the strong coupling limit. Thus, it implies even a weakly coupled theory should exhibit the supercharge chaos behavior.  In other words, if indeed that appropriate notion of supercharge chaos holds in $\mathcal{N}=4$ SYM near its fortuitous states in the large $N$ limit, we expect that the classical supercharge should already exhibit this property. 
A consequence of this conjecture would be that the fortuitous BPS spectra should not flow with the coupling constant since they should all represent the same random matrix ensemble, which is consistent with the conjectured one-loop exactness of the $Q$-cohomology \cite{Grant:2008sk}.

\section{Fortuitous and monotonous BPS states in SYK}
\label{sec:defns}

In this section, following \cite{Chang:2024zqi}, we use a long exact sequence to provide precise definitions of the fortuitous and monotonous cohomology classes, as well as BPS states, in the context of SYK models. Our approach, however, differs slightly from that in \cite{Chang:2024zqi}, which would require defining an infinite $N$ cohomology. Defining an infinite $N$ Hilbert space and a supercharge $Q$ acting on it presents several challenges.\footnote{The basic difficulties of defining an infinite $N$ limit are tied to the fact that the bulk dual of SYK models resemble theories that contain order $N$ number of light fields, as opposed to a sparse light spectrum one expects for a conventional holographic system. Therefore we expect this to be not merely a technical difficulty, see Section \ref{sec:discussion} for more discussion.} Consequently, instead of directly relating the monotonous cohomology to the infinite $N$ cohomology, we will relate it progressively to monotonous cohomologies at successively larger $N$.

Let us start with a sequence of fermion systems
\ie\label{eqn:seqT}
{\cal T}_1\,,\quad {\cal T}_2\,,\quad {\cal T}_3\,,\quad\cdots\,,
\fe
where the theory ${\cal T}_N$ contains $MN$ complex fermions $\psi_i$ for some positive integer $M$ (for regular $\mathcal{N}=2$ SYK, $M=1$). The Hilbert space of the theory ${\cal T}_N$ is a tensor product
\ie
{\cal H}_{N}=\bigotimes_{i=1}^{MN}V^{[i]}\,.
\fe
where $V^{[i]}\cong {\mathbb C}^2 = {\rm span}(\ket 0,\ket 1)$ with $\bar{\psi}_i\ket 0=0$ and $\ket 1=\psi_i\ket 0$ is the Hilbert space for the $i$-th fermion $\psi_i$. The theory ${\cal T}_N$ is supersymmetric with the supercharge
\ie\label{eqn:M_flavor_Q}
Q_N=\sum_{i_1,i_2,\cdots,i_{q}=1}^{MN}C^{N}_{i_1i_2\cdots i_{q}}\psi_{i_1}\psi_{i_2}\cdots \psi_{i_{{q}}}\,,
\fe
for $q$ odd and the coefficients $C^N_{i_1,i_2,\cdots,i_{q}}$ are randomly chosen but with some constraints that will be specified later in different examples. The Hamiltonian of the theory ${\cal T}_N$ is given by the anticommutator
\ie
H_N = \{Q_N,Q^\dagger_N\}\,.
\fe
The supercharges in the sequence are related by
\ie
C^{N+1}_{i_1\cdots i_{q}}\big|_{i_1\cdots i_{q}\le MN} = C^{N}_{i_1,\cdots,i_{q}}\,,
\fe
such that the theories ${\cal T}_N$ with larger $N$ are extensions of the theories with smaller $N$.

The theory has ${\rm U}(1)$ fermion number symmetry. We decompose the fermion number operator ${\bf N}_\psi$ as 
\ie\label{eqn:f_num_sym}
{\bf N}_\psi = \sum_{i=1}^{MN}\psi_{i}\bar\psi_{i} = n q + q_f\,,
\fe
where $n\in{\mathbb Z}_{\ge0}$ is the degree for the cochain complex that we introduce presently, and $q_f$ is a ${\mathbb Z}_q$ charge that commutes with the supercharge $Q$. The Hilbert space ${\cal H}_N$ can be written as a direct sum
\ie{\cal H}_N=\bigoplus^{q-1}_{q_f=0}{\cal H}_{N,q_f}\,,\quad {\cal H}_{N,q_f}=\bigoplus^{\lfloor (MN-q_f)/q \rfloor }_{n=0} {\cal H}_{N,q_f}^n\,.
\fe
For a fixed $q_f$ charge sector, we have a cochain complex ${\cal H}^{\bullet}_{N,q_f}$ of the supercharge $Q_N$.
We define the supercharge $Q_N$-cohomology $H^n({\cal H}_{N,q_f})$ as 
\ie
H^n({\cal H}_{N,q_f}) = \frac{\big\{\ket\Psi\big|\ket\Psi\in{\cal H}^n_{N,q_f},\, Q\ket\Psi=0\big\}}{\big\{Q\ket\Psi\big|\ket\Psi\in{\cal H}^{n-1}_{N,q_f}\big\}}\,,\quad  H^{\bullet}({\cal H}_{N,q_f})=\bigoplus_{n=0}^{\lfloor (MN-q_f)/q \rfloor }H^n({\cal H}_{N,q_f})\,.
\fe
By the standard Hodge theory argument, there is a one-to-one correspondence between the cohomology class of the supercharge $Q$ and the BPS states (the ground states) $\ket \Psi$ satisfying $H\ket \Psi=0$ or equivalently $Q\ket\Psi=0=Q^\dagger\ket\Psi$.

We are interested in relating fermion systems ${\cal T}_N$ with different $N$ in the sequence \eqref{eqn:seqT}. The Hilbert spaces of theories ${\cal T}_N$ and ${\cal T}_{N+1}$ are related by projecting all the qubits $V^{[i]}$ with $i>MN$ to the vacuum state $\ket0$,
\ie\label{eqn:projection}
\pi{\cal H}_{(N+1)M} = {\cal H}_{N}\otimes \ket{\omega_{N+1}}\,,\quad\ket{\omega_{N+1}} = \bigotimes_{i=MN+1}^{M(N+1)} \ket0^{[i]}
\fe
where the projector $\pi$ is
\ie
\pi= \prod_{i=MN+1}^{M(N+1)}\bar\psi_{i}\psi_{i}\,.
\fe
The projection \eqref{eqn:projection} can be rewritten as a short exact sequence
    \ie\label{eqn:SES}
        \begin{tikzcd}
        0\arrow[r]  &  {\cal I}_{N+1}\arrow[hookrightarrow,r,"\iota"] & \cH_{N+1}\arrow[r,"\pi"] & \cH_{N}\otimes\ket{\omega_{N+1}} \arrow[r] & 0.
        \end{tikzcd}
    \fe
where $\iota$ is an inclusion map and the space ${\cal I}_{N+1}$ is the kernel of $\pi$, or more explicitly 
\ie
{\cal I}_{N+1}={\cal H}_{N}\otimes \ket{\omega_{N+1}}^{\perp}\,,
\fe
where $\ket{\omega_{N+1}}^{\perp}$ denotes the orthogonal complement of $\ket{\omega_{N+1}}$ inside $\bigotimes^{M(N+1)}_{i=MN+1}V^{[i]}$. Let us decompose the supercharge $Q_{N+1}$ in the $(N+1)$-th theory ${\cal T}_{N+1}$ as
\ie\label{eqn:Q_decomp}
Q_{N+1} = Q_{N}+\delta_N\,,
\fe
where the operator $\delta_N$ contains all the terms in $Q_{N+1}$ that involves fermions $\psi_i$ with $i>MN$. It is easy to see that 
\ie\label{eqn:pi_Q_comm}
\pi Q_{N+1} = Q_{N}\pi\,,\quad Q_{N+1} \iota=\iota Q_{N+1}\,, \quad\{Q_N,\delta_N\}=0\,.
\fe
Hence, we could take the $Q$-cohomology of the short exact sequence \eqref{eqn:SES}, and obtain the long exact sequence
\ie\label{eqn:LES}
    \begin{tikzcd}
    \cdots \arrow[r]  & H^n( {\cal I}_{N+1}) \arrow[r,"\iota_*"] & H^n(\cH_{N+1}) \arrow[r, "\pi_*"] & H^n(\cH
    _{N}) \arrow[r, "\delta_N"] & H^{n+1}( {\cal I}_{N+1}) \arrow[r]& \cdots\,.
    \end{tikzcd}
\fe
where we suppressed the subscript $q_f$ of the ${\mathbb Z}_{q}$ charge, which plays no role in the following discussion. $\iota_*$ and $\pi_*$ are the maps induced from $\iota$ and $\pi$.

Note that we use the notation $\delta_N$ for both the map $\delta_N:H^n({\cal H}_N)\to H^{n+1}({\cal I}_{N+1})$ in the long exact sequence \eqref{eqn:LES} and the operator $\delta_N$ in \eqref{eqn:Q_decomp}, because the map is induced by the operator.
To see this, we first note that $\delta_N$ defines a map $\delta_N:{\cal H}_N\otimes \ket{\omega_{N+1}}\to {\cal I}_{N+1}$, since for any state $\ket{\Psi_N}\in {\cal H}_N$, the state $\delta_N\ket{\Psi_N}\otimes \ket{\omega_{N+1}}$ is inside the space ${\cal I}_{N+1}$. Next, let us consider a state $\ket{\Psi_{N+1}}\in {\cal H}_{N+1}$, which can be decomposed as
\ie\label{eqn:PsiN2PsiN-1}
\ket{\Psi_{N+1}} = \ket{\Psi_{N}}\otimes \ket {\omega_{N+1}}+ \ket{\alpha_{N+1}}\quad {\rm for}~\ket{\alpha_{N+1}}\in {\cal I}_{N+1}\,.
\fe
$Q_{N+1}$ acts on $\ket{\Psi_{N+1}}$ as
\ie\label{eqn:QN+1_act}
Q_{N+1}\ket{\Psi_{N+1}} &=(Q_{N}+\delta_N)(\ket{\Psi_{N}}\otimes \ket{\omega_{N+1}}+ \ket{\alpha_{N+1}})
\\
&=Q_{N}\ket{\Psi_{N}}\otimes \ket{\omega_{N+1}} + (Q_N+\delta_N)\ket{\alpha_{N+1}}+\delta_N\ket{\Psi_{N}}\otimes \ket{\omega_{N+1}}\,.
\fe
Now, if the state $\ket{\Psi_{N+1}}$ represents a cohomology class $[\Psi_{N+1}]\in H^{\bullet}({\cal H}_{N+1})$, we have
\ie\label{eqn:QN_PsiN}
Q_{N}\ket{\Psi_{N}}=0\,,\quad Q_{N+1}\ket{\alpha_{N}}+\delta_N\ket{\Psi_{N}}\otimes \ket{\omega_{N+1}}=0\,.
\fe
By the first equation in \eqref{eqn:QN_PsiN}, the state $\ket{\Psi_{N}}$ represents a cohomology class $[\Psi_{N}]\in H^{\bullet}({\cal H}_{N})$, and we have established the map $\pi_*$ in \eqref{eqn:LES}. This map might not be subjective, because the state $\ket{\Psi_{N}}$ is not only closed but also needs to satisfy the second equation in \eqref{eqn:QN_PsiN} stating that $\delta_N\ket{\Psi_{N}}\otimes \ket\omega$ is $Q_{N+1}$-exact. Hence, we have established
\ie
{\rm im}\,(\pi_*) = {\rm ker}\,(\delta_N)\,,
\fe
and the equivalence between the $\delta_N$'s in \eqref{eqn:Q_decomp} and \eqref{eqn:LES}. 

The equations \eqref{eqn:PsiN2PsiN-1} and \eqref{eqn:QN_PsiN} provide a way to uplift cohomology classes in \( H^{\bullet}({\cal H}_{N}) \) to cohomology classes in \( H^{\bullet}({\cal H}_{N+1}) \). More precisely, given a cohomology class \( [\Psi_N] \in H^{\bullet}({\cal H}_{N}) \), if there exists a state \( \ket{\alpha_N} \) satisfying the second equation in \eqref{eqn:QN_PsiN}, we can construct a cohomology class \( [\Psi_{N+1}] \in H^{\bullet}({\cal H}_{N+1}) \), represented by \( \ket{\Psi_{N+1}} \) in \eqref{eqn:PsiN2PsiN-1}. In the long exact sequence \eqref{eqn:LES}, \( [\Psi_{N+1}] \) serves as a lift of \( [\Psi_N] \) via the map \( \pi_* \), i.e. \( \pi_*[\Psi_{N+1}] = [\Psi_N] \), which ensures that \( [\Psi_{N+1}] \) is nontrivial if \( [\Psi_N] \) is nontrivial.

In Section~\ref{sec:twoflavor}, we will encounter situations where the projection map \( \pi \) does not commute with the supercharge \( Q \). More precisely, the first equation in \eqref{eqn:pi_Q_comm} is not satisfied, see equation (\ref{pitwof}). Hence, there is no induced map \( \pi_* \) and the long exact sequence \eqref{eqn:LES}. However, the lifting procedure given by \eqref{eqn:PsiN2PsiN-1} and \eqref{eqn:QN_PsiN} remains valid. The only caveat is that the lift \( [\Psi_{N+1}] \) may turn out to be trivial. This motivates the following definition:

\begin{defn}[Monotonous cohomology]\label{def:monotone}
A cohomology class in $H^{\bullet}(\cH_{N})$ is called monotonous if it could be uplifted to a nontrivial cohomology class in $H^{\bullet}(\cH_{N'})$ for all positive integer $N'>N$ by iteratively applying \eqref{eqn:PsiN2PsiN-1} and \eqref{eqn:QN_PsiN}.

\end{defn}

\noindent By the relation \eqref{eqn:PsiN2PsiN-1}, the uplift $\ket{\Psi_{N+1}}$ has the same fermion number as the state $\ket{\Psi_{N}}$. This property would allow us later to quickly recognize which states must be fortuitous.

Let us consider the $Q^\dagger$-cohomology.
By the same Hodge theory argument, the $Q^\dagger$-cohomology classes correspond one-to-one to the BPS states and also to the $Q$-cohomology classes. Does the one-to-one correspondence preserve the notion of monotony? To address this, we define the lifting procedure for $Q^\dagger$-cohomology classes using the projection map $\tilde\pi$ and the state $\ket{\tilde\omega_N}$,
\ie
\tilde\pi = \prod_{i=MN+1}^{M(N+1)}\psi_{i}\bar\psi_{i}\,,\quad \ket{\tilde\omega_N}=\bigotimes_{i=MN+1}^{M(N+1)} \ket1^{[i]}\,.
\fe
The lifting rule for a $Q^\dagger$-cohomology class $[\Psi_{N}]$ is
\ie\label{eqn:QNdagger_PsiN}
&\ket{\tilde\Psi_{N+1}} = \ket{\Psi_{N}}\otimes \ket {\tilde\omega_{N+1}}+ \ket{\tilde\alpha_{N+1}}\quad {\rm for}~\ket{\tilde\alpha_{N+1}}\in \tilde{\cal I}_{N+1}\,,
\\
&Q_N^\dagger\ket{\Psi_N}=0\,,\quad Q_N^\dagger\ket{\tilde\alpha_N}+\delta_N^\dagger \ket{\Psi_N}\otimes \ket{\tilde\omega_{N+1}}=0\,,
\fe
where $\tilde {\cal I}_{N+1}$ denotes the kernel of the projector $\tilde \pi$. However, it is not guaranteed that, for a given $[\Phi_N]$, the final equations of \eqref{eqn:QN_PsiN} and \eqref{eqn:QNdagger_PsiN} can be solved simultaneously.
With this in mind, we define the monotonous and fortuitous states as follows.

\begin{defn}[Monotonous and fortuitous state]\label{def:monotone_state}
A BPS state is monotonous if it corresponds one-to-one to a monotonous $Q$-cohomology class or monotonous $Q^\dagger$-cohomology class. A BPS state that is not a monotonous state is called fortuitous.
\end{defn}

\section{Two-flavor models with monotonous states}\label{sec:twoflavor}

As we discussed in Section \ref{sec:sykfortuity}, generic $\mathcal{N}=2$ SUSY SYK only contains fortuitous states, which is a consequence of the supercharge chaos of the model. This differs the SYK model from other more intricate holographic supersymmetric field theories, such as the $\mathcal{N}=4$ SYM theory, which also contains monotonous states that describe light BPS gravitons or heavy horizonless geometries. 

Given the many advantages of the SYK model in its tractability, both analytically and numerically, and potentially experimental realization in the future, it is natural to seek a simple model that retains this solvability while exhibiting \emph{both} monotonous \emph{and} fortuitous states, analogous to what occurs in the 
$\mathcal{N}=4$ SYM theory. In particular, such a model would provide an ideal framework for studying the differences between monotonous and fortuitous states and for extracting meaningful lessons from these distinctions.

In search for monotonous states, one might attempt to modify the supercharge of $\mathcal{N}=2$ SUSY SYK, by introducing additional structure into the coupling constants. However, we have not found a simple modification that allows for the existence of monotonous states while preserving key features of the
$\mathcal{N}=2$ SYK model, such as its superconformal solutions. Some naive attempts could indeed lead to states that formally satisfy the definition of monotonous states, but nonetheless carry some undesired features that are absent in conventional holographic field theories. We will in fact encounter an example of these states near (\ref{badmonotone}).

Instead, we will start with a related model, known as the two-flavor model, proposed by Heydeman, Turiaci and Zhao in \cite{Heydeman:2022lse}. We will first review the properties of their model and describe the fortuitous states in it, then we will modify the model such that it contains also desirable monotonous states.

\subsection{The ordinary two-flavor model and its fortuitous states}\label{sec:twoflavorreview}

The two-flavor model \cite{Heydeman:2022lse} contains two types of complex fermions, $\psi_{i}$ and $\chi_{i}$, where $i$ runs from $1$ to $N$ as usual. 
The supercharge of the two-flavor model is given by:\footnote{Note that we chose a different convention compared to \cite{Heydeman:2022lse}, where instead of $\bar{\chi}$, the creation operators $\chi$ appears in the supercharge. The two models are physically equivalent up to a charge conjugation acting on the $\chi$ fermions. We choose this convention for the simplicity of latter presentation.}
\begin{equation}\label{eqn:2_flavor_Q}
	Q=\sum_{1\leq i<j\leq N,k = 1}^N C_{ijk}\psi_{i}\psi_{j}\bar\chi_{k}
\end{equation}
where the $C_{ijk}$ are independent and identically distributed complex Gaussian random variables:
\begin{equation}
    \langle C_{ijk} \bar{C}_{lmn}\rangle={J\over N^2}\delta_{il}\delta_{jm}\delta_{kn}.
\end{equation}
The supercharge \eqref{eqn:2_flavor_Q} could be obtained by specializing the coefficients $C^N_{i_1\cdots i_q}$ of the supercharge \eqref{eqn:M_flavor_Q} in Section~\ref{sec:defns} with $M=2$ and $q=3$.
We choose the projection map to be
\ie\label{pitwof}
\pi= \bar\psi_{N}\psi_{N}\bar\chi_{N}\chi_{N}\,,
\fe
which does not satisfy the first equation in \eqref{eqn:pi_Q_comm}. However, Definitions~\ref{def:monotone} and \ref{def:monotone_state} remain valid for defining monotonous and fortuitous states. 
Due to the specialization, the two-flavor model possesses more symmetries than the model in Section~\ref{sec:defns}.  There are two conserved U(1) charges, associated with the fermionic numbers of $\psi$ and $\chi$:
  \begin{equation}
      \textbf{N}_{\psi}=\sum_{i=1}^N \psi_i\bar\psi_i,~~~ \textbf{N}_{\chi}=\sum_{i=1}^N \chi_i\bar\chi_i.
  \end{equation}
  One linear combination of the U(1) charges, denoted as $J$, commutes with $Q$, while the other is the $R$ charge:
\begin{equation}\label{JR}
J=\textbf{N}_{\psi}+2\textbf{N}_{\chi}, \quad [J,Q]=0\,;\quad R=-\textbf{N}_{\chi},\quad  [R,Q]=Q\,. 
\end{equation}
We can write the Hilbert space ${\cal H}$ as a direct sum of different charge sectors
\ie
{\cal H}=\bigoplus_{J}{\cal H}_{J}\,,\quad {\cal H}_{J}=\bigoplus_{R} {\cal H}_{J}^R\,.
\fe
Note that, unlike the ${\cal N}=2$ SUSY SYK model, both the supercharge $Q$ and the $\chi$ fermion carry unit charge under the $R$-symmetry, and the $R$-charge can be regarded as the grading of the $Q$-cohomology. 
As a result, there is a single irreducible cochain complex ${\cal H}^{\bullet}_J$ within a fixed $J$ charge sector. The Schwarzian theory implies that all the fortuitous states are concentrated in the $R_{\textrm{IR}}=0$ charge sectors, where $R_{\textrm{IR}}$ is the IR $R$-charge, which can differ from the UV $R$-charge in (\ref{JR}) by additional mixing terms involving $J$ and the identity operator.

The two-flavor model shares several important features with the
$\mathcal{N}=2$ SYK model. It is a strongly quantum chaotic system and has emergent conformal symmetry at low energies. In a two-dimensional plane of various charge sectors $(\textbf{N}_{\psi},\textbf{N}_{\chi})$, using the large $N$ $G-\Sigma$ approach, one generically finds conformal solutions governing the low energy physics \cite{Heydeman:2022lse}.

We will be interested in the BPS states in this model. The Witten index of the model with a flavor chemical potential $y$ turned on is
   \ie
   I(y)=\Tr\left[(-1)^{{\bf N}_\chi}e^{\i y J}\right]=\left[(1+e^{\i y})(1-e^{2\i y})\right]^N\,.
   \fe
   The Witten index for the cochain complex with a fixed flavor charge $J$ is
   \ie\label{eqn:index}
   I_J=\int^{2\pi}_0\frac{\d y}{2\pi}e^{-\i y J}I(y)=\int^{2\pi}_0\frac{\d y}{2\pi}e^{-\i y J+N \log\left((1+e^{\i y})(1-e^{2\i y})\right)}\,.
   \fe

As it is, the two-flavor model already contains some number of monotonous states. The monotonous states have either $\textbf{N}_\chi=0$ or $\textbf{N}_\chi=N$, and correspond to the $Q$-cohomology classes represented by 
\ie\label{badmonotone}
\psi_{i_1}\cdots\psi_{i_n}\ket{\Omega}\,,
\fe
or the $Q^\dagger$-cohomology classes represented by
\ie\label{badmonotone2}
\bar{\psi}_{i_1} \cdots\bar{\psi}_{i_n}
\ket{\bar{\Omega}}\,, \quad \ket{\bar{\Omega}}\equiv \psi_1\cdots\psi_N \chi_1 \cdots \chi_N \ket{\Omega}.
\fe
Within (\ref{badmonotone}) and (\ref{badmonotone2}), those cohomology classes that can be iteratively lifted to larger $N$ are monotonous, based on Definition \ref{def:monotone}.\footnote{We note that since this model does not satisfy (\ref{eqn:pi_Q_comm}), some of these states may be lifted into the trivial cohomology classes.} 
We will not focus on this class of states further since they carry some undesirable features that distinguish them from the monotonous states in conventional holographic systems such as $\mathcal{N}=4$ SYM. The main feature is that in (\ref{badmonotone}) and (\ref{badmonotone2}), the indices are exposed. As a consequence, the number of such cohomology classes grows unboundedly as we take $N$ to infinity.  This feature is absent in $\mathcal{N}=4$ SYM due to the $\textrm{SU}(N)$ gauging, where we only have an order one number of monotonous states at fixed charge when we take $N$ to infinity. This feature is crucial for the bulk dual of such states being non-interacting BPS gravitons in the large $N$ limit, so we would like to maintain it in the toy models. 
As we will see, without too much extra effort, one can have better toy models of monotonous states where the indices are contracted rather than exposed. 

Apart from (\ref{badmonotone}) and (\ref{badmonotone2}), we expect all the other BPS states in this model are fortuitous. The main evidence of this comes from the large $N$ analysis of the Schwinger-Dyson equations \cite{Heydeman:2022lse}. Despite that conformal solutions exist generically in the $(\textbf{N}_{\psi},\textbf{N}_{\chi})$ plane, the \emph{superconformal} solutions, which describe sectors with supersymmetric ground states, only exist along a one-parameter line in the $(\textbf{N}_{\psi},\textbf{N}_{\chi})$ plane. One could view this as an analogue of the non-linear charge constraint known for supersymmetric black holes in AdS$_5 \times S^5$ \cite{Chong:2005hr}, 
\begin{equation}\label{nonlinearc}
   \textrm{ ``non-linear charge constraint": } \quad f( n_{\psi},n_{\chi} ) = 0\,
\end{equation}
where we defined $n_{\psi,\chi} \equiv \textbf{N}_{\psi,\chi}/ N$ in the large $N$ limit. 

We will determine the form of the constraint $f$ momentarily. For now, let's emphasize that even though (\ref{nonlinearc}) is only a large $N$ statement, one should really view it as a coarsed-grained version of a sharp constraint in a theory with fixed large $N$, due to the $R$-charge concentration predicted from the Schwarzian theory.  To be more concrete, $R$-charge concentration implies that, along a cochain complex with fixed $J$, the width of the constraint is one, even in the large $N$ limit.

\begin{figure}[t]
\begin{center}
\includegraphics[width=0.9\textwidth]{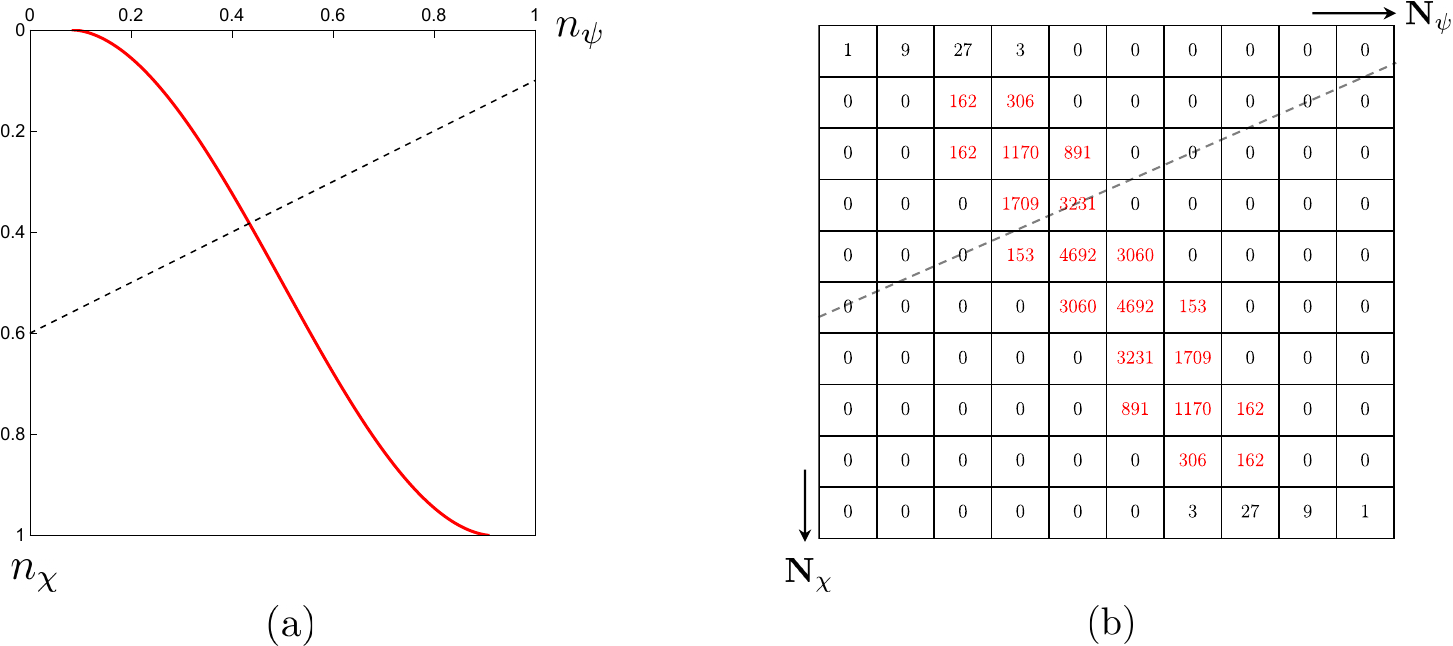}
\caption{(a) In red, we display the one-dimensional line where superconformal solutions exist in the large $N$ limit. The same line can be derived either from the index or the Schwinger-Dyson equations. The dashed line, which labels a cochain complex with fixed $J$, crosses the red line at one point. (b) We show a table containing the number of BPS states in various charge sectors, computed by exact diagonalization of the $N=9$ theory. $\textbf{N}_\psi$ ($\textbf{N}_\chi$) increases from $0$ to $9$ along the horizontal (vertical) direction. We highlight the fortuitous states in red. The dashed line denotes an example of a cochain complex with $J = 10$, along which we find BPS spectra $\{0,0,3231,0,0\}$, which indeed exhibits $R$-charge concentration. }
\label{fig:twofconcentration}
\end{center}
\vspace{-1em}
\end{figure}

Let's now discuss how to compute the constraint (\ref{nonlinearc}). As mentioned, one way to determine it is through solving the large $N$ collective field equations and finding superconformal solutions. This way of deriving the constraint is analogues to the discussion in gravity, i.e. finding for what charges the supersymmetric black holes exist. We review this computation of \cite{Heydeman:2022lse} in Appendix \ref{app:conf}. Here we simply adopt the method of \eqref{index} and \eqref{rc} in the introduction, where we use the index to pin down where the BPS states are concentrated at. We have
\ie\label{IJ}
I_J&=\Tr_J(-1)^{{\bf N}_\chi}=\sum_{{\bf N}_\chi=0}^{\min(N,\frac J2)}(-1)^{{\bf N}_\chi}\binom{N}{{\bf N}_\chi}\binom{N}{J-2{\bf N}_\chi}
\\
&\approx \int_0^{\min(1,\frac j2)} \d n_{\chi}\, e^{ N\left[\pi \i n_{\chi}+(n_{\chi}-1)\log(1-n_{\chi})-n_{\chi}\log n_{\chi}+(j-2n_\chi-1)\log(1-j+2n_\chi)-(j-2n_\chi)\log(j-2n_\chi)\right]}\,,
\fe
where $j = J/N = n_\psi + 2 n_\chi$. There are two saddle points
\ie
n^\pm_{\chi,*}(j)=\frac1{16}\left(6j-1\pm\sqrt{1+4j(j-3)}\right)\,.
\fe
Comparing with the result from numerical integration suggests that the saddle $n^+_{\chi,*}$ gives the main contribution. By $R$-charge concentration, we have $I_J=D_{{\rm BPS},J}e^{\i \pi N n_{\chi,c}}$. Matching the phases, we find the concentrated value of $n_{\chi}$:
\ie\label{nchiconstraint}
n_{\chi,c}=&{\rm Im}\,\bigg\{ \i n^+_{\chi,*}(j)+\frac1\pi\Big[ \left(n^+_{\chi,*}(j)-1\right)\log\left(1-n^+_{\chi,*}(j)\right)-n^+_{\chi,*}(j)\log n^+_{\chi,*}(j)
\\
&+\left(j-2n^+_{\chi,*}(j)-1\right)\log\left(1-j+2n^+_{\chi,*}(j)\right)-\left(j-2n^+_{\chi,*}(j)\right)\log\left(j-2n^+_{\chi,*}(j) \right)\Big] \bigg\}\,.
\fe
The expression (\ref{nchiconstraint}) determines $n_\chi$ as a function of $j = n_{\psi}+ 2 n_\chi$ and therefore determines the non-linear constraint in (\ref{nonlinearc}) implicitly. In Figure \ref{fig:twofconcentration} (a), we display the line determined by (\ref{nchiconstraint}) in the $n_\psi - n_\chi$ plane. The real part of the exponent in (\ref{IJ}) at the saddle point also gives the entropy of the fortuitous states and we find it to be always positive for $0\leq j\leq 3$, implying an $e^{\mathcal{O}(N)}$ amount of fortuitous states.

Given the non-linear constraint (\ref{nchiconstraint}), it is easy to understand why these states are fortuitous. The idea is that, once expressed in terms of the charges $\textbf{N}_\psi, \textbf{N}_\chi$, the constraint depends on $N$ explicitly. Therefore, starting from a sector allowed by the constraint, fix $\textbf{N}_\psi, \textbf{N}_\chi$ and increase $N$, we will eventually fall out of the region allowed by the coarse-grained constraint.\footnote{Here we are using the constraint derived in the large $N$ limit, which might make it seem that we would need to vary $N$ by an $\mathcal{O}(N)$ amount such that the ratio $n_\psi,n_\chi$ change by order one for the states to disobey the constraint. However, recall that in the microscopic theory, the constraint only has order one width, which suggests that we just need to change $N$ by an order one amount for the microscopic constraint to be dissatisfied.} This means that these states will be lifted and cannot represent monotonous $Q$-cohomology classes. Similarly, they also cannot represent monotonous $Q^\dagger$-cohomology classes and are therefore fortuitous.

We can compare the large $N$ analysis with finite $N$ answers from the exact diagonalization of the model. In Figure \ref{fig:twofconcentration} (b), we show the BPS spectra for the case of $N=9$. We find that the distribution of fortuitous states qualitatively agrees with the large $N$ analysis. What is more striking is perhaps, even for this moderate value of $N$, the property of $R$-charge concentration already holds, namely the fortuitous states are concentrated at a single space in the complex with a fixed $J$. In summary, we conclude that the supercharge of a generic two-flavor model satisfies our Conjecture \ref{conj:Qchaos}.

In Section~\ref{sec:N=4}, we will explore the $R$-charge concentration in ${\cal N}=4$ SYM and review how to use the superconformal index to identify the $R$-charge value where the BPS states are concentrated. However, we adopt a variant of the approach in \eqref{index} and \eqref{rc} in the Introduction. Rather than applying the saddle point approximation to the integral over the $R$-charge, we apply it to the integral over the fugacities of flavor charges. To demonstrate the equivalence of these two methods, we rederive the non-linear charge relation \eqref{nchiconstraint} using this different approach. Let us evaluate the integral \eqref{eqn:index} by saddle point approximation with the saddle at
\ie \label{ystar}
y^\pm_{*}(j)=-\i \log\left(\frac{1\pm \sqrt{1+4j(j-3)}}{6-2j}\right)\,,
\fe
where $j = J/N = (\textbf{N}_{\psi} + 2 \textbf{N}_\chi)/N$. Comparing with the result from numerical integration suggests that the saddle $y^+_{*}$ gives the main contribution. Together with $R$-charge concentration, this implies that 
\ie \label{nchic}
n_{\chi,c}=\frac1\pi \,{\rm Im}\,\left[-\i j y_{*}^+(j) + \log\left((1+e^{\i y_{*}^+(j)})(1-e^{2\i y_{*}^+(j)})\right)\right]\,,
\fe
which agrees with \eqref{nchiconstraint} in the range $0\le j\le 3$.\footnote{Here we are determining the UV charges where the BPS states concentrate. In terms of the IR $R$-charge, the BPS states are concentrated at $R_{\textrm{IR}} = 0$. The IR  $R$-charge can be determined through $I$-extremization \cite{Benini:2024cpf}.}

\begin{figure}[t]
\begin{center}
\includegraphics[width=0.4\textwidth]{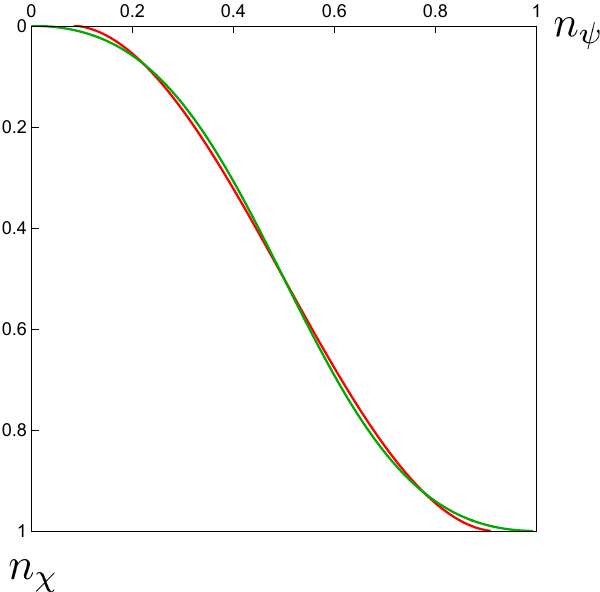}
\caption{We compare the location of the BPS states (in red) with the location of the maximal dimension spaces (in green). We see the two lines are close to each other but do not overlap apart from some crossings.}
\label{fig:comparemax}
\end{center}
\vspace{-1em}
\end{figure}

In the case of $\mathcal{N}=2$ SUSY SYK, discussed in Section \ref{sec:sykfortuity}, the BPS states are concentrated in the maximal-dimension space of the cochain complexes. This is in general not true. We can compare the location of the BPS states, determined by (\ref{nchiconstraint}), with the location of the maximum dimension spaces along different cochain complexes with different values of $J$. The latter comes from maximizing
\begin{equation}
\begin{aligned}
 D(J- 2\textbf{N}_\chi,  \textbf{N}_\chi) =    & \binom{N}{J- 2\textbf{N}_\chi} \binom{N}{ \textbf{N}_\chi} \\
  \approx  &  e^{N \left( -n_\chi \log n_\chi + ( n_\chi - 1) \log (1-n_{\chi}) +(j-2n_\chi-1)\log(1-j+2n_\chi)-(j-2n_\chi)\log(j-2n_\chi) \right)  }
\end{aligned}
\end{equation}
with respect to $n_\chi$, which gives an expression $n_\chi (j)$ that is different from (\ref{nchiconstraint}). We omit the expression for simplicity, but in Figure \ref{fig:comparemax}, we compare it with (\ref{nchiconstraint}). We see that the two lines are quite close to each other but do not overlap apart from some crossings.

\subsection{Models containing monotonous states}

Apart from some monotonous states associated with cohomology classes of the form (\ref{badmonotone}), (\ref{badmonotone2}), which we discard as discussed earlier, the two-flavor model does not contain other monotonous states. This can be traced to the genericity of the supercharge, which contains random coupling constants that are structureless. 
In order to find models that contain other monotonous states, we can introduce some further structure into the couplings. 

As a simple example, we can modify the model by symmetrizing the last two indices of $C_{ijk}$, leading to a construction that admits monotonous states:
\begin{equation}\label{Qexample}
    Q= \sum_{i,j,k=1}^N C_{ijk}\psi_{i}\psi_{(j}\bar\chi_{k)}.
\end{equation}
The symmetrization ensures that the supercharge commutes with a fermion bilinear operator 
\begin{equation}\label{QVcomm}
   [Q, V] = 0, \quad \quad V \equiv \sum_{i=1}^N \psi_i \chi_i\,,
\end{equation}
which we will refer to as a monotonous operator.
We will explain why it is monotonous momentarily. 
We will be interested in a family of states built from acting $V$ multiple times on the Fock vacuum $\ket{\Omega}= |\textbf{N}_{\psi} = 0, \textbf{N}_{\chi} = 0\rangle$,
\begin{equation}\label{Vk}
    \ket{V^k} \equiv V^k \ket{\Omega} , \quad k = 0, 1,..., N.
\end{equation}
All such states are $Q$-closed, as
\begin{equation}\label{VQclosed}
     Q\ket{V^k} = Q \left(V^k \ket{\Omega}\right) = V^k Q   \ket{\Omega} = 0\,.
\end{equation}
We can understand this family of states better by observing that the operators \(V\), \(V^{\dagger}\) together with a third generator forms an $\mathfrak{su}(2)$ algebra, given by
\begin{equation}
[V,V^{\dagger}] =\textbf{N}_{\psi} + \textbf{N}_{\chi} - N   \equiv 2\textbf{N}_V, \quad [\textbf{N}_V, V] = V, \quad [\textbf{N}_V, V^{\dagger}] = -V^{\dagger}.
\end{equation}
In the usual notation of the $\mathfrak{su}(2)$ algebra, $V, V^\dagger, \textbf{N}_V$ correspond to $J_+, J_-, J_z$, respectively. The Fock vacuum is a lowest weight state, $V^\dagger \ket{\Omega} = 0$, with $ \textbf{N}_V \ket{\Omega} = - \frac{N}{2} \ket{\Omega}$. 
Therefore, the states in (\ref{Vk}) form a spin-\({N\over2}\) representation of the $\mathfrak{su}(2)$ algebra.  

The highest weight state in the representation $V^N \ket{\Omega}$ has $\textbf{N}_{\psi} = N, \textbf{N}_{\chi} = N$, and there is only one state with such property, namely the state $\ket{\bar{\Omega}}$ with all fermions excited. Since (\ref{QVcomm}) also implies $[Q^\dagger , V^\dagger]=0$, we naturally have
\begin{equation}\label{VQdaggerclosed}
    Q^\dagger \left((V^\dagger)^k V^N \ket{\Omega} \right) =   (V^\dagger)^k Q^\dagger V^N \ket{\Omega}  = 0\,. 
\end{equation}
Therefore, we reach the conclusion that all the states in the representation are not only $Q$-closed due to (\ref{VQclosed}), but also $Q^\dagger$-closed due to (\ref{VQdaggerclosed}). In other words, all the states in (\ref{Vk}) are BPS states.

We now explain why these states are monotonous. Heuristically, once the index $i$ in $V$ is contracted, the form of the operator is formally $N$ independent. The state $\ket{V}$ is BPS regardless of the value of $N$.  More precisely, we can show they are monotonous using the machinery developed in Section \ref{sec:defns}. Let $V_N$, $\ket{\Omega_N}$ and $V_{N+1}$, $\ket{\Omega_{N+1}}$ be the monotonous operators and the Fock vacuum states of the $N$ and $N+1$ models, respectively. Consider the state
\ie\label{eqn:monstates}
\ket{\Psi_{N}}=V_{N}^n\ket{\Omega_N}\,,
\fe
where $n$ is a positive integer and $n\le N$. This state can be lifted to a state $\ket{\Psi_{N+1}}$ using \eqref{eqn:PsiN2PsiN-1} with 
\ie
\ket{\alpha_{N+1}}=(V_{N+1}^n-V_{N}^n)\ket{\Omega_{N+1}}\in {\cal I}_{N+1}\,.
\fe
The condition \eqref{eqn:QN_PsiN} is satisfied because of \eqref{eqn:QN+1_act} and $Q_{N+1}V_{N+1}^n\ket{\Omega_{N+1}}=0$. Hence, the state $\ket{\Psi_N}$ is monotonous.
We note that these monotonous states evade the problem of the states discussed in (\ref{badmonotone}) and (\ref{badmonotone2}), as here all the indices inside $V$ are contracted. The number of such states with fixed charges stabilizes in the large $N$ limit.
We expect they are the only monotonous states in the model (\ref{Qexample}) that satisfy such properties.\footnote{There could be other monotonous cohomology classes, such as those built from the product of powers of $V$ and $\psi_i$, i.e. $V^m \psi_{i_1}... \psi_{i_n}$. However, they again have open indices and do not lead to monotonous states that have a nice large $N$ limit. We do not focus on them in the following, though importantly, their existence implies that $R$-charge concentration does not hold for the monotonous states. In fact, since the monotonous states built out of $V$'s could easily be lifted under perturbations of the supercharge that are not symmetric with respect to the last two indices, there must be nearby BPS states in the cochain complex so they can join into long multiplets.}

\begin{figure}[t]
\begin{center}
\includegraphics[width=0.9\textwidth]{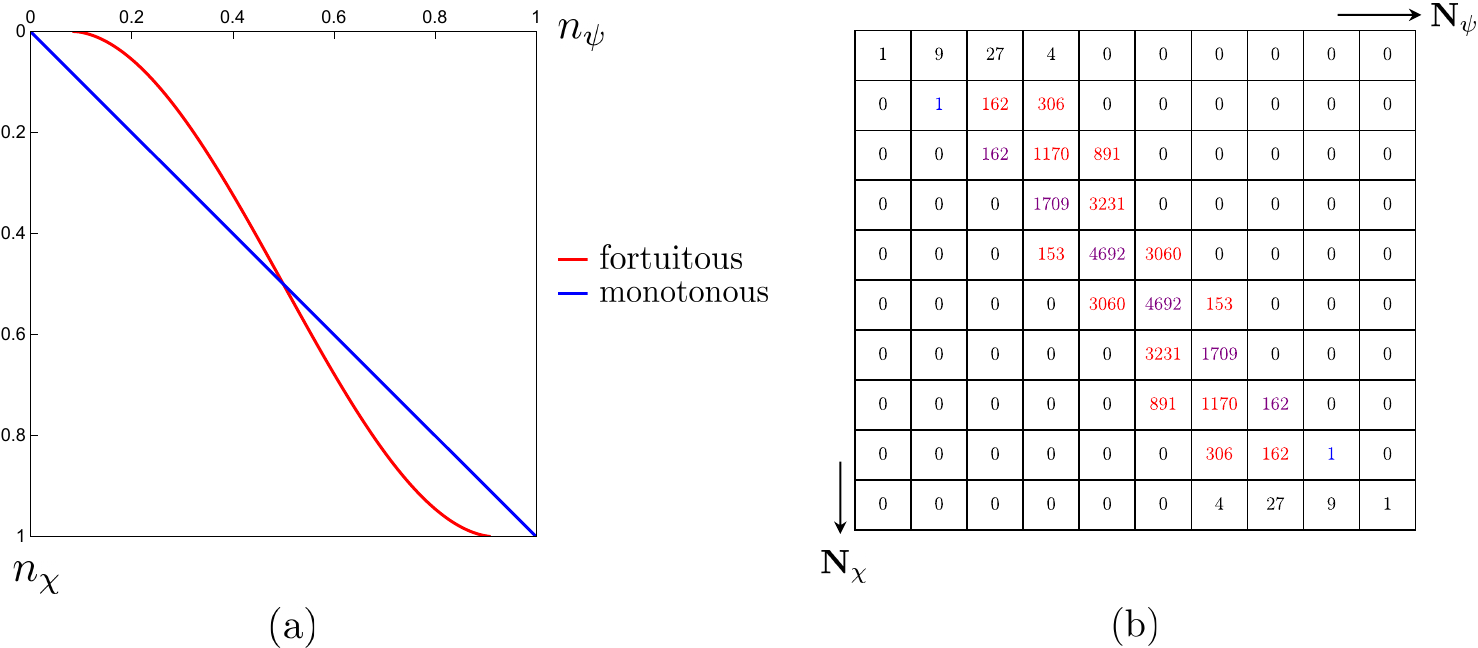}
\caption{(a) With additional structure in the coupling $C_{ijk}$, we introduce a family of monotonous states with $\textbf{N}_{\psi} = \textbf{N}_{\chi}$, denoted by the blue line. The location of the fortuitous states, denoted by the red line, is not modified compared to the generic two-flavor model. The two lines are cleanly separated in the large $N$ limit, only crossing in the middle. (b) We show the BPS spectra of the model (\ref{Qexample}) at $N=9$, which can be compared with Figure \ref{fig:twofconcentration} (b). At $\textbf{N}_{\psi} = \textbf{N}_{\chi}=1$, we find a single monotonous state $\ket{V}$, shown in blue. As opposed to (a), for $N=9$, the fortuitous and monotonous states are not yet cleanly separated. In the sectors that are in purple, we have a mixture of fortuitous and monotonous states built from $V$.}
\label{fig:twofmonotonous}
\end{center}
\vspace{-1em}
\end{figure}

As we will show in Section \ref{sec:SDequation}, the introduction of extra structure into the random coupling and the existence of monotonous states do not alter the properties of the model in the large $N$ limit in a significant way.  In particular, the model still contains a large number of fortuitous states, satisfying $R$-charge concentration determined by the same equation (\ref{nchiconstraint}) as in the generic two-flavor model. In Figure \ref{fig:twofmonotonous}, we illustrate the situation both in the large $N$ limit and at finite $N$.

In the simplest construction (\ref{Qexample}), with fixed $\textbf{N}_\psi = \textbf{N}_\chi = n$, we only get a single monotonous state $\ket{V^n}$ in this model. We can improve the situation and construct models with more monotonous states.
To achieve this, we introduce an extra index \(\alpha \in \{1, \dots, K\}\) to the fermions, which gives us a more general family of two-flavor models with additional structure:
\begin{equation}\label{eqn:monmodel}
        Q = \sum_{i,j,k=1}^N \sum_{\alpha,\beta=1}^K C^{\alpha}_{ijk} \psi^{\alpha}_{i} \psi^{\beta}_{(j} \bar{\chi}^{\beta}_{k)}\,, \quad\quad\quad  \langle C^{\alpha}_{ijk} \bar{C}^{\beta}_{lmn}\rangle={J\over 4N^2 K}\delta^{\alpha\beta}\delta_{il}\delta_{jm}\delta_{kn}.
\end{equation}
The Hamiltonian of the model is still given by $H = \{Q,Q^\dagger\}$. 
This model fits into the general discussion in Section \ref{sec:defns} by identifying $M=2K$. We will later refer to this class of models by the value of $K$. For example, in Section \ref{sec:num2f} and \ref{sec:followmono} we will be studying the $K=2$ case, which we will refer to as the ``$K=2$ two-flavor model".

The corresponding monotonous operators in these models are also fermion bilinears, now symmetric in the \(\alpha, \beta\) indices. Consequently, there are a total of \(\frac{K(K+1)}{2}\) monotonous operators:
\begin{equation}\label{Valphabeta}
        V_{\alpha,\beta} = \sum_{i=1}^N \psi_i^{(\alpha} \chi_i^{\beta)}, \quad 1\leq\alpha \leq \beta\leq K.
\end{equation}
These monotonous operators, together with $V^\dagger_{\alpha,\beta}$ and their commutators, form an \(\mathfrak{sp}(2K)\) Lie algebra (see appendix \ref{app:malg}). By acting $V$'s on the Fock vacuum $\ket{\Omega}$, we get a family of states
\begin{equation}\label{moremono}
    \prod_{1\leq\alpha \leq \beta\leq K} (V_{\alpha,\beta})^{n_{\alpha,\beta}}  \ket{\Omega}\,.
\end{equation}
Following the same analysis as before, we find that all the states in (\ref{moremono}) are monotonous BPS states. We note that this way of building monotonous states is similar in spirit to $\mathcal{N}=4$ SYM, where one can build heavy monotonous operators by taking product of light single trace monotonous operators dual to gravitons. In this sense, the $V$ operators in (\ref{Valphabeta}) are the analogy of ``gravitons" in our model.

The existence of these new monotonous BPS states is the main novel feature of these models in comparison to the two-flavor model in \cite{Heydeman:2022lse}. The other features associated with the fortuitous states remain unchanged. Before we present a more detailed analysis, we highlight several 
distinct features of the monotonous states and the fortuitous states as follows:
\begin{itemize}
    \item Fixing $\textbf{N}_{\chi}, \textbf{N}_\psi \sim \mathcal{O}(N)$, the number of monotonous states as constructed from (\ref{moremono}) grows only polynomially with \(N\), whereas the number of fortuitous states grows exponentially with \(N\). Therefore
    \begin{equation}
        D_{\textrm{monotonous}} \ll  D_{\textrm{fortuitous}}\,. 
    \end{equation}
    \item The energy gap above the monotonous states is parametrically larger than that of the fortuitous states.\footnote{This property could be useful for quantum memories, we thank Hui Zhai for suggesting this.} The energy gap above the fortuitous states is of order $1/N$ \cite{Heydeman:2022lse}. In the generic two-flavor model, we expect the gap in the sectors with $\textbf{N}_{\chi}= \textbf{N}_\psi\sim \mathcal{O}(N)$ (and sufficiently far away from $\textbf{N}_{\chi}= \textbf{N}_\psi= N/2$) will be of order $N$. This is the natural scaling one would expect from the large $N$ $G-\Sigma$ analysis and also agrees with the extrapolation of Schwarzian prediction $E_{\textrm{gap}}\sim \frac{1}{N}(R-R_c)^2$ when $R-R_c\sim\mathcal{O}(N)$.
    Since we will find in Section \ref{sec:SDequation} that the large $N$ conformal solutions are not modified in the new models, we expect the gap above monotonous states to be still of order $N$.\footnote{More carefully, we mean the gap between the monotonous states to the continuum of states with order $N$ entropy. We do not exclude the possibility that there are small number of low-lying states that are not captured by the large $N$ analysis.}   
    \item The wavefunctions of the fortuitous states resemble random states and have strong dependence on $N$. On the other hand, the wavefunctions of the monotonous states are simple and have straightforward large \(N\) limits. We will demonstrate their differences numerically in Section \ref{sec:num2f}.
    \item The fortuitous cohomology classes can be seen to exhibit a ``no-hair" property \cite{Choi:2023vdm}. Starting from a fortuitous state $\ket{O_{\textrm{f}}}$, we can act on it with $V$. The resulting state $V\ket{O_{\textrm{f}}}$ is guaranteed to be $Q$-closed but may not lead to new BPS states if it is $Q$-exact. The phenomenon that sometimes one does not get new BPS states this way was interpreted as the resistance of black holes against being dressed by hair in \cite{Choi:2023vdm}. 
     In the large $N$ limit, this can be seen from Figure \ref{fig:twofmonotonous} (a) that the slope of the red line at generic points is not equal to one. In the example of Figure \ref{fig:twofmonotonous} (b) with $N=9$, we find this property to hold partially. For example, start with the sector with $\textbf{N}_\psi=4,\textbf{N}_{\chi}=2$ that contains 891 BPS states, by acting with $V$, we go to a sector with zero BPS states. Notice that at $N=9$, it is not true that all the states of the form $V\ket{O_{\textrm{f}}}$ are $Q$-exact. It would be interesting to study the dressings in this model in more detail. 
\end{itemize}

In the next two subsections, we provide further evidence for these claims using two different approaches. In Section \ref{sec:SDequation}, we study the large $N$ $G$-$\Sigma$ description of the modified models (\ref{eqn:monmodel}). 
In section \ref{sec:num2f}, we will carry out a finite $N$ numerical study of these models, where we contrast the fine-grained properties of the monotonous states and fortuitous states.

\subsection{Large $N$ analysis of the modified two-flavor models}\label{sec:SDequation}

In this section, we will carry out a large $N$ analysis of our newly proposed models in (\ref{eqn:monmodel}), following the conventional $G$-$\Sigma$ techniques of SYK models (see e.g. \cite{Maldacena:2016hyu,Fu:2016vas,Heydeman:2022lse}).  The central result is that, in the large \( N \) limit, these models reproduce the behavior of the original two-flavor system. To proceed, we use the the superfield formalism for analyzing the large \( N \) Schwinger-Dyson equations, as outlined in Section 2.1 of \cite{Heydeman:2022lse} which provides a useful parallel for the discussion below.

Consider our models in the Euclidean signature, with $\tau$ being Euclidean time. We begin by introducing the superfield coordinates \( Z(\tau, \theta, \bar{\theta}) \). The infinitesimal supersymmetry transformations, parametrized by the complex Grassmann variable \( \eta \), and $R$-symmetry transformations, parametrized by a phase variable \( a \), act on the superfield coordinates as follows:
\begin{align}
\tau \rightarrow \tau + \theta \bar{\eta} + \bar{\theta} \eta, \quad \theta \rightarrow \theta + \eta, \quad \bar{\theta} \rightarrow \bar{\theta} + \bar{\eta};
\end{align}
\begin{equation}
 \theta \rightarrow e^{\i a } \theta, \quad \bar{\theta} \rightarrow e^{-\i a } \bar{\theta}.   
\end{equation}
With these superfield coordinates, we define the chiral fermionic superfields \( \Psi_i^{\alpha}(\tau, \theta, \bar{\theta}) \) and \( X_i^{\alpha}(\tau, \theta, \bar{\theta}) \) to represent the fields \( \psi \) and \( \bar{\chi} \) and their bosonic partners, satisfying the chiral conditions:\footnote{We acknowledge the potential confusion that may arise from our notation, as the field component \( \bar{\chi} \) appears within the chiral superfield definition of $X$. We make this choice such that the following discussion aligns with the convention in \cite{Heydeman:2022lse}.}
\begin{equation}
    D_{\bar{\theta}} \equiv \partial_{\bar{\theta}} + \theta \partial_{\tau}, \quad D_{\bar{\theta}} \Psi_i^{\alpha}(\tau, \theta, \bar{\theta}) = D_{\bar{\theta}} X_i^{\alpha}(\tau, \theta, \bar{\theta}) = 0;
\end{equation}
\begin{equation}
    \Psi_i^{\alpha} = \psi_i^{\alpha}(\tau + \theta \bar{\theta}) + \sqrt{2} \, \theta \, b^{\alpha}_{\psi, i}(\tau), \quad X_i^{\alpha} = \bar{\chi}_i^{\alpha}(\tau + \theta \bar{\theta}) + \sqrt{2} \, \theta \, b^{\alpha}_{\chi, i}(\tau).
\end{equation}
Similarly, we introduce the anti-chiral fermionic superfields \( \bar{\Psi}_\alpha^{i}(\tau, \theta, \bar{\theta}) \) and \( \bar{X}_\alpha^{i}(\tau, \theta, \bar{\theta}) \) to represent the fields \( \bar{\psi} \) and \( \chi \) and their bosonic partners.

The supersymmetric Lagrangian density associated with the Hamiltonian in equation (\ref{eqn:monmodel}) can be expressed in terms of the chiral superfields, where for simplicity we adopt the Einstein summation convention for repeated indices,
\begin{equation}
\begin{aligned}
    \mathcal{L} & = \frac{1}{2} \int \mathrm{d}\bar\theta\mathrm{d}\theta \, \big( \bar{\Psi}^i_{\alpha} \Psi_i^{\alpha} + \bar{X}^i_{\alpha} X_i^{\alpha} \big) + \mathcal{L}_{\text{int}}, \\
    \mathcal{L}_{\text{int}} & = \i \int \mathrm{d}\theta \, C^{\alpha}_{ijk} \Psi^{\alpha}_{i} \Psi^{\beta}_{(j} X^{\beta}_{k)} + \i \int \mathrm{d}\bar{\theta} \, \bar{C}^{\alpha}_{ijk} \bar{\Psi}_{\alpha}^{i} \bar{\Psi}_{\beta}^{(j} \bar{X}_{\beta}^{k)}.
\end{aligned}
\end{equation}
By integrating out the random couplings using (\ref{eqn:monmodel}), \( \mathcal{L}_{\text{int}} \) simplifies to:
\begin{equation}\label{Lint}
    \mathcal{L}_{\text{int}} = -\frac{J}{2N^2 K} \int \mathrm{d} \bar{\theta}_1 \, \mathrm{d} \theta_2 \, \bar{\Psi}_{\alpha}^{i}\Psi^{\alpha}_{i}  \big(  \bar{\Psi}_{\beta'}^{j}\Psi^{\beta}_{j} \bar{X}_{\beta'}^{k} X^{\beta}_{k}  - \bar{X}_{\beta'}^{k} \Psi^{\beta}_{k}  \bar{\Psi}_{\beta'}^{j} X^{\beta}_{j}  \big).
\end{equation}
As a comparison, the interacting term in the Lagrangian of the ordinary two-flavor model in \cite{Heydeman:2022lse} is:
\begin{equation}
    \mathcal{L}_{\text{int,\,ordinary}} = -\frac{J}{2 N^2} \int \mathrm{d} \bar{\theta}_1 \, \mathrm{d}\theta_2 \, \bar{\Psi}^i\Psi_i \bar{\Psi}^j \Psi_j  \bar{X}^k X_k .
\end{equation}
Thus, our Lagrangian includes additional terms involving:
\begin{equation}
  \bar{\Psi}_{\beta'}^{j}  \Psi^{\beta}_{j} , \,\, \text{for}~\beta \neq \beta'; \quad \quad \bar{X}_{\beta'}^{k} \Psi^{\beta}_{k},\,\,\text{for any}~ \beta ~\text{and}~ \beta'.
\end{equation}
In the large \( N \) limit, \( N \gg K \), we use a mean-field approximation to replace the nonlocal interaction terms with two point function of fermions, averaged over the index $i$. Under this, the extra terms in (\ref{Lint}) involve mixed correlators:
\begin{equation}\label{twoextra}
    \langle \bar{\Psi}_{\beta'}^{j}(Z_1) \Psi^{\beta}_{j}(Z_2) \rangle, \,\, \text{for}~\beta \neq \beta';  \quad\quad \langle \bar{X}_{\beta'}^{k}(Z_1) \Psi^{\beta}_{k}(Z_2) \rangle, \,\, \text{for any}~ \beta ~\text{and}~ \beta'.
\end{equation}
The first correlator \( \langle \bar{\Psi}_{\beta'} \Psi^\beta \rangle, \beta \neq \beta'\) breaks the averaged \( \mathrm{U}(K) \) symmetry of (\ref{Lint}), while the second correlator \( \langle \bar{X}\Psi \rangle  \) breaks the \( \mathrm{U}(1)_{\chi} \) symmetry. Under the assumption of no spontaneous symmetry breaking, these additional terms can be set to zero in the large \( N \) action.\footnote{In Appendix \ref{app:SD}, we study the action without assuming $\textrm{U}(1)_\chi$ being unbroken; numerically, we find no solutions indicating symmetry breaking.} 

With these assumptions in place, we can proceed to write down the large $N$ $G$-$\Sigma$ action by introducing the $G$ and $\Sigma$ variables:
\begin{equation}\label{Gfields}
    \mathcal{G}_{\bar{\Psi} \Psi} (Z_1, Z_2) = \frac{1}{N K} \langle \bar{\Psi}^i_{\alpha}(Z_1) \Psi_i^{\alpha}(Z_2) \rangle, \quad \mathcal{G}_{\bar{X} X} (Z_1, Z_2) = \frac{1}{N K} \langle \bar{X}^i_{\alpha}(Z_1) X_i^{\alpha}(Z_2) \rangle.
\end{equation}
We can impose (\ref{Gfields}) by inserting the following identity into the path integral
\begin{equation}
    1 = \int \mathrm{D} \mathcal{G}_{\bar{\Psi} \Psi}  \mathrm{D} \Sigma_{\bar{\Psi} \Psi} \exp\left[-NK \int \mathrm{d}\bar{Z}_1  \mathrm{d} Z_2  \Sigma_{\bar{\Psi} \Psi}(Z_1, Z_2) \left( \mathcal{G}_{\bar{\Psi} \Psi}(Z_1, Z_2) - \frac{1}{NK} \bar{\Psi}^i_{\alpha}(Z_1) \Psi_i^{\alpha}(Z_2) \right) \right].
\end{equation}
Applying the same identity to \( \mathcal{G}_{\bar{X} X} \), we get the Lagrangian
\begin{align}
   \mathcal{L} &= \frac{1}{2} \int \mathrm{d} \bar{\theta} \mathrm{d} \theta\, \left( \bar{\Psi}^i_{\alpha} \Psi_i^{\alpha} + \bar{X}^i_{\alpha} X_i^{\alpha} \right) -  \int \mathrm{d} \bar{Z}_1 \, \mathrm{d} Z_2 \, \Sigma_{\bar{\Psi} \Psi}(Z_1, Z_2) \bar{\Psi}^i_{\alpha}(Z_1) \Psi_i^{\alpha}(Z_2) \notag \\
    &\quad -  \int \mathrm{d} \bar{Z}_1 \, \mathrm{d} Z_2 \, \Sigma_{\bar{X} X}(Z_1, Z_2) \bar{X}^i_{\alpha}(Z_1) X_i^{\alpha}(Z_2) + \mathcal{L}_{\text{int}},
\end{align}
where the interaction term \(\mathcal{L}_{\text{int}}\) is given by
\begin{align}
    \mathcal{L}_{\text{int}} &= NK \int \mathrm{d} \bar{Z}_1 \, \mathrm{d} Z_2 \left( \Sigma_{\bar{\Psi} \Psi}(Z_1, Z_2) \mathcal{G}_{\bar{\Psi} \Psi}(Z_1, Z_2) + \Sigma_{\bar{X} X}(Z_1, Z_2) \mathcal{G}_{\bar{X} X}(Z_1, Z_2) \right) \notag \\
    &\quad - \frac{J N K}{2} \int \mathrm{d} \bar{Z}_1 \, \mathrm{d} Z_2 \, \mathcal{G}_{\bar{\Psi} \Psi}(Z_1, Z_2)^2 \mathcal{G}_{\bar{X} X}(Z_1, Z_2),
\end{align}
which, upon a redefinition of $NK$ to $N$, reduces to the effective action of the two-flavor model in \cite{Heydeman:2022lse}. Therefore, under the assumption of no spontaneous symmetry breaking, we find that the large $N$ properties of the modified two flavor models are exactly the same as the original model.

For completeness, in Appendix \ref{app:SD}, we present the component form of the action, as given in equation (\ref{eqn:fullaction1},\ref{eqn:fullaction2}). We also present the corresponding Schwinger-Dyson equations (\ref{eqn:SDfull}-\ref{eqn:SDfull2}).

\subsection{Numerical comparison of fortuitous and monotonous states}\label{sec:num2f}

A benefit of our toy model is that we can access the fortuitous and monotonous states easily through numerics and compare their properties. As opposed to $\mathcal{N}=4$ SYM, where the fortuitous states wavefunction are hard to access numerically, here we have an abundance of them, as described in Section \ref{sec:twoflavorreview}; it's the monotonous states that are hard to find. For concreteness, we will study the model in (\ref{eqn:monmodel}) but with $K = 2$, whose supercharge we reproduce below:
\begin{equation}
    \textrm{$K=2$ two-flavor model:} \quad\quad Q =  \sum_{i,j,k = 1}^{N} \sum_{\alpha,\beta = 1}^2 C_{ijk}^{\alpha} \psi_i^\alpha \psi_{(j}^\beta \bar{\chi}_{k)}^\beta \,.
\end{equation}
The model contains three monotonous BPS operators with $\textbf{N}_\psi = \textbf{N}_\chi = 1$ that can be organized using Pauli matrices
\begin{equation}\label{3grav}
    V^{1} = \sum_{i} \psi_i^{\alpha} (\mathbb{1})_{\alpha\beta} \chi_{i}^{\beta},\quad  V^{2} = \sum_{i} \psi_i^{\alpha} (\sigma^x)_{\alpha\beta} \chi_{i}^{\beta},\quad  V^{3} = \sum_{i} \psi_i^{\alpha} (\sigma^z)_{\alpha\beta} \chi_{i}^{\beta}.
\end{equation}
As we discussed, with these simple $V$'s, we can then form monotonous states with larger fermion numbers by successively act $V$'s on the Fock vacuum $\ket{\Omega}$, 
\begin{equation}\label{monotoneform}
    \ket{\textrm{monotone}} = (V^{1})^{m_1} (V^{2})^{m_2} (V^{3})^{m_3} \ket{\Omega} , 
\end{equation}
where $\textbf{N}_{\psi} = \textbf{N}_\chi = n = m_1 + m_2 + m_3$. We emphasize that these states are not only $Q$-closed, but also $Q^\dagger$-closed, so they are valid BPS states. The number of such monotonous states is upper bounded by the number of ways of dividing $n\in \mathbb{Z}$ into three natural numbers, which is given by 
\begin{equation}\label{numbermono}
    D_{\textrm{monotonous}} (n) \leq \binom{n+2}{2}  .
\end{equation}
Therefore, for large fermion number $n\sim \mathcal{O}(1)N$, the number of monotonous states (\ref{monotoneform}) only grows as $\sim N^2$, which is in sharp contrast with the exponential growth $e^{\mathcal{O}(N)}$ of the fortuitous states. We note that in (\ref{numbermono}) we only have an inequality instead of equality since some of the states (\ref{monotoneform}) can be linearly dependent as $N$ becomes small. This is analogous to the fact in $\mathcal{N}=4$ SYM that the monotonous subspace shrinks as $N$ is decreased, due to trace relations. We discuss the precise counting of the number of monotonous states in Appendix \ref{app:malg}, utilizing the $\mathfrak{sp}(2K)$ algebra.

Given these monotonous states, a natural question is to ask how their properties compare with the fortuitous states. To have a fair comparison, we would like to consider fortuitous states that have similar fermion numbers as the monotonous states. As we can see in Figure \ref{fig:twofmonotonous} (a), in the large $N$ limit, the fortuitous states and the monotonous states are separated in different charge sectors, with the exception near $\textbf{N}_{\psi} = \textbf{N}_{\chi} \sim N/2$. 

However, for relatively small values of $N$ that are numerically accessible, along the diagonal $\textbf{N}_{\psi} = \textbf{N}_{\chi}$ where the monotonous states reside, we also have a large number of fortuitous states. To separate them, we can first find the full BPS subspace with $\textbf{N}_{\psi} = \textbf{N}_{\chi}=n$. We denote the projection operator into the full BPS subspace as $P_{\textrm{BPS}}$. Then, we can separately construct the monotonous subspace explicitly using (\ref{monotoneform}), which we denote by $P_{\textrm{m}}$. Given $P_{\textrm{BPS}}$ and $P_{\textrm{m}}$, the projection operator $P_{\textrm{f}}$ into the fortuitous subspace is simply given by
\begin{equation}
 P_{\textrm{f}}  =P_{\textrm{BPS}}-    P_{\textrm{m}}.
\end{equation}

Below, we will discuss several methods to numerically compare the properties of $P_{\textrm{f}}$ and $P_{\textrm{m}}$. One such method is given by the LMRS criterion, proposed and explored in \cite{Lin:2022rzw,Lin:2022zxd,Chen:2024oqv}. The simplicity of the SYK models also allow us to explore some other quantities, such as the information entropy \cite{Budzik:2023vtr,McLoughlin:2020zew} and entanglement entropy. We will find that the fortuitous subspace $P_{\textrm{f}}$ is much more \emph{chaotic} compared to the monotonous subspace $P_{\textrm{m}}$ under all criteria. We remind the readers that here by chaotic, we mean the notion of BPS chaos - whether the BPS subspace behaves like a random subspace with respect to simple operators in the theory.

\subsubsection*{LMRS criterion}

One sharp notion of chaos in the BPS subspace is provided by the LMRS criterion \cite{Lin:2022rzw,Lin:2022zxd}, in which one projects a simple operator $O$ into the BPS subspaces, i.e. 
\begin{equation}
    \hat{O}_{\textrm{f,m}} \equiv P_{\textrm{f,m}} O P_{\textrm{f,m}}
\end{equation}
and examine the random matrix properties of the projected operators $\hat{O}_{\textrm{f}},\hat{O}_{\textrm{m}}$. In \cite{Chen:2024oqv}, the LMRS criterion was examined for various monotonous subspaces in the $\mathcal{N}=4$ SYM theory and it was found that the projected operators $\hat{O}_{\textrm{m}}$ are only \emph{weakly chaotic}, as quantified by a long Thouless time. In contrast, one expect that the operator $\hat{O}_{\textrm{f}}$ that is projected into the fortuitous subspace exhibits \emph{strong chaos}, as suggested by a ``chaos invasion" picture.\footnote{We will explore this picture in more detail in Section \ref{sec:followNsyk}.} 

\begin{figure}[t]
\begin{center}
\includegraphics[width=1\textwidth]{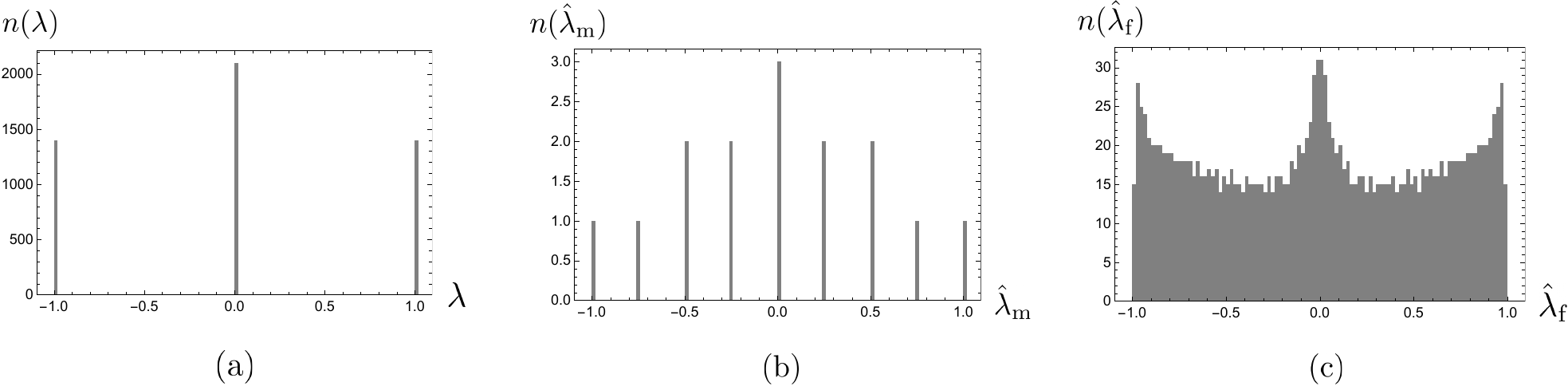}
\caption{We display the histogram for the eigenvalues of (a) $O$, (b) $\hat{O}_{\textrm{m}}$ and (c) $\hat{O}_{\textrm{f}}$\,.}
\label{fig:LMRS}
\end{center}
\vspace{-1em}
\end{figure}

\begin{figure}[t]
\begin{center}
\includegraphics[width=0.8\textwidth]{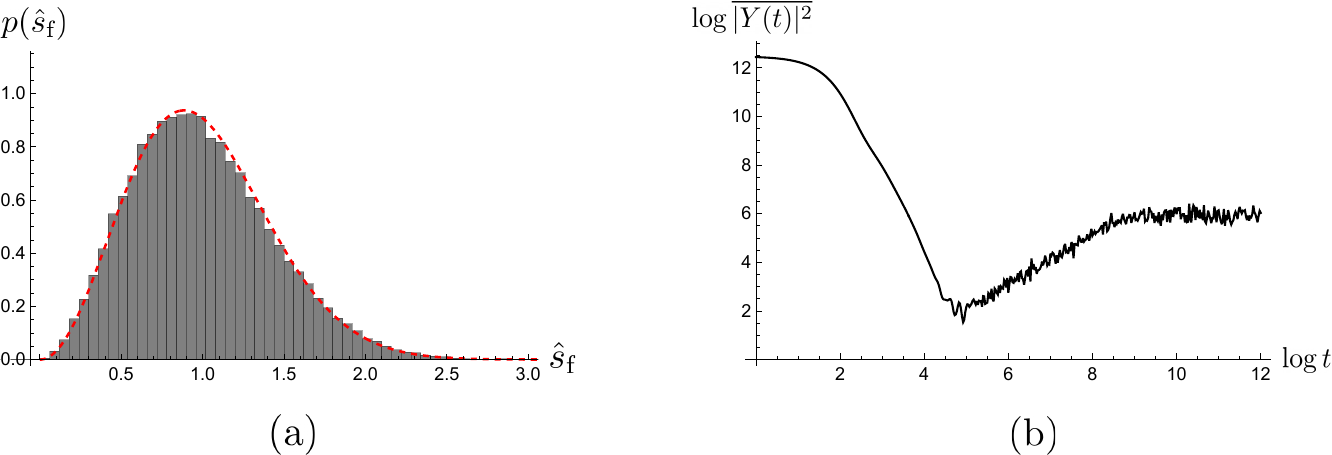}
\caption{(a) We look at the distribution of nearest-neighbor spacings $\hat{s}_{\textrm{f},i} = \hat{\lambda}_{\textrm{f},i+1}-\hat{\lambda}_{\textrm{f},i}$ in the (unfolded) spectrum $\hat{\lambda}_{\textrm{f},1}< \hat{\lambda}_{\textrm{f},2}< ...$ of operator $\hat{O}_\textrm{f}$. In the red dashed line, we plot the Wigner surmise of the GUE ensemble, $p_{\textrm{GUE}}(s) \approx \frac{32}{\pi^2} s^2 e^{-\frac{4}{\pi}s^2}$.   (b) We compute the (Gaussian-filtered) spectral form factor $Y(t) = \sum_{i} \exp(- \hat{\lambda}_{\textrm{f},i}^2/ (2 \times 0.2^2 ) - i\hat{\lambda}_{\textrm{f},i} t  )$. Both (a) and (b) are results of averaging over a sample size of $50$.}
\label{fig:LMRSchaos}
\end{center}
\vspace{-1em}
\end{figure}

Let's consider our two-flavor model with $K=2$ and $N=4$ (in total $2\times K\times N = 16$ complex fermions), and focus on the charge sector $\textbf{N}_{\psi} = \textbf{N}_{\chi } = 4$. In this charge sector, we have in total $4900$ states, among which $1820$ are BPS. Within the BPS states, we have $15$ monotonous states as given by the right hand side of (\ref{numbermono}) with $n=4$, with the rest being fortuitous states, i.e.
\begin{equation}\label{2P}
    \textrm{dim} (P_{\textrm{m}}) = 15, \quad \textrm{dim} (P_{\textrm{f}}) = 1805.
\end{equation}
In SYK, a natural choice of simple operator $O$ would be an operator that only contains an order one number of fermions in the large $N$ limit. For concreteness, here we focus on a particular choice of simple operator
\begin{equation}
    O = \psi_1^1 \bar{\psi}_1^2 + \psi_1^2 \bar{\psi}_1^1 \,.
\end{equation}
Without performing any further projection, the spectrum of $O$ in the charge sector is very simple - the eigenvalues are either $0$ or $\pm 1$, see Figure \ref{fig:LMRS} (a). In Figure \ref{fig:LMRS} (b) and (c), we show the histogram of the projected operator in the monotonous subspace and the fortuitous subspace, respectively. We see that, the spectrum of $\hat{O}_{\textrm{m}}$ is evenly spaced by $1/4$ and has degeneracies, therefore does not exhibit random matrix statistics. On the other hand, the spectrum of $\hat{O}_{\textrm{f}}$ does not contain degeneracies. To verify that it indeed exhibits random matrix behavior,
we can look at the statistics of nearest-neighbor spacings in the spectrum of $\hat{O}_{\textrm{f}}$, as seen in Figure \ref{fig:LMRSchaos} (a). We find that the distribution agrees well with the Wigner surmise of a Gaussian unitary ensemble (GUE). Furthermore, in Figure \ref{fig:LMRSchaos} (b), we compute the (Gaussian-filtered) spectral form factor \cite{stanfordunpublished,Gharibyan:2018jrp}, where we find that it exhibits a long period of linear ramp -  a sharp signature of the long-range universal level repulsion in the spectrum.\footnote{In \cite{Chen:2024oqv}, the authors used the $N$-scaling of the Thouless time (where the linear ramp starts) as a criterion for strong/weak chaos. We have not performed a careful numerical study of the $N$-scaling of the Thouless time of $\hat{O}_{\textrm{f}}$. Notice that, since the LMRS wormhole \cite{Lin:2022zxd} involves infinite Euclidean time evolution, matter corrections due to one-loop determinant on the wormhole \cite{Chen:2023hra} will be suppressed. Therefore, from the gravity picture, it is likely that $\hat{O}_\textrm{f}$ can have an order one Thouless time.} 

\subsubsection*{Information entropy of typical states}

The simplicity of the SYK model allows us to explore some other quantities that could diagnose BPS chaos. One such quantity is called the information entropy, which has been applied to BPS states in $\mathcal{N}=4$ SYM in \cite{Budzik:2023vtr} (see also \cite{McLoughlin:2020zew}). To define the information entropy, one first chooses a ``simple" orthonormal basis $\{\ket{1}, ..., \ket{d}\}$ for the Hilbert space, in which one can expand a pure state $\ket{\psi}$,
\begin{equation}
    \ket{\psi} = \sum_{i=1}^{d} c_i \ket{i}.
\end{equation}
The information entropy for pure state $\ket{\psi}$ is then defined as
\begin{equation}\label{Sinfo}
    S_{\textrm{info}} (\ket{\psi}) = -\sum_{i=1}^{d} |c_{i}|^2 \log |c_{i}|^2\,,
\end{equation}
which measures how random the coefficients $c_i$ are. As an example, for a random state, we can easily estimate the value of $S_{\textrm{info}}$ to be
\begin{equation}
    S_{\textrm{info, random}} = \log d + \gamma_E - 1 + \mathcal{O}\left(\frac{1}{d}\right)\,,
\end{equation}
where $\gamma_E$ is the Euler constant. 

There are two main issues when applying (\ref{Sinfo}) to the BPS subspaces. First, we want to study a degenerate subspace instead of a particular pure state. To circumvent this issue, we can choose to study the information entropy of a \emph{typical} state in the subspace, defined as $\int \d U\, S_{\textrm{info}} (U\ket{\psi})$ where $U$ is a unitary in the subspace and $\int \d U = 1$ is the Haar measure. In practice, we can estimate this quantity by averaging over a finite ensemble of randomly sampled unitaries. The second issue is to choose a suitable simple basis. Fortunately, a particular natural choice of simple basis in SYK models is provided by the Fock basis, namely the eigenbasis where each fermion is either excited or not. We will work with this basis in the following analysis. 

Let's focus on the same set of states we have studied using the LMRS criterion, see around (\ref{2P}). By averaging over an ensemble of randomly chosen unitaries, we find that
\begin{equation}\label{typicalinfo}
    S_{\textrm{info, typical}}(\textrm{monotonous}) \approx 5.19, \quad S_{\textrm{info, typical}}(\textrm{fortuitous}) \approx 8.06.
\end{equation}
As a comparison, we have that $
    S_{\textrm{info, random}} \approx  \log 4900 + \gamma_E - 1 \approx 8.074 $. Therefore, we find that a typical fortuitous states is very close to a random state in terms of the information entropy,\footnote{A preliminary finite $N$ scaling analysis suggests that the $( S_{\textrm{info, typical}}(\textrm{fortuitous} )- S_{\textrm{info, random}} ) / S_{\textrm{info, random}}$ goes to zero in the large $N$ limit.} while the monotonous states are separated by a big gap. We've looked at other charge sectors and found similar behavior. This agrees with the supercharge chaos conjecture and may also be related to the pseudorandom properties associated with black hole horizons \cite{Kim:2020cds,Engelhardt:2024hpe}.

We note that, for the monotonous subspace, there exists a ``preferred" basis given by $V$'s acting on the Fock vacuum (\ref{monotoneform}). We expect the averaged information entropy for these basis states to be smaller compared to the typical value in (\ref{typicalinfo}). Indeed, we find that 
\begin{equation}
     S_{\textrm{info, $V$-basis}}(\textrm{monotonous}) \approx 4.46 <  S_{\textrm{info, typical}}(\textrm{monotonous}) \,.
\end{equation}
However, for the fortuitous subspace, there doesn't exist an obvious choice of basis that would reduce the information entropy compared to the typical value (\ref{typicalinfo}).

\subsubsection*{Entanglement entropy of typical states}

Another quantity one can study is the entanglement entropy between different subsets of fermions. We expect that the fortuitous states should be highly entangled, while the monotonous states should exhibit less entanglement. Again, similar to the information entropy, here we consider the entanglement entropy of a typical state in the corresponding BPS subspace.

We consider the same set of states as studied previously. For concreteness, let's focus on a particular choice of subset, where we divide the fermions into two halves, the first half being $\psi_i^{\alpha},\chi_i^{\alpha}, i = 1,2, \alpha= 1,2$, and the second half being its complement.  
By averaging over an ensemble of randomly chosen unitaries, we find that\footnote{Note that here we used $\log_2$ in the definition of the entropy, so the numerical value of the entropy has the interpretation of the effective number of qubits being entangled.}
\begin{equation}
    S_{\textrm{EE, typical}}(\textrm{monotonous}) \approx 2.70, \quad S_{\textrm{EE, typical}}(\textrm{fortuitous}) \approx 6.97.
\end{equation}
Therefore, we confirm the expectation that the fortuitous states are much more entangled than the monotonous states. 

Similar to the discussion of information entropy, in the monotonous subspace, we can compare the averaged entanglement entropy for the basis of states given by $V$'s acting on the Fock vacuum and that of a typical state. We find 
\begin{equation}
     S_{\textrm{EE, $V$-basis}}(\textrm{monotonous}) \approx 2.36 <  S_{\textrm{EE, typical}}(\textrm{monotonous}) \,.
\end{equation}

\section{Following $N$ and chaos invasion in SYK models}\label{sec:followN}

A particularly illuminating perspective to understand the fortuitous and monotonous states is by viewing $N$ as a continuous parameter, through which one can interpolate smoothly between the energies and wavefunctions of states in theories with different $N$. In \cite{Budzik:2023vtr}, this idea was applied to the $1/16$-th BPS sector of $\mathcal{N}=4$ $\textrm{SU}(N)$ SYM, where in the charge sector containing the lightest fortuitous state, one sees explicitly that the anomalous dimension of a single fortuitous state decreases smoothly as $N$ is decreased, and hits the BPS bound at $N=2$. 

The following-$N$ picture provides a perspective to understand why the fortuitous BPS states are strongly chaotic, through a ``chaos invasion" mechanism \cite{Chen:2024oqv}. One can think of the chaos of the BPS subspace as coming from the chaotic non-BPS states which enters the BPS subspace as $N$ is decreased.  On the other hand, for the sectors that only contain monotonous states, the gap between BPS states and non-BPS states never closes as one decreases $N$. This 
leaves the possibility for the BPS states to exhibit distinct properties, such as being only weakly chaotic \cite{Chen:2024oqv}.

\subsection{Scheme of following $N$ in SYK models}\label{sec:genefollow}

In this section, we study the scheme of following $N$ in the SYK models. In comparison to the case of $\mathcal{N}=4$ SYM where only one fortuitous state was seen invading the BPS subspace, due to the simplicity of the SYK models, here we are able to access the regime where a large number (exponential in $N$) of fortuitous states enter the BPS subspace at the same time. We can further follow each individual microstate and study their properties in great detail. 

We shall first explain the precise meaning of following $N$ in the SYK model. A way to interpolate between SYK model from $N = N_*$ to $N = N_* - 1$ is to simply decouple the $N$-th fermion gradually. In other words, given the random couplings $C_{ijk}$ in the $N = N_*$ theory, we can define a one-parameter family of model parametrized by continuous $N$, with random couplings
\begin{equation}\label{naiveint}
    C_{ijk} (N) \equiv \left\{ \begin{aligned} 
  &   C_{ijk} , \quad  & i,j,k< N_* , \\
  & (N - (N_* - 1)) C_{ijk}   , \quad   & \textrm{one of\, $i,j,k$} = N_*,
    \end{aligned} \right. \quad \quad N \in [N_* - 1,N_*].
\end{equation}
This model therefore interpolates between the $N_*$ theory and the $N_* - 1$ theory. At $N = N_* - 1$, the decoupled fermion $\psi_{N_*}$ would lead to a two-fold degeneracy in the spectrum, half of them would be redundant since they contain $\psi_{N_*}$ excitation and does not really make sense in the $N= N_*-1$ theory. One can filter out the physical states in the $N= N_* -1$ theory by simultaneously diagonalizing the number operator $n_{N_*} \equiv\psi_{N_*} \bar{\psi}_{N_*}$ and focus on all the states with $n_{N_*} = 0 $.\footnote{Here we choose to only modify the couplings, without modifying the inner product between states. We can find out whether a state is physical or redundant at integer $N$ by examining the number operators of decoupled fermions. This is slightly different from the $\mathcal{N}=4$ SYM discussion in \cite{Budzik:2023vtr} where the inner product is modified and states can become null or have a negative norm. One could also choose to modify the inner product in the current SYK discussion, though we do not think this distinction plays a significant role in the purpose of our discussion.} 

The interpolation (\ref{naiveint}) suffices for the purpose of interpolating between $N_*$ to $N_* - 1$. However, for the purpose of the discussion in the following sections, we would like to follow the states across multiple different values of $N$. One could decouple multiple fermions successively, but this would lead to non-analytic behavior in the spectrum at $N \in \mathbb{Z}$. Instead, one could try to find a way to analytically interpolate different models by varying $N$ while still landing on the desired theories at integer $N$. Such interpolation is non-unique, while one possibility is the following
\begin{equation}\label{CinN}
    C_{ijk} (N) = w_{i} w_{j} w_{k} C_{ijk},
\end{equation}
where
\begin{equation}\label{wa}
    w_a (N)\equiv \frac{1}{N_*} \sum_{\ell=1}^{N_*}  e^{\frac{2\pi \i \ell a }{N_*} }  \frac{  1 -  e^{-\frac{2\pi \i \ell N  }{N_*} }    }{   e^{\frac{2\pi \i  \ell  }{N_*} } - 1  }\,. 
\end{equation}
In the sum of (\ref{wa}), the $\ell = N_*$ term needs to be treated with care, which contributes $N$ to the sum. The expression (\ref{wa}) has the nice feature that it is analytic in $N$, and when $N \in \mathbb{Z}$, one can show that it evaluates to
\begin{equation}
    w_{a} (N) = \left\{ \begin{aligned} 
  &   1 , \quad  a = 1 , ..., N ,\\
  & 0 , \quad a  = N+1 , ..., N_*, 
\end{aligned} \right. \quad \textrm{when} \quad N \in \mathbb{Z}
\end{equation}
and therefore reproduce the desired integer $N$ theory.
We discuss in Appendix \ref{app:interpolation} how the formula (\ref{wa}) can be derived. 

Note that when $N \notin \mathbb{Z}$, all the components of $w_a$ are nonzero, and therefore via (\ref{CinN}) the model is effective a generic $N= N_*$ model. The genericity of the model at non-integer $N$ implies that no level crossings should occur, making it possible to follow individual microstates in $N$ unambiguously in between integer $N$'s. On the other hand, at integer $N$, there could be level crossings. These level crossings, or extra degeneracies, originate from the fact that some fermions are decoupled. If we wish to follow individual microstates further, we could circle around integer values of $N$ by deforming the contour of $N$ slightly into the complex plane.

\subsection{Following $N$ in the $\mathcal{N}=2$ SUSY SYK model}\label{sec:followNsyk}

We now study the problem of following $N$ in the $\mathcal{N}=2$ SUSY SYK model with $q = 3$. We would like to examine the following: how does an exponential large number of states come into the BPS subspace, and, whether they retain their chaotic features in the process.

To keep the discussion concrete, let's consider an explicit example where we follow $N$ from $N=12$ down to $N=9$.\footnote{We choose these moderately small values of $N$ to avoid clutter in the figures, though numerically one can easily access larger values of $R$ and $N$.} In Table \ref{table1}, we list various charge sectors in these theories and whether they contain BPS states or not. Note that when $N$ is an odd integer, there are special charge sectors with
\begin{equation}
    R = \frac{N}{2} \pm \frac{q}{2} = \frac{N}{2} \pm \frac{3}{2}
\end{equation}
that are \emph{gapless} \cite{Stanford:2017thb}.\footnote{See also recent discussion in \cite{Heydeman:2024fgk} on appearances of such gapless sectors for AdS$_4$ black holes in M-theory.} In other words, even though these sectors do not contain (or when $N=1$ mod $4$ contain only a single) BPS states, there is a large number ($\sim e^{\mathcal{O}(N)}$) of states right above zero energy. These sectors show up in the following discussion in an interesting way.

\begin{table}[t]
\begin{center}
\begin{tabular}{|c|c|c|c|c|c|}
    \hline
    \diagbox{$N$}{$R$} & 3 & 4 & 5 & 6 & 7 \\
    \hline
    12 & \color{dred}{$\cross$} & {\color{dred}{$\cross$}} & \color{dgreen}{$\checkmark$} & \color{dgreen}{$\checkmark$} & \color{dgreen}{$\checkmark$} \\
    \hline
     11 & \color{dred}{$\cross$} & {\color{orange}{$\cross$}} & \color{dgreen}{$\checkmark$} & \color{dgreen}{$\checkmark$}  & \color{orange}{$\cross$} \\
     \hline
      10 & \color{dred}{$\cross$} & \color{dgreen}{$\checkmark$} & \color{dgreen}{$\checkmark$} & \color{dgreen}{$\checkmark$} & \color{dred}{$\cross$}   \\
    \hline
     9 & \color{orange}{$\cross$} & \color{dgreen}{$\checkmark$} & \color{dgreen}{$\checkmark$} & \color{orange}{$\cross$} & \color{dred}{$\cross$}  \\
    \hline
\end{tabular}
\caption{We look at various $R$ charge sectors for $9\leq N\leq 12$ and label them by the features of the low energy spectrum. We have {\color{dgreen}{$\checkmark$}}: contain exponentially many BPS states and there is a gap above them; {\color{dred}{$\cross$}}: no BPS states and there is a gap; {\color{orange}{$\cross$}}: no BPS states (or sometimes only one) but the spectrum is gapless, i.e. there are large number of states with energy scale as $e^{- N}$.}
\label{table1}
\end{center}
\end{table}

Given Table \ref{table1}, we can focus on a particular $R$-charge sector and see how different $N$'s are interpolated as we follow $N$. In the following, let's focus on the sector $R=4$, i.e. the second column in Table \ref{table1}, which should exhibit the chaos invasion phenomenon.

\begin{figure}[t]
\begin{center}
\includegraphics[width=1\textwidth]{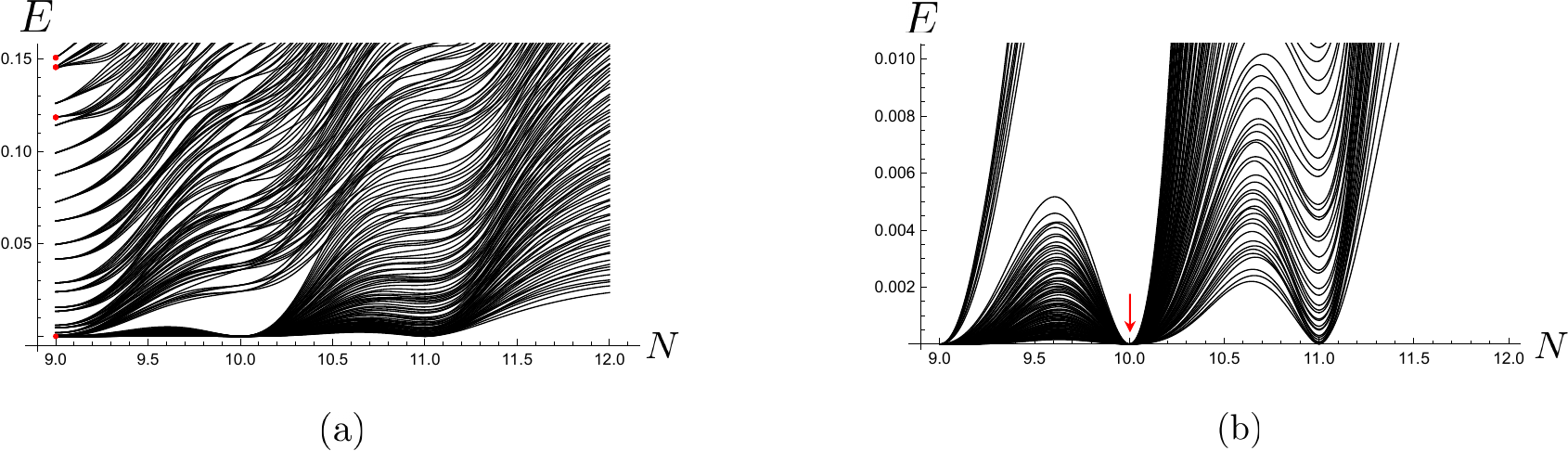}
\caption{(a) We display the spectrum (in units of $J$, same below) in the charge sector $R=4$ as we follow $N$. At $N=9$, most of the states have become unphysical. We use red dots to label the energies that still contain physical states. (b) We zoom into the low energy part of the spectrum and focus only on the fortuitous states. A large number of states enter the BPS subspace at $N=10$, indicated by the red arrow. See main text for more discussion.  }
\label{fig:followN}
\end{center}
\vspace{-1em}
\end{figure}

In Figure \ref{fig:followN}, we show the result of following $N$ in this sector, using the approach as described in (\ref{CinN}), (\ref{wa}) with $N_* = 12$. In Figure \ref{fig:followN} (a) we show how the spectrum including relatively high energy states evolve with $N$. At $N=12$, the spectrum is gapped, but the gap closes as we approach $N=11$ as many states come down. However, even though it is hard to see from Figure \ref{fig:followN} (b), despite there being many low lying states at $N=11$, there aren't states with exact zero energy. A total number of $3^{\frac{10}{2} - 1} = 81$ of low energy states eventually become BPS states at $N=10$, while at the same time, a gap above BPS states opens up again.

Further decreasing $N$ from $N=10$, we see that all $81$ BPS states at $N=10$ are lifted and rejoin the BPS subspace at $N=9$, together with some amount of new fortuitous states.
At $N=9$, we have again $3^{\frac{9-1}{2}} = 81$ BPS states, which implies that at least some number of states should become unphysical. In the specific example here, we can check that $78$ out of $81$ BPS states at $N=10$ remain physical at $N=9$, while the remaining $3$ physical states belong to those new fortuitous states that weren't BPS at $N=10$.    

In Figure \ref{fig:followN}, one can indeed see that there are extra degeneracies at integer $N$, due to the fact that some fermions are being decoupled.  As we have remarked in Section \ref{sec:genefollow}, one can avoid these degeneracies by deforming $N$ slightly into the complex plane around integer points. Apart from these, we indeed do not find level crossings for non-integer $N$. 

\begin{figure}[t]
\begin{center}
\includegraphics[width=0.5\textwidth]{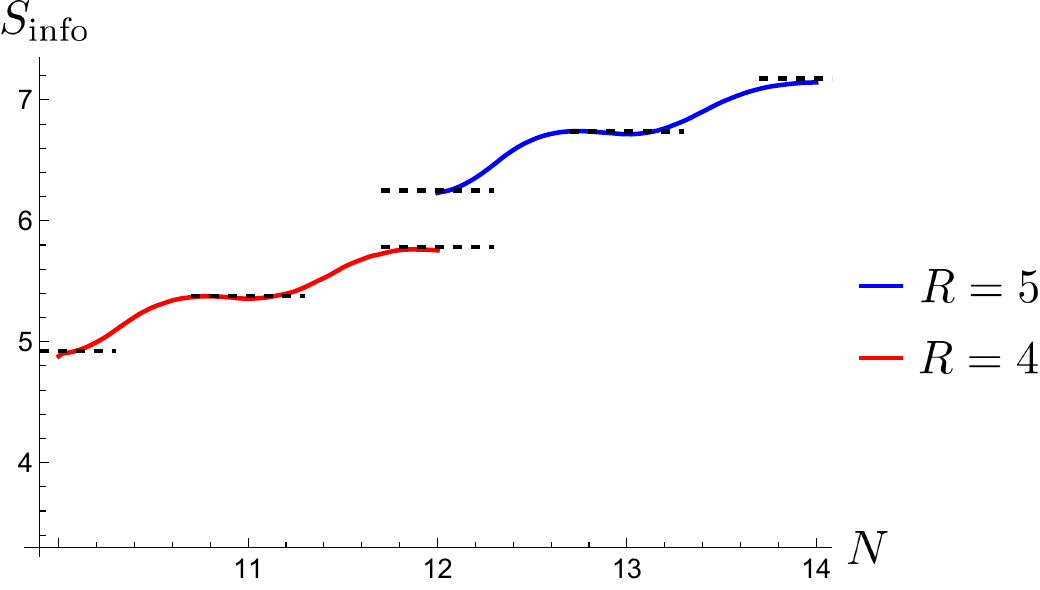}
\caption{We track the averaged information entropy for the fortuitous states with $R=4$ and $R=5$, which become BPS states at $N=10$ and $N=12$, respectively. We use dashed lines to denote the information entropy corresponding to a random state at integer values of $N$.}
\label{fig:followinfo}
\end{center}
\vspace{-1em}
\end{figure}

We can further follow the properties of individual fortuitous states as we vary $N$. Here let's focus on the information entropy of states, similar to the discussion of \cite{Budzik:2023vtr}.  In Figure \ref{fig:followinfo}, we show the results of following the averaged information entropy of the fortuitous states with $R=4$ through its journey from $N=12$ to $N=10$. We also display an analogous sector with $R=5$, following from $N=14$ to $N=12$. In both cases, we find that the information entropy is always close to saturate the value for random states $S_{\textrm{info,random}}(N,R)$ at integer $N$, indicated by the dashed lines in the plot. We note that as $N$ is decreased, the information entropy also decreases. This is inevitable since the size of the physical Hilbert space shrinks; the important feature that Figure \ref{fig:followinfo} demonstrates is that the information entropy of the fortuitous states is always close to that of a random state. Therefore, here we can see explicitly that the non-BPS states carry chaos with them into the BPS subspace through fortuity \cite{Chen:2024oqv}.

\subsection{Following $N$ in the $K=2$ two-flavor models} \label{sec:followmono}

We can generalize the discussion of following $N$ in Section \ref{sec:genefollow} to the two-flavor models in (\ref{eqn:monmodel}), by defining a continuous family of models with couplings
\begin{equation}\label{followmono}
    C^{\alpha}_{ijk} (N) = w_{i} w_{j} w_k  C_{ijk}^{\alpha}\, ,
\end{equation}
where $w_i$ is still given by (\ref{wa}). Since $   C^{\alpha}_{ijk} (N)$ still has the property that it is symmetric in exchanging $j,k$, we still have the same monotonous cohomology classes represented by (\ref{Valphabeta}) even for $N$ that is non-integer. 
Therefore, as opposed to the rich behavior for the fortuitous states in Figure \ref{fig:followN} (b) when we follow $N$, the monotone states simply have $E=0$ independent of $N$.\footnote{This is a feature of our specific interpolation (\ref{followmono}) rather than a general feature, as one could in principle go through generic two-flavor models for non-integer $N$.}

\begin{figure}[t]
\begin{center}
\includegraphics[width=0.6\textwidth]{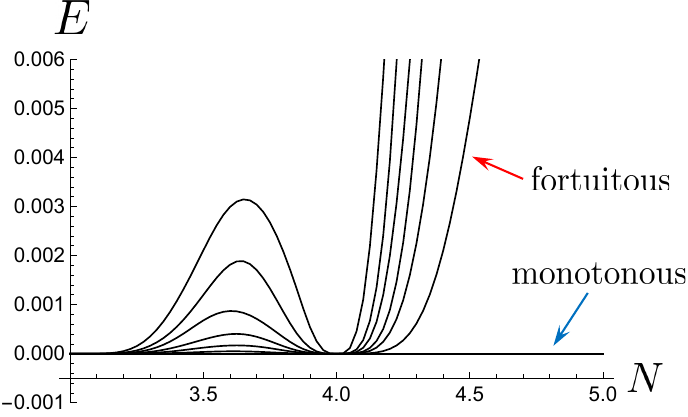}
\caption{In the sector with ${\bf N_\psi} = {\bf N_\chi} = 1$, we have three monotonous BPS states that remain BPS as we follow $N$, as indicated by the blue arrow. We further have some amount of fortuitous BPS states, indicated by the red arrow. We only plotted a subset of the fortuitous states to avoid visual clutter.}
\label{fig:followmonotone}
\end{center}
\vspace{-1em}
\end{figure}

 As a concrete example, we consider the $K=2$ two flavor model. We focus on the charge sector with ${\bf N_\psi} = {\bf N_\chi} = 1$ and follow $N$ from $5$ to $3$. When $N=5$, this charge sector only contains three monotonous states $V^{a} \ket{\Omega}$, discussed in (\ref{3grav})  (as can be seen in (\ref{PauliBPS})), while when $N=3,4$, there are also fortuitous BPS states in this sector. In Figure \ref{fig:followmonotone}, we show the results of following $N$. As we can see, there are indeed two distinct behaviors of states as we vary $N$, agreeing with our discussion. The monotonous states remain BPS as we vary $N$, while the fortuitous states are only BPS when $N = 3,4$.

\section{$R$-charge concentration and supercharge chaos}\label{sec:concentration}

In Conjecture \ref{conj:Qchaos}, we have proposed universal features for generic $q$-local supercharges called ``supercharge chaos". One of the key consequences of supercharge chaos is $R$-charge concentration. In previous sections, we've seen that both the $\mathcal{N}=2$ SUSY SYK model and the ordinary two-flavor model exhibit $R$-charge concentration. In our modified two-flavor models, the supercharge is not fully generic due to the existence of monotonous states. However, locally around the fortuitous states we still have $R$-charge concentration. 

In this section, we study some further examples to illustrate the idea of supercharge chaos. In Section \ref{sec:sparse}, we consider tuning the supercharge away from genericity and study how the $R$-charge concentration breaks down. In Section \ref{sec:N=4}, we discuss how our discussion could apply to the $\mathcal{N}=4$ SYM theory.

\subsection{Sparse supercharge and the breakdown of concentration}\label{sec:sparse}

To illustrate the connection in Conjecture \ref{conj:Qchaos} between the concentration of the BPS states and the genericity of the supercharge, we consider a sparse-version of the $\mathcal{N}=2$ SUSY SYK model \cite{Xu:2020shn}, where one introduces a sparsity parameter $p_s \in (0,1]$, which corresponds to the probability with which the random coupling $C_{ijk}$ takes non-zero value. The closer $p_s$ is to zero, the sparser the random coupling is. We would like to see how robust the concentration of BPS states is with respect to the sparseness parameter $p_s$. 

\begin{figure}[t]
\begin{center}
\includegraphics[width=0.6\textwidth]{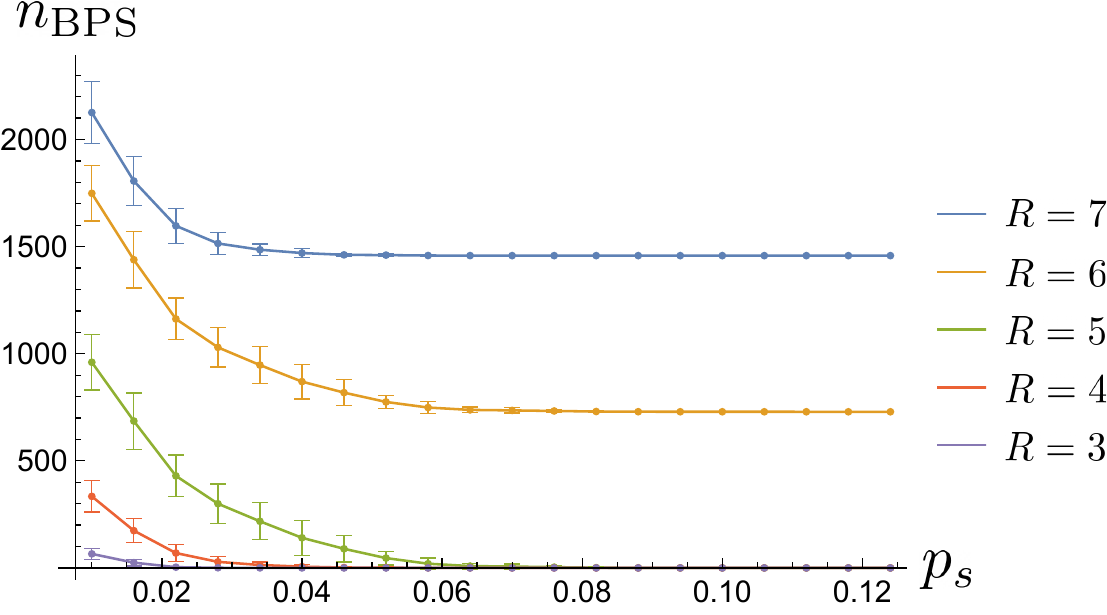}
\caption{We look at the number of BPS states in various charge sectors as a function of the sparseness parameter $p_s$. Here $N=14$ and $q=3$. Each dot represents the average of a sample size of $100$; the error bars indicate the standard deviation within each sample.}
\label{fig:deconcentration}
\end{center}
\vspace{-1em}
\end{figure}

Using exact diagonalization, we find that the concentration is highly robust against making the random couplings sparse. In Figure \ref{fig:deconcentration}, we show the results from exact diagonalization in the model with $N=14$ fermions. We see that the deconcentration of BPS states only starts to happen when $p_s$ is rather small, here $\sim 10^{-2}$. From the numerics, one also notices that the deconcentration in different charge sectors happens at different $p_s$ - the charge sectors that are closer to $R= N/2$ first start to deconcentrate. Importantly, the deconcentration is not at the cost of decreasing the number of BPS states in the original sectors, but rather, the (refined-)index is no longer saturated. Another interesting feature in Figure \ref{fig:deconcentration} is that, after the deconcentration happens, the number of BPS states $n_{\textrm{BPS}}$ has large fluctuations, depending on the different random choices of subsets of $C_{ijk}$ to be non-zero. 

We would like to see how the critical sparseness parameter $p_{s}^{\textrm{crit}}$ for deconcentration scales with $N$ as we take $N\rightarrow \infty$. However, from Figure \ref{fig:deconcentration} we can already see that one could in principle define $p_{s}^{\textrm{crit}}$ in various ways as different charge sectors deconcentrate at different points. In principle, we also might not be too interested in the regime where we only have an order one number of extra BPS states in the large $N$ limit, rather we would like to look at $p_{s}^{\textrm{crit}}$ such that the deconcentrated BPS states have an entropy of $\mathcal{O}(N)$. Taking these into account, we propose the following working definition for $p_{s}^{\textrm{crit}}$ (for $N$ even and $q=3$)
\begin{equation}
    p_s^{\textrm{crit}}(N,\alpha)  \equiv \textrm{$p_s$ such that on average the charge sector $R = \frac{N}{2} - 2$ has $\log n_{\textrm{BPS}}= \alpha N$.}
\end{equation}
Of course, since $n_{\textrm{BPS}}$ is upper bounded by the total Hilbert space dimension, we need to consider $\alpha < \frac{1}{N}\log \binom{N}{N/2-2} \sim \log 2$.  
In the larger $N$ limit, we expect that $p_s^{\textrm{crit}}$ should go to zero as a power-law,
\begin{equation}\label{pspower}
      p_s^{\textrm{crit}}(N,\alpha) \sim \frac{c(\alpha)}{N^{\gamma}}
\end{equation}
where $c(\alpha)$ is a constant that depends on $\alpha$. 

\begin{figure}[t]
\begin{center}
\includegraphics[width=0.6\textwidth]{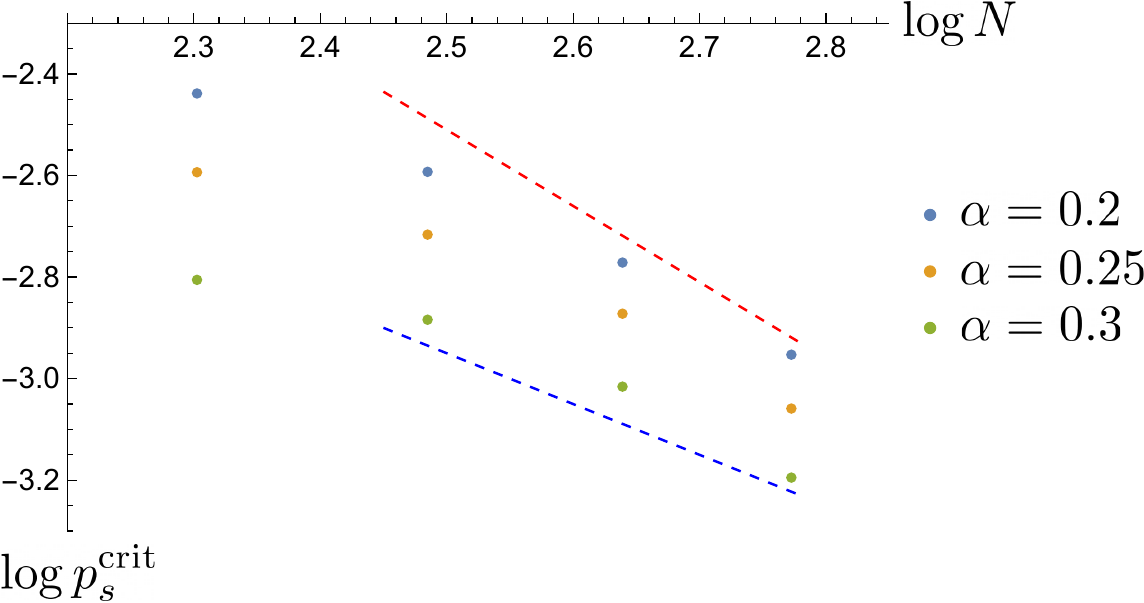}
\caption{We numerically determine $p_s^{\textrm{crit}}(N,\alpha)$ for $N = \{ 10,12,14,16\}$ and $\alpha = \{ 0.2,0.25,0.3\}$. The sample size, for different $N$ but with fixed $\alpha$ and $p_{s}$ are $\{1000,500,200,80$\}, respectively. We also overlap two dashed lines with slope $-1$ (in blue) and $-\frac{3}{2}$ (in red) for comparison.}
\label{fig:sparseness}
\end{center}
\vspace{-1em}
\end{figure}

In Figure \ref{fig:sparseness}, we show the numerical results of $p_s^{\textrm{crit}}(N,\alpha)$ for a range of $N$ and $\alpha$. We see that at the range of $N$ that is accessible with our numerics, the power-law ansatz (\ref{pspower}) is not yet a very good approximation, though a preliminary estimate for $\gamma$ would be greater than one.

The main conclusion of our numerical analysis is that $R$-charge concentration is rather robust against making the couplings sparse. We note that as the couplings become sparse, the SYK models also undergo a chaotic-to-integrable transition \cite{Orman:2024mpw}. It will be interesting to compare the onset of this transition to (\ref{pspower}), which would provide a comparison of the notion of supercharge chaos and many-body chaos.

\subsection{Comments on the $\mathcal{N}=4$ $\textrm{SU}(N)$ SYM theory}\label{sec:N=4}

The phenomenon of $R$-charge concentration is not specific to the SYK models. In the $\mathcal{N}=4$ SYM theory, we know from gravity that (pure) supersymmetric black hole solutions only exist when the angular momenta $J_1, J_2$ and charges $Q_1, Q_2, Q_3$ of the black holes satisfy a non-linear constraint \cite{Chong:2005hr}. This has been puzzling since it appears to suggest that the BPS states in the dual field theory are concentrated in charge sectors satisfying the constraint. At the level of effective theory, we can explain the concentration using the Schwarzian theory governing the AdS$_2$ part of the geometry, as discussed in \cite{Boruch:2022tno} and reviewed in the Introduction. However, a boundary explanation of the constraint in the UV theory has not been understood (see however \cite{Cabo-Bizet:2024gny}).

We should note that recently, there has been potentially an important update to this picture. Supersymmetric black hole solutions that \emph{do not} satisfy the charge constraints have been proposed \cite{Minwallatalk,Choi:2025lck}. However, these solutions are not ``pure" black hole solutions, but rather are black hole cores dressed by gravitons or D-branes that are far away from the horizon \cite{Kim:2023sig,Choi:2024xnv}. On the boundary side, dressing a black hole core could be understood as taking the product of ``core" fortuitous cohomology classes with monotone cohomology classes \cite{Choi:2023znd}. The existence of these solutions is therefore intimately connected to the existence of monotonous operators in the $\mathcal{N}=4$ SYM theory. However, if we focus on the black hole core, or the core fortuitous operators, we are still left with the puzzling feature of concentration.

Our SYK story suggests a potential microscopic perspective of the non-linear charge constraint. Let's consider the problem of finding $Q$-cohomology classes in the $1/16$-th BPS sector of $\mathcal{N}=4$ SYM. It is conjectured that the number of cohomology classes is one-loop exact \cite{Grant:2008sk} and it suffices to consider the action of the tree level supercharge $Q$ on the classical BPS states, i.e. all the states that saturate the BPS bound in the free theory.

Now, consider a cochain complex, if we are completely blind to the details of the supercharge and assume it to be completely generic, Conjecture~\ref{conj:Qchaos} predicts that the BPS states should be all concentrated completely in a specific space in the cochain complex. Even though the true $Q$ is not quite generic for the BPS states to be completely concentrated, we still expect it to be generic in the sector of core fortuitous states.
Assuming the $R$-charge concentration in this sector, we determine where the core fortuitous states are concentrated using the superconformal index
\ie
I(\omega_1,\Delta_a)=\Tr\left[(-1)^{2J_2}e^{-\omega_1 (J_1-J_2)-\Delta_1 (Q_1+J_2)-\Delta_2 (Q_2+J_2)-\Delta_3 (Q_3+J_2)}\right]\,,
\fe
where we set our convention that the supercharge that has $(J_1,J_2,Q_1,Q_2,Q_3)=(-\frac12,-\frac12,\frac12,\frac12,\frac12)$, and the $R$-charge is $R=2J_2$. The combinations $J_1-J_2$ and $Q_a+J_2$ commute with the supercharge $Q$, and are regarded as flavor charges. In the large $N$ limit, the index becomes \cite{Cabo-Bizet:2018ehj,Choi:2018hmj,Benini:2018ywd}
\ie
\log I(\omega_1,\Delta_a)\approx N^2\frac{\Delta_1\Delta_2\Delta_3}{2\omega_1(\Delta_1+\Delta_2+\Delta_3-\omega_1-2\pi i)}\,.
\fe
The index with fixed flavor charges is given by the inverse Laplace transform,
\ie
I_{J_1-J_2,Q_a+J_2}&={\rm Tr}_{J_1-J_2,Q_a+J_2}\,(-1)^{2J_2}
\\
&\approx\int \d\omega_1 \d\Delta_1\d\Delta_2 \d\Delta_3\,e^{\omega_1(J_1-J_2)+\sum_{a=1}^3\Delta_a (Q_a+J_2)+N^2\frac{\Delta_1\Delta_2\Delta_3}{2\omega_1(\Delta_1+\Delta_2+\Delta_3-\omega_1-2\pi i)}}\,.
\fe
In the large $N$ limit, we evaluate the integral by saddle point approximation that is equivalent to extremizing the exponent of the integrand
\ie
S(j_1-j_2,r_a+j_2)&=\underset{\omega_1,\Delta_a}{\rm Ext}\,\left[\frac{\Delta_1\Delta_2\Delta_3}{2\omega_1(\sum \Delta_a - \omega_1-2\pi i)}+\omega_1 (j_1-j_2)+\sum\Delta_a (r_a+j_2)\right]\,,
\fe
where $j_i=J_i/N^2$ and $q_a=Q_a/N^2$. The $R$-charge concentration implies $I_{J_1-J_2,Q_a+J_2}=D^{\rm BPS}_{J_i,Q_a} e^{2\pi \i N^2 j_2}$. Matching the phases, we find
\ie
 j_{2,c}=\frac{1}{2\pi}\,{\rm Im}\,S(j_1-j_2,r_a+j_2)\,,
\fe
which is exactly the non-linear charge relation satisfied by the pure black hole solutions or the black hole cores of the dressed black holes.\footnote{See also recent discussion in \cite{Larsen:2024fmp} which derived an analogous but different constraint from free field countings. }
Of course, our calculation here is identical to \cite{Cabo-Bizet:2018ehj,Choi:2018hmj}, what's new is really the microscopic interpretation.

So far, our discussion of $R$-charge concentration in $\mathcal{N}=4$ SYM and the non-linear charge constraint assumes that we are focusing on the core fortuitous cohomology classes, or the black hole core. If we instead choose to focus on the full geometry, then one might naively interpret the results of \cite{Minwallatalk} as saying that $R$-charge concentration breaks down completely. However, we shouldn't rush to such conclusions so fast. We note that, what the results in \cite{Minwallatalk} suggest is that along a cochain complex that contains the pure black hole solution, there also exists black hole solutions dressed by gravitons or dual diants. It is in fact possible that these other solutions are far apart from the pure black hole along the cochain complex, such that at order one degree away from the space describing the pure black hole we do not have other BPS states.  
This is consistent with the fact that in the construction of \cite{Minwallatalk} and related work \cite{Kim:2023sig,Choi:2024xnv}, the dressings carry large charge and are far separated from the horizon. Instead, dressing the black hole by a simple scalar hair has been found to not leading to new SUSY solutions \cite{Dias:2024edd}.
Further evidence for this picture comes from the partial ``no-hair" property of the core fortuitous operator discussed in \cite{Choi:2023znd}, where it was found in some cases that the fortuitous states can only be dressed by graviton operators carrying large enough quantum numbers. 
It would be useful to understand this from the Schwarzian point of view, perhaps by trying to couple Schwarzian to some additional BPS particles.\footnote{Our discussion here only applies to gray galaxy dressed solutions and dual giant graviton dressed solutions. As was pointed out in \cite{Kim:2023sig,Choi:2025lck}, there exists another family of dressed black holes that are ``revolving black holes" - black holes that revolve around in the AdS space. We don't expect our comments to apply for them. We thank Shiraz Minwalla for reminding us about them. }

We've argued that the phenomenon of $R$-charge concentration might not only apply to SYK models, but also to $\mathcal{N}=4$ SYM. In fact, there are some further schematic similarities between SYK and SYM.  The expression for the classical supercharge in the $1/16$-BPS sector of $\mathcal{N}=4$ SYM can be cast into the following form
\begin{equation}\label{Qoneloop}
  Q =  \textrm{Tr}\left[ \Psi \Psi \frac{\delta}{\delta \Psi}\right]
\end{equation}
where $\Psi$ is an $N\times N$ ``superfield". We refer the readers to \cite{Chang:2013fba,Chang:2022mjp} for more discussion of the formalism. For now, we simply notice that if we expand the trace expression into components, it looks schematically like that of SYK models, with the differences being that the couplings are very sparse (and non-random), and we also have dynamical bosons. We note that we have $\tilde{N} \equiv N^2$ components of $\Psi$, while by expanding out (\ref{Qoneloop}) we get $N^3 = \tilde{N}^{\frac{3}{2}}$ terms, so we can estimate the sparseness parameter as
$ p_s \sim \tilde{N}^{\frac{3}{2}}/ \tilde{N}^3 = 1/ \tilde{N}^{\frac{3}{2}}.$ 
Of course, this cannot be compared directly with (\ref{pspower}) due to other differences of the two models.

As another toy model whose supercharge possesses algebraic structure that is analogous to  (\ref{Qoneloop}), we might consider the model mentioned briefly around (\ref{eqn:BRST_Q}). For simplicity, we set the linear term in $\psi$ to be zero and we have
\begin{equation}
    Q = \sum_{i,j,k} f_{ij}^k \, \psi^i \psi^j \bar{\psi}_k.
\end{equation}
The nilpotency condition of $Q$ demands 
the couplings \( f_{ij}^k \) obeying the Jacobi identity, therefore they can be chosen as the structure constants of a Lie algebra, such as \( \mathfrak{su}(N) \). This construction lends itself naturally to a reformulation in terms of a fermionic matrix model by introducing the \( \text{SU}(N) \) generators \( T_i \) with the following properties:
\begin{equation}
    \psi = \psi^i T_i, \quad \bar{\psi} = \bar{\psi}_i T^i, \quad [T_i, T_j] = \i f_{ij}^k T_k, \quad \text{Tr}(T_i T_j) = \delta_{ij}.
\end{equation}
With these definitions, we can rewrite the supercharge and Hamiltonian in terms of single-trace operators as follows:
\begin{equation}
    Q = -\frac{\i}{2} \, \text{Tr}\left[\psi \psi \bar{\psi}\right], \quad H = \{ Q,Q^{\dagger}\} = \text{Tr}\left[\psi \psi \bar{\psi} \bar{\psi}\right] + \text{Tr}\left[\{\psi, \bar{\psi}\}^2\right].
\end{equation}
As we discussed in Section \ref{sec:Qchaos}, this model is non-generic compared to the standard $\mathcal{N}=2$ SUSY SYK model. Nonetheless, it would be interesting to understand whether even this model exhibits $R$-charge concentration. The cohomology of this model has interesting connections to Lie algebra cohomology, which we leave for future study.

\section{Discussion}\label{sec:discussion}

In this paper, we studied the notions of fortuity and monotony in SYK models. By studying $\mathcal{N}=2$ SUSY SYK as a toy model, we uncover a close connection between fortuity and the universal Schwarzian description of the near-BPS black holes, both of which are tied to a ``smoking-gun" feature we term $R$-charge concentration - all the BPS states are concentrated in a single space along an irreducible cochain complex. 

We conjecture $R$-charge concentration as a universal property for a generic $q$-local supercharge, which is also approximated by the Turiaci-Witten random matrix ensemble and is governed by the Schwarzian theory near the BPS states. We summarize these features of a generic supercharge as ``supercharge chaos" (see Conjecture \ref{conj:Qchaos}).  We expect that for holographic large $N$ CFTs, the action of the supercharge near their fortuitous states should behave as a generic supercharge.

Utilizing the flexibility of SYK models, we also generalized the two-flavor model in \cite{Heydeman:2022lse} such that it contains both fortuitous states and monotonous states. We find that they display sharply different fine-grained properties. Using various measures including the LMRS criterion, information entropy and entanglement entropy, we find that the fortuitous states resemble random states while the monotonous states do not.

In the following, we point out some open questions that are worth further investigation.
  
\paragraph{Testing the supercharge chaos conjecture}

In our Conjecture \ref{conj:Qchaos}, we conjectured  the universal properties of a generic $q$-local supercharge. Clearly, further work is required in understanding the validity of the conjecture, and, to what extent such properties are realized in non-disordered theories, such as the ${\cal N}=4$ SYM, ABJM theory and the D1-D5 CFT.\footnote{For the quarter BPS states in the D1-D5 CFT,  one would need to apply $\mathcal{N}=4$ JT, whose precise random matrix description is not fully understood \cite{Turiaci:2023jfa}. It would be interesting to understand how various aspects of our story generalizes to the $\mathcal{N}=4$ context. The supercharge $Q$-cohomology in the D1-D5 CFT was recently studied in \cite{Chang:2025rqy}.}

To test our conjecture for generic supercharges, it would be useful to study various models with disorder, where the notion of genericity can be established more concretely. These models may exhibit distinct features from the SYK models we have studied, potentially introducing tensions with our conjecture. 
Such models are often solvable using similar large \( N \) techniques as SYK, which is helpful to check not only the aspect of $R$-charge concentration, but also the domination of the Schwarzian mode as well as random matrix properties (such as strong chaos of BPS states).  For instance, the \( \mathcal{N}=2,4 \) SYK models with dynamical bosons  \cite{Anninos:2016szt,Biggs:2023mfn} display non-conformal and non-chaotic IR phases, despite having a large index. The Murugan-Stanford-Witten model \cite{Murugan:2017eto,Bulycheva:2018qcp} and its generalizations to disordered Landau-Ginzburg models \cite{Chang:2023gow} and three-dimensional models \cite{Chang:2021fmd,Chang:2021wbx} exhibit a sub-maximal chaos exponent, but one could wonder whether maximal chaos and Schwarzian dominance is restored in some larger charge sectors.

For the non-disordered theories, the first and foremost task is to establish the notion of fortuity and locate the fortuitous states, where we expect the supercharge to behave as generic. For $\mathcal{N}=4$ SYM theory, where the notion of fortuity has been understood \cite{Chang:2022mjp}, one might hope to numerically verify the $R$-charge concentration phenomenon, for core fortuitous states, at reasonably large $N$ and charge. One can also hope to directly verify the prediction from the Turiaci-Witten ensemble through studying the statistics of the one-loop anomalous dimensions of classical $1/16$-th BPS states, by improving upon the study of \cite{Chang:2023zqk}. More generally, it is important to develop further tools in studying the BPS sectors of non-disorder theories in the large charge and large $N$ limit and compare their properties with disordered theories \cite{Benini:2022bwa,turiaciwip}. 

\paragraph{Features of $q$-locality in the fortuitous states}
The supercharge chaos conjecture states that the fortuitous states of a $q$-local supercharge should be approximated by random states. It is interesting to understand how well this approximation holds.
This amounts to estimating the corrections to the random matrix prediction for physical quantities, arising from the constraint of $q$-locality, such as in the supercharge of the SYK model.

From the bulk perspective, this question is translated to estimating various corrections to the pure gravity prediction on various spacetime geometries.
One possible correction arises from bulk matter fields.
In the leading disk geometry, the super-Schwarzian effect dominates over matter corrections in the near-BPS limit.
On a wormhole geometry, there are additional integrals over the moduli space, whose measure could depend on the matter. Here the most important correction comes from wormholes containing short geodesic cycles, where one-loop effects of matter can lead to potential tachyonic divergence. We anticipate that a UV-complete theory would regularize this tachyonic divergence while still leaving a nontrivial measure on the moduli that is theory-dependent.  In the non-supersymmetric case, a further low energy limit takes the JT matrix model into the Kontsevich model\cite{kontsevich1992intersection}. This is the strip limit of the wormhole geometries and there will be no short geodesic cycles\cite{do2008intersection,Saad:2022kfe}.
This means the matter one-loop corrections should be negligible.
For the BPS states, the spectrum does not fluctuate and instead one examines the chaotic behaviour of the projector into the BPS subspace \cite{Lin:2022zxd}.  The wormholes involved in this case are the LMRS type wormholes for the projected simple operators, where the boundary involves an infinite Euclidean time evolution which again leads to large size wormholes. The analysis of \cite{Lin:2022zxd} shows that the typical length of the Einstein-Rosen bridge is of order $\log S_{\textrm{BPS}}$ due to quantum fluctuations, indicating that the one-loop effect from matter will be polynomially decaying in terms of $S_{\textrm{BPS}}$.  In SYK, there are order $N$ light particles, and we expect the matter correction to be negligible after an order-one time scale for the LMRS wormhole, compared to the $\log N$ scaling for the Thouless time of the regular spectral form factor of SYK \cite{Saad:2018bqo}. This suggests that, in some sense, the random matrix behaviour is more prominent in the LMRS analysis.

However, in SYK, there are other corrections not captured by the above analysis, which scale as $N^{-q}$\cite{Maldacena:2016hyu,Garcia-Garcia:2018ruf, Jia:2019orl,Berkooz:2020fvm}.\footnote{We thank Douglas Stanford for reminding us about this.}
The leading correction arises from sample-to-sample fluctuations of the average spectral density. In the low energy limit, such corrections to the edge spectrum were recently investigated in \cite{Altland:2024ubs}. The study indicates convergence between the near edge spectral density of the SYK model and the Airy spectral density after incorporating such corrections. It would be interesting to study such $N^{-q}$ corrections to the eigenvectors, for instance, the variation in the range of eigenvalues of the projected operator in LMRS analysis, which is linked to the ensemble fluctuation of its two-point function. This could shed light on the distinction between fortuitous states and Haar random states. Of course, this question is not limited to BPS states, one can also ask the same question for the other energy eigenstates.

\paragraph{Fragility of monotonous states} An interesting difference between monotonous and fortuitous states in SYK models is their behavior under small perturbations of the coupling constants $C_{ijk}$. The existence of monotonous states is highly sensitive to the fine-tuned structures in the coupling constants.  For example, for the model in (\ref{Qexample}), if we perturb the couplings $C_{ijk}$ slightly such that it is no longer symmetric under exchanging $j,k$, we will immediately lose the monotonous BPS states. In this sense, monotonous states are rather fragile. In contrast, fortuitous states are highly robust since they exist for generic choices of couplings. A difficulty in generalizing this distinction to higher dimensions is that for superconformal field theories, the minimal SUSY is larger than the ${\cal N}=2$ supersymmetry algebra. Therefore, they are more rigid and do not allow generic deformations while preserving only the $\mathcal{N}=2$ SUSY.     
Nevertheless, it would be interesting to see whether there is an analogue of this feature in higher dimensions and understand how it is reflected in the bulk descriptions.

\paragraph{Monotony and gauge principle} 
In non-disordered holographic theories, such as the  ${\cal N}=4$ SYM and the D1-D5 CFTs, one can define an infinite $N$ Hilbert space and compute its $Q$-cohomology. These $Q$-cohomology classes are, by definition, monotonous and match perfectly with the bulk BPS spectrum of light fields around vacuum ${\rm AdS}$. In contrast, in the disordered models, we are unable to define an infinite $N$ Hilbert space. This difficulty is tied to the absence of a gauge principle in these models. In the non-disordered examples, the gauge principle reduces the number of non-gauge-invariant light states, which grows with $N$, to the number of gauge-invariant light states, which grow as order one. In the disordered models, however, the number of light states grows with $N$, reflecting an ${\cal O}(N)$ number of light fields in the bulk. From this perspective, the disordered models are not suited as holographic descriptions of gravity around the AdS vacuum.

However, in Section~\ref{sec:defns}, we proposed an alternative definition of monotonous states that does not rely on the existence of an infinite \( N \) Hilbert space. In the models we studied, this definition yields a finite spectrum of monotonous states in certain charge sectors, such as the ${\bf N}_\psi={\bf N}_\chi$ sector in the two-flavor models. From the bulk perspective, this implies that while there are many non-BPS light fields, the number of BPS light fields could remain of order one if we restrict to appropriate charge sectors. In the two-flavor model, we further argued that there is an \({\cal O}(N)\) energy gap above such monotonous states. Thus, in disordered models, the monotonous sector could have a well-defined bulk dual as BPS excitations around the AdS vacuum. The concept of monotony replaces the role of the gauge principle in the BPS case, ensuring a finite spectrum of light states.

\paragraph{Other solutions in the Turiaci-Witten ensemble}
The random matrix ensemble proposed by Turiaci and Witten \cite{Turiaci:2023jfa} (see also further discussion in \cite{Johnson:2023ofr,Johnson:2024tgg}) describes generic supercharges constrained by the $\mathcal{N}=2$ supersymmetric algebra, particularly the nilpotency condition $Q^2 = 0$ of the supercharge $Q$. 
In general, the family of solutions to the $Q^2 = 0$ constraint is parametrized by its BPS spectrum, which can be more general than the concentrated spectrum for JT gravity discussed in their paper. Genericity only ensures that there are no BPS states in two adjacent spaces in the cochain complex, as the presence of such states would allow them to combine into a long supermultiplet and be lifted under generic perturbations. 
Once such a distribution of BPS state is given, it is stable under small deformations. This is because, in order to change the BPS spectrum, at least some of the long multiplets must become BPS. This is non-generic because under small deformations, the number of nonzero eigenvalues of the supercharge acting on each space in the cochain complex, corresponding to the number of long supermultiplets, could only increase.\footnote{The discussion around equation (\ref{eqn:polyconstaint}) is a specific example of this property in the SYK case.} 
This means in the space of all possible supercharges, there are different classes of generic supercharges labeled by their BPS spectrum. The path of deformation from a generic supercharge in one class to a different class must pass through points associated to non-generic supercharges. 
Turiaci-Witten shows that in each class the supermultiplets in different $R$-charge sectors are statistically independent, each governed by an Altland-Zirnbauer class with parameters $(\alpha, \beta) = (1 + 2\nu, 2)$, where $\nu$ denotes the number of BPS states in the corresponding space.

In other words, the random matrix ensemble dual to $\mathcal{N}=2$ JT gravity, which exhibits the peculiar feature of $R$-charge concentration, is only a special family of solution allowed by the general Turiaci-Witten $\mathcal{N}=2$ ensemble. 
The supercharge chaos conjecture basically is saying that by imposing $q$-locality, the supercharge belongs to this family.
But, there are other matrix ensembles obey the $\mathcal{N}=2$ SUSY algebra that do {\it not} satisfy $R$-charge concentration.

A natural question arises: given a vast families of solutions to $Q^2 = 0$ as parametrized by different distributions of BPS states, can one define a probability distribution and determine which is the most likely? A naive approach would assign a probability distribution based on the number of independent components in $Q$ after specifying the distribution of the BPS states. Such an approach would favor having the least amount of BPS states as demanded by the index. Given that the index is saturated, such an approach would however favor families where the BPS states are spread out rather than concentrated.
This preference arises because, crudely speaking, the number of independent components in $Q$ in each space in the cochain complex scales quadratically as the number of non-BPS states. Therefore, starting from a concentrated distribution, by moving BPS states into other charge sectors, the increase of independent components in the previously concentrated sector wins over the decrease in other charge sectors.

This discrepancy presents an interesting puzzle, as the simplest prediction from random matrix theory deviates from the gravitational answer. The compelling question then is whether there exists a natural random matrix measure that leads to $R$-charge concentration without imposing it as an additional input as in \cite{Turiaci:2023jfa}.
The space of solutions to the quadratic equation $Q^2 = 0$ with an $R$-symmetry is a subvariety of a nilpotent cone.
Another interesting question is to study the topological properties of this subvariety. 

Another intriguing question concerns the bulk dual of the non-concentrated solutions.\footnote{We emphasize that the deconcentration effect here should not be confused with the case of sparse $\mathcal{N}=2$ SYK discussed in Section \ref{sec:sparse} or the case of grey galaxies in AdS$_5$, where there are BPS states in adjacent spaces in the cochain complex that could be lifted under perturbation.} What gravitational interpretation, if any, corresponds to these configurations?

\paragraph{Fortuity and null states}

An intriguing aspect of fortuity is that it connects two seemingly different topics: the physics of BPS black hole microstates and the discussion of null states in quantum gravity. To make this connection more explicit, let's revisit the fortuity phenomenon in $\mathcal{N}=4$ SYM theory. A fortuitous operator $O_\textrm{f}$ that becomes BPS at $N=N_*$ satisfies the equation $Q O_\textrm{f} \in \mathcal{I}_{N_*}$ while $Q O_\textrm{f} \notin \mathcal{I}_{N_*+1}$, where $\mathcal{I}_{N}$ is the set of trace relations in the $\textrm{SU}(N)$ theory. One way of interpreting this phenomenon is the following. When $N>N_*$, $\{  O_\textrm{f}, Q O_\textrm{f}  \}$ are two non-BPS black hole microstates in the same supermultiplet. They have exactly the same anomalous dimension but opposite statistics. When $N$ is decreased towards $N_*$, the energies of these two states decrease towards the BPS  bound. Eventually, at $N=N_*$, as the operator $O_\textrm{f}$ becomes BPS, the superpartner black hole microstate $Q O_\textrm{f}$ becomes a null state.\footnote{A trace relation is a null state from the point of view of the infinite $N$ theory - a state whose norm is zero if one uses the finite $N$ inner product.}
This perspective shows that the trace relation that is associated with fortuitous operator $O_\textrm{f}$ should be highly complex, since it can be smoothly interpolated into a black hole microstate by varying $N$ using the prescription of \cite{Budzik:2023vtr}. Notice that due to $R$-charge concentration, we expect an exponential number of fortuitous states becoming BPS at $N=N_*$, and as a consequence, all of their partner black hole microstates should become null at the same ``time" ($N=N_*$).

It is interesting to contrast this discussion with other occurences of trace relations in AdS/CFT. Previous studies of trace relations have mostly focused on the regime where they involve $\mathcal{O}(N)$ letters. In the bulk, they have been associated with the physics of D-branes, particularly in the context of giant gravitons \cite{Lee:2023iil}. Here we see that the trace relations associated with fortuitous operators, which contain order $\mathcal{O}(N^2)$ letters, appear to have a different bulk representation as ``phantom black holes" - bulk null states that come from extrapolation of black hole microstates. It would be very interesting if one can understand these null states using the collective field variables, such as the $G-\Sigma$ variables in the SYK models. It would be useful to understand the connection, if any, between these null states and other null states that have been discussed in the context of black holes and spacetime wormholes (see for example \cite{Penington:2019kki,Marolf:2020xie,Akers:2022qdl,Dong:2024tjx}).

\paragraph{Fortuity and average in gravity}

The fortuitous phenomenon suggests the large $N$ limit of BPS black hole states does not exist. If we wish to consider larger $N$, in order to preserve SUSY, we are forced to look at states in sectors with charges varying with $N$ and the states will change erratically. On the other hand, semiclassical gravity provides a smooth bulk description of such states in the large $N$ limit, which seems in tension with the boundary behavior. 
This is in contrast to the monotonous states, which admit a smooth extrapolation to the large $N$ limit both in the bulk and the boundary. 

The sharp distinction between the two classes of states is reflected in their bulk descriptions. For the monotonous states, which are described by horizonless geometries, one can expect a direct correspondence even at finite $N$, as everything is causally connected to the boundary. As an example, for the Lin-Lunin-Maldacena geometry in asymptotically AdS$_5 \times  S^5$ spacetime \cite{Lin:2004nb},  quantization of the classical moduli space can be found to recover the $1/2$-BPS subspace of $\mathcal{N}=4$ SYM theory, even at finite $N$ \cite{Grant:2005qc,Maoz:2005nk,Chang:2024zqi}.

On the other hand, the known bulk description for the fortuitous states appears to contain inherent average. Semiclassical gravity provides a smooth bulk description of such states in terms of supersymmetric black holes, but fails to capture their erratic dependence of them in $N$ \cite{Schlenker:2022dyo}. In lower dimensional toy models, such an average can be made explicit by ensemble averaging. For example, in the double-scaled supersymmetric SYK model, averaging over the ensemble of coupling constants yields an effective ``chord Hilbert space"\cite{Berkooz:2018jqr, Berkooz:2020xne}, which corresponds to the bulk length Hilbert space in JT gravity \cite{Lin:2022rbf,Boruch:2023bte}. In higher dimensional models, however, the precise notion of the averaging is unclear, though we expect that the random matrix ensemble of Turiaci-Witten can be used as an approximate ensemble for coarse-grained quantities.   
The difficulty in establishing a precise bulk to boundary map, as opposed to the case of monotonous states, is signified by the existence of a horizon. Indeed, operators behind the horizon do not necessarily correspond to linear operators on the boundary \cite{Papadodimas:2013jku,Marolf:2015dia}. Instead, their boundary duals, if exist, may be highly non-linear, with complexity being a plausible candidate\cite{Susskind:2014rva, Bouland:2019pvu}.

Of course, these issues are not unique to fortuitous BPS states but reflect more general difficulties in understanding the non-perturbative bulk description of black hole microstates. Nonetheless, given the sharp connection of fortuity to finite $N$ effects on the boundary side, and with the help of extra machinery from supersymmetry, fortuitous states provide a valuable playground in trying to address these deep questions about black holes.

\section*{Acknowledgements} 

We would like to thank Douglas Stanford for many insightful discussions at various stages of this project. We would like to thank Iosif Bena, Jan Boruch, Matthew Heydeman, Luca Iliesiu, Seok Kim, Henry Lin, Jingru Lu, Juan Maldacena, Henry Maxfield, Shiraz Minwalla, Manqian Ou, Cheng Peng, Mukund Rangamani, Stephen Shenker, Gustavo Turiaci, Jinzhao Wang, Yifan Wang, Hui Zhai, Ziruo Zhang for helpful discussions. CC is partly supported by the National Key R\&D Program of China (NO. 2020YFA0713000). YC acknowledges support from DOE grant DE-SC0021085. ZY acknowledges support from NSFC12342501. CC and ZY thank the hospitality of Southeast University and Peng Huanwu Center for Fundamental Theory, Hefei, where part of the work was done during the visit.

\appendix

\section{Details of the full Schwinger-Dyson equations}\label{app:SD}

In this appendix, we present some further detail on the analysis in Section \ref{sec:SDequation} of the large $N$ properties of the modified two-flavor models without assumptions that the mixed correlator $\langle \bar X \Psi\rangle$ vanishes, so we need to introduce another $\mathcal{G}$-$\Sigma$ variable to impose: 
\begin{equation}\label{Gfields2}
    \mathcal{G}_{\bar{X} \Psi} (Z_1, Z_2) = \frac{1}{N K} \langle \bar{X}^i_{\alpha}(Z_1) \Psi_i^{\alpha}(Z_2) \rangle,\, ~~~\text{and}~~~\mathcal{G}_{\bar{\Psi} X} (Z_1, Z_2) = \frac{1}{N K} \langle \bar{\Psi}^i_{\alpha}(Z_1) X_i^{\alpha}(Z_2) \rangle.
\end{equation}

\subsection{Components of the full Schwinger-Dyson equations}\label{SD}
We retain the assumption of 
$\textrm{U}(K)$-invariance, which leads to a full action with off-diagonal bilinear terms activated. The Lagrangian is given by:
\begin{align}
   \mathcal{L} &= \frac{1}{2} \int \d^2\theta \, \left( \bar{\Psi}^i_{\alpha} \Psi_i^{\alpha} + \bar{X}^i_{\alpha} X_i^{\alpha} \right) -  \int \mathrm{d} \bar{Z}_1 \, \mathrm{d} Z_2 \, \Sigma_{\bar{\Psi} \Psi} \bar{\Psi}^i_{\alpha} \Psi_i^{\alpha} +\, \Sigma_{\bar{X} X} \bar{X}^i_{\alpha} X_i^{\alpha}\notag \\
    &\quad - \int \mathrm{d} \bar{Z}_1 \, \mathrm{d} Z_2 \, \Sigma_{\bar{X} \Psi} \bar{X}^i_{\alpha} \Psi_i^{\alpha}+\Sigma_{\bar{\Psi} X} \bar{\Psi}^i_{\alpha} X_i^{\alpha}+\mathcal{L}_{\text{int}},
\end{align}
where the interaction term is expressed as:
\begin{equation}
\begin{aligned}   \mathcal{L}_{\text{int}} &= NK \int \mathrm{d} \bar{Z}_1 \, \mathrm{d} Z_2 \left( \Sigma_{\bar{\Psi} \Psi} \mathcal{G}_{\bar{\Psi} \Psi} + \Sigma_{\bar{X} X}\mathcal{G}_{\bar{X} X} + \Sigma_{\bar{\Psi}X} \mathcal{G}_{\bar{\Psi}X} + \Sigma_{\bar{X}\Psi}\mathcal{G}_{\bar{X} \Psi}\right) \notag \\
    &\quad - \frac{J NK}{2} \int \mathrm{d} \bar{Z_{1}} \, \mathrm{d} Z_{2} \, \left(  \mathcal{G}_{\bar{\Psi} \Psi}^2 \mathcal{G}_{\bar{X} X}  - \mathcal{G}_{\bar{X}\Psi}\mathcal{G}_{\bar{\Psi} X} \mathcal{G}_{\bar{\Psi} \Psi}\right)\,.
\end{aligned}
\end{equation}
Assuming there are no mixed correlators between bosons and fermions, the $\mathcal{G}-\Sigma$ fields admit a component expansion:
\begin{align}
    \mathcal{G}_{\bar{\Gamma}\Gamma'} (Z_{1}, Z_{2} )= G_{\bar{\gamma} \gamma'} (\tau_{1}- \theta_{1} \bar{\theta}_{1}, \tau_{2} + \theta_{2} \bar{\theta}_{2}) + 2 \overline{\theta}_{1}\theta_{2}G_{\bar{b}_{\gamma}b_{\gamma'}} (\tau_{1}, \tau_{2});\\
     \Sigma_{\bar{\Gamma}\Gamma'} (Z_{1}, Z_{2} )= {1\over 2}\Sigma_{\bar{b}_{\gamma}b_{\gamma'}} (\tau_{1}- \theta_{1} \bar{\theta}_{1}, \tau_{2} + \theta_{2} \bar{\theta}_{2}) +  \overline{\theta}_{1}\theta_{2} \Sigma_{\bar{\gamma} \gamma'}(\tau_{1}, \tau_{2}).
\end{align}
for $\Gamma=\Psi,X$ and correspondingly $\gamma=\psi,\bar\chi$.
After expanding into components, the Lagrangian separates into 
$\mathcal{L}=\mathcal{L}_0+\mathcal{L}_{\text{int}}$, The quadratic term, normalized as $\mathcal{L}_0/ NK$, becomes \footnote{We also include possible chemical potential terms $\mu_\psi$ and $\mu_\chi$ for the fermions.}:
\begin{align}\label{eqn:fullaction1}
&\iint \, 
\begin{pmatrix}
\bar\psi & \chi
\end{pmatrix}
\begin{pmatrix}
(\partial_{\tau_1}+\mu_{\psi}) \delta(\tau_1 - \tau_2) - \Sigma_{\bar\psi \psi}(\tau_1, \tau_2) 
& -\Sigma_{\bar\psi \bar\chi}(\tau_1, \tau_2) \\
-\Sigma_{\chi \psi}(\tau_1, \tau_2) 
& (\partial_{\tau_1}-\mu_{\chi}) \delta(\tau_1 - \tau_2) - \Sigma_{\chi \bar\chi}(\tau_1, \tau_2)
\end{pmatrix}
\begin{pmatrix}
\psi \\ 
\bar\chi
\end{pmatrix}\notag \\
&\quad
+\iint  \, 
\begin{pmatrix}
\bar b_\psi & \bar b_\chi
\end{pmatrix}
\begin{pmatrix}
- \delta(\tau_1 - \tau_2)  - \Sigma_{\bar b_\psi b_\psi}(\tau_1, \tau_2) 
& -\Sigma_{\bar b_\psi b_\chi}(\tau_1, \tau_2) \\
-\Sigma_{\bar b_\chi b_\psi}(\tau_1, \tau_2) 
& - \delta(\tau_1 - \tau_2)  - \Sigma_{\bar b_\chi b_\chi}(\tau_1, \tau_2)
\end{pmatrix}
\begin{pmatrix}
b_\psi \\ 
b_\chi
\end{pmatrix}.
\end{align}
The interaction term $\mathcal{L}_{\text{int}}/ NK$ is given by:
\begin{align}\label{eqn:fullaction2}
   & \iint \Sigma_{\bar{\psi} \psi} G_{\bar{\psi} \psi} + \Sigma_{\chi \bar{\chi}}G_{\chi \bar{\chi}} + \Sigma_{\bar{b}_{\psi} b_{\psi}}G_{\bar{b}_{\psi} b_{\psi}} + \Sigma_{\bar b_{\chi} b_{\chi}}G_{\bar b_{\chi} b_{\chi}}+\Sigma_{\chi \psi}G_{\chi\psi}+\Sigma_{\bar\psi\bar\chi}G_{\bar\psi\bar\chi}\notag\\
  &\quad +\iint\Sigma_{\bar b_{\chi},b_{\psi}}G_{\bar b_{\chi} b_{\psi}}+\Sigma_{\bar b_{\psi},b_{\chi}}G_{\bar b_{\psi} b_{\chi}}-J\iint G_{\bar\psi\psi}^2G_{\bar b_{\chi}b_{\chi}}+2G_{\bar\psi\psi}G_{\bar b_{\psi}b_{\psi}}G_{\chi\bar\chi}\notag\\
  &\quad+J\iint G_{\bar b_{\chi}b_{\psi}}G_{\bar\psi\bar\chi}G_{\bar\psi\psi}+G_{\bar b_{\psi}b_{\chi}}G_{\chi\psi}G_{\bar\psi\psi}+G_{\bar b_{\psi}b_{\psi}}G_{\chi\psi}G_{\bar\psi\bar\chi}.
\end{align}
where certain arguments have been omitted for clarity in obvious contexts.

From variation of the action, we can get the components of the Schwinger Dyson equations:
\begin{align}\label{eqn:SDfull}
\Sigma_{\bar{\psi} \psi} &= 2 J \big( G_{\bar{\psi} \psi} G_{\bar{b}_{\chi} b_{\chi}} + G_{\bar{b}_{\psi} b_{\psi}} G_{\chi \bar{\chi}} \big) 
- J \big( G_{\chi \psi} G_{\bar{b}_{\psi} b_{\chi}} + G_{\bar{b}_{\chi} b_{\psi}} G_{\bar{\psi} \bar{\chi}} \big), \\
\Sigma_{\bar{b}_{\psi} b_{\psi}} &= 2 J G_{\bar{\psi} \psi} G_{\chi \bar{\chi}} - J G_{\chi \psi} G_{\bar{\psi} \bar{\chi}}, \quad
\Sigma_{\chi \bar{\chi}} = 2 J G_{\bar{\psi} \psi} G_{\bar{b}_{\psi} b_{\psi}}, \quad 
\Sigma_{\bar{b}_{\chi} b_{\chi}} = J G_{\bar{\psi} \psi}^2, \\
\Sigma_{\bar{b}_{\chi} b_{\psi}} &= - J G_{\bar{\psi} \psi} G_{\bar{\psi} \bar{\chi}}, \quad
\Sigma_{\chi \psi} = - J \big( G_{\bar{b}_{\psi} b_{\psi}} G_{\bar{\psi} \bar{\chi}} + G_{\bar{\psi} \psi} G_{\bar{b}_{\psi} b_{\chi}} \big), \\
\Sigma_{\bar{b}_{\psi} b_{\chi}} &= - J G_{\bar{\psi} \psi} G_{\chi \psi}, \quad
\Sigma_{\bar{\psi} \bar{\chi}} = - J \big( G_{\bar{b}_{\psi} b_{\psi}} G_{\chi \psi} + G_{\bar{\psi} \psi} G_{\bar{b}_{\chi} b_{\psi}} \big), \\
\begin{pmatrix}
G_{\bar{\psi} \psi} & G_{\bar{\psi} \bar{\chi}} \\[10pt]
G_{\chi \psi} & G_{\chi \bar{\chi}}
\end{pmatrix}^T  &=  \begin{pmatrix}
-(\partial_{\tau_1} + \mu_\psi) \delta(\tau_1 - \tau_2) + \Sigma_{\bar{\psi} \psi}(\tau_1, \tau_2) 
& \Sigma_{\bar{\psi} \bar{\chi}}(\tau_1, \tau_2) \\[10pt]
\Sigma_{\chi \psi}(\tau_1, \tau_2) 
& -(\partial_{\tau_1} - \mu_\chi) \delta(\tau_1 - \tau_2) + \Sigma_{\chi \bar{\chi}}(\tau_1, \tau_2)
\end{pmatrix}^{-1}, \\
\begin{pmatrix}
G_{\bar{b}_{\psi} b_\psi} & G_{\bar{b}_{\psi} b_{\chi}} \\[10pt]
G_{\bar{b}_{\chi} b_{\psi}} & G_{b_\chi \bar{b}_\chi}
\end{pmatrix}^T &= -\begin{pmatrix}
\delta(\tau_1 - \tau_2) + \Sigma_{\bar{b}_\psi b_\psi}(\tau_1, \tau_2) 
&  \Sigma_{\bar{b}_{\psi} b_{\chi}}(\tau_1, \tau_2) \\[10pt]
 \Sigma_{\bar{b}_{\chi} b_{\psi}}(\tau_1, \tau_2) 
&  \delta(\tau_1 - \tau_2) + \Sigma_{b_{\chi} \bar{b}_{\chi}}(\tau_1, \tau_2)
\end{pmatrix}^{-1}.\label{eqn:SDfull2}
\end{align}
where the transpose acts both on the $\psi,\chi$ indices and the coordinate $\tau$.
One can numerically solve these equations using standard iteration method. Numerically, we do not find solutions with nonzero off-diagonal correlators.

\subsection{Review of superconformal solutions of the two-flavor model}\label{app:conf}
In this section, we will give a brief review the method of deriving the locations of fortuitous BPS states using the Schwinger-Dyson equation in the grand canonical ensemble, where the chemical potentials for the two  U(1)  charges are introduced. This is analogous of solving the bulk supergravity solutions and finding the charge constraints. We refer \cite{Heydeman:2022lse} for more detailed explanations.
To proceed, we take the theory to its infrared conformal limit. In this regime, the solutions are parametrized by the scaling dimension \( \Delta \) and the spectral asymmetry \( \mathcal{E} \) associated with each fermionic and bosonic mode. Consequently, there are a total of \( 2 \times 4 = 8 \) parameters. The Schwinger-Dyson equations then impose six constraints on these eight variables, leaving two independent parameters, which can be identified as the spectral asymmetries \( \mathcal{E}_{\psi} \) and \( \mathcal{E}_{\chi} \) for the fields \( \psi \) and \( \chi \).\footnote{Requiring conformal solution dominance at IR imposes further constraints: $ 0 < \Delta_{\psi} < \frac{1}{2(q-1)} ,\quad 0 < \Delta_\chi < \frac{1}{2} - (q-2)\Delta_\psi < \frac{1}{2}.$} The six constraints are\footnote{Notice that (\ref{typo}) contains an extra $(q-1)$ factor in the first term on the right hand side compared to (3.36) of \cite{Heydeman:2022lse}, which we believe to be a typo.}
\begin{align}\label{eqn:confsoln}
-        \mathcal{E}_{b_\psi} = -(q-2) \mathcal{E}_\psi - \mathcal{E}_\chi ,&\quad \mathcal{E}_{b_\chi}= -(q-1) \mathcal{E}_{\psi},\\
            \Delta_{b_\chi} + (q-1)\Delta_\psi = 1 ,&\quad \Delta_{b_\psi} + (q-2) \Delta_{\psi} + \Delta_\chi = 1,\\
    \frac{(1-2\Delta_{\chi})\sin 2 \pi \Delta_{\chi}}{4\pi \prod_{\pm} \cos \pi(\Delta_{\chi} \pm \i \mathcal{E}_{\chi})} &=\frac{(1-2\Delta_{b_\psi})\sin 2 \pi \Delta_{b_\psi}}{4\pi \prod_{\pm} \sin \pi(\Delta_{b_\psi} \pm \i \mathcal{E}_{b_\psi})},\\
    \frac{(1-2\Delta_{\psi})\sin 2 \pi \Delta_{\psi}}{4\pi \prod_{\pm} \cos \pi(\Delta_{\psi} \pm \i \mathcal{E}_{\psi})} &=\frac{(q-1)(1-2\Delta_{b_\chi})\sin 2 \pi \Delta_{b_\chi}}{4\pi \prod_{\pm} \sin \pi(\Delta_{b_\chi} \pm \i \mathcal{E}_{b_\chi})} + \frac{(q-2)(1-2\Delta_{\chi})\sin 2 \pi \Delta_{\chi}}{4\pi \prod_{\pm} \cos \pi(\Delta_{\chi} \pm \i \mathcal{E}_{\chi})}.\label{typo}
\end{align}
To relate these spectral asymmetries to the physical charges \( n_{\psi} \) and \( n_{\chi} \), we utilize the Luttinger-Ward relation which expressed the  U(1) charges in terms of the conformal dimensions and spectral asymmetries.
In terms of the fermion charges, defined by \( Q_\psi = \sum_i \psi_i \bar{\psi}_i - N/2 \) and \( Q_\chi = \sum_i \bar{\chi}_i\chi_i  - N/2 \),\footnote{Notice a relative charge conjugation of $\chi$ between our notation and the notation in \cite{Heydeman:2022lse}.}  the Luttinger-Ward relation reads:
\begin{align}\label{eqn:Lutt}
    \frac{Q_\psi}{N} &= \mathfrak{q}_f(\Delta_\psi, \mathcal{E}_\psi) + (q-2) \mathfrak{q}_b(\Delta_{b_\psi}, \mathcal{E}_{b_\psi}) + (q-1) \mathfrak{q}_b(\Delta_{b_\chi}, \mathcal{E}_{b_\chi}),  \\
    \frac{Q_\chi}{N} &= \mathfrak{q}_f(\Delta_\chi, \mathcal{E}_\chi) + \mathfrak{q}_b(\Delta_{b_\psi}, \mathcal{E}_{b_\psi}), 
\end{align}
where
\begin{align}
    \mathfrak{q}_f(\Delta, \mathcal{E}) &= \frac{(\frac{1}{2}-\Delta) \sinh 2\pi \mathcal{E}}{\cosh 2\pi \mathcal{E} + \cos 2 \pi \Delta} + \frac{\i}{2\pi} \log \left( \frac{\cos \pi (\Delta + \i \mathcal{E})}{\cos \pi (\Delta - \i \mathcal{E})} \right), \label{eqn:LuttQf} \\
    \mathfrak{q}_b(\Delta_b, \mathcal{E}_b) &= \frac{(\frac{1}{2}-\Delta_b) \sinh 2\pi \mathcal{E}_b}{\cosh 2\pi \mathcal{E}_b - \cos 2 \pi \Delta_b} + \frac{\i}{2\pi} \log \left( \frac{\sin \pi (\Delta_b + \i \mathcal{E}_b)}{\sin \pi (\Delta_b - \i \mathcal{E}_b)} \right). \label{eqn:LuttQb}
\end{align}
This family of solutions describes the low-energy dynamics of the two-flavor model across various charge sectors, that contains both supersymmetric and non-supersymmetric configurations.

In order to determine the location of the fortuitous states, we can now impose supersymmetric condition that the superspace two-point function to be chiral and antichiral in the respective superspace coordinates. This, together with translational invariance, uniquely determines the relationship between the fermionic correlator and its bosonic partner:
\begin{equation}
    \mathcal{G}_{\bar{\Psi} \Psi} (Z_1, Z_2) = f(\tau_1 - \tau_2 - \theta_1 \bar\theta_1 - \theta_2 \bar\theta_2 - 2 \bar\theta_1 \theta_2), \quad \rightarrow \quad G_{b_{\psi}, b_{\psi}} = -\partial_{\tau} G_{\psi\psi}.
\end{equation}
This imposes an additional linear relation between the spectral asymmetries \( \mathcal{E}_{\psi} \) and \( \mathcal{E}_{\chi} \):
\begin{align}
    \mathcal{E}_{\chi} + (q - 1) \mathcal{E}_{\psi} = 0.
\end{align}
Mapping this relation through the Luttinger-Ward relation \cite{Gu:2019jub,Heydeman:2022lse} yields the constraint on the fortuitous (or black hole) states, correlating the two  U(1) charges \( n_{\psi} \) and \( n_{\chi} \). This analysis gives rise to the red curve depicted in figure \ref{fig:twofconcentration}. The discrete symmetry (rotation by $\pi$) in this figure is due to the time reversal symmetry $\mathcal{T}$ which maps \( n_{\psi} \) and \( n_{\chi} \) to \( 1-n_{\psi} \) and \( 1-n_{\chi} \).

\section{Exact diagonalization results in the $K=2$ two-flavor model}\label{app:ED}

We numerically diagonalized the Hamiltonian for the $K=2$ two-flavor model in Section \ref{sec:twoflavor}, for the case of $N=5$ (in total $2\times K \times N = 20$ complex fermions). We record the number of BPS states in various charge sectors in a matrix $(D_\textrm{BPS})_{{\bf{N}}_\chi + 1,{\bf{N}}_\psi + 1  }$, which is given by
\begin{equation}\label{PauliBPS}
   D_\textrm{BPS} = 
   {\footnotesize{  \begin{pmatrix}
1 & 10 & 35 & 23 & 0 & 0 & 0 & 0 & 0 & 0 & 0 \\
0 & \textbf{{\color{blue}{3}}} & 195 & 498 & 0 & 0 & 0 & 0 & 0 & 0 & 0 \\
0 & 0 & 15 & 1800 & 2205 & 0 & 0 & 0 & 0 & 0 & 0 \\
0 & 0 & 0 & 2150 & 7001 & 2260 & 0 & 0 & 0 & 0 & 0 \\
0 & 0 & 0 & 0 & 9785 & 10830 & 0 & 0 & 0 & 0 & 0 \\
0 & 0 & 0 & 0 & 4845 & 15504 & 4845 & 0 & 0 & 0 & 0 \\
0 & 0 & 0 & 0 & 0 & 10830 & 9785 & 0 & 0 & 0 & 0 \\
0 & 0 & 0 & 0 & 0 & 2260 & 7001 & 2150 & 0 & 0 & 0 \\
0 & 0 & 0 & 0 & 0 & 0 & 2205 & 1800 & 15 & 0 & 0 \\
0 & 0 & 0 & 0 & 0 & 0 & 0 & 498 & 195 & 3 & 0 \\
0 & 0 & 0 & 0 & 0 & 0 & 0 & 23 & 35 & 10 & 1
\end{pmatrix}   }}
\end{equation}
In blue, we highlight the $3$ BPS states with ${\bf{N}}_\psi  ={\bf{N}}_\chi  = 1$ (in a generic model, it would instead be $0$). These are the monotonous graviton states which would be absent in a generic two-flavor model. We note that even though there are also monotonous states in other charge sectors with ${\bf{N}}_\psi  ={\bf{N}}_\chi  > 1$, at this relatively small value of $N$ they are not cleanly separated from the fortuitous states. We have checked that for $N=6$, the sector with ${\bf{N}}_\psi  ={\bf{N}}_\chi  = 2$ will begin to only contain monotonous states.

\section{Deriving the analytic interpolation between $N$}\label{app:interpolation}
Our approach to the analytic continuation of \(N\) involves moving to the ``momentum space" of the fermion indices. We start by fixing a large but finite \(N_*\) and then perform the analytic continuation of \(N\) from \(N_*\) to smaller values.

In the \(N_*\)-dimensional space, we consider the fermion operators \(\psi_n\), which can be expressed in their Fourier basis:
\begin{align}
    \psi_p &= \sum_{n=1}^{N_*} \psi_n e^{-2\pi \i \frac{p n}{N_*}}, \quad p \in \mathbb{Z}_{N_*}.
\end{align}
The inner product between the momentum space fermions are given by:
\begin{equation} \label{eqn:innerproduct}
    \langle \psi_p | \psi_q \rangle = \sum_{n=1}^{N_*} e^{2\pi \i \frac{(p-q)n}{N_*}} = r \frac{1 - r^{N_*}}{1 - r}, \quad r = e^{2\pi \i \frac{p-q}{N_*}}.
\end{equation}
The idea is to analytically continue the power of $r$, while keeping the definition of $r$ fixed. In other words, we take
\begin{equation} 
    \langle \psi_p | \psi_q \rangle  = r \frac{1 - r^{N}}{1 - r}, \quad r = e^{2\pi \i \frac{p-q}{N_*}}.
\end{equation}
This expression is analytic in $N$ and introduces null states at integer values of \(N < N_*\).

Next, we consider transforming the momentum basis back to the original basis. The inner product in the original basis, which turns out to be diagonal, is:
\begin{equation} \label{innerSYK}
    \langle \psi_a | \psi_b \rangle = \frac{1}{N_*} \sum_{\ell=1}^{N_*}  e^{\frac{2\pi \i \ell a }{N_*} }  \frac{  1 -  e^{-\frac{2\pi \i \ell N  }{N_*} }    }{   e^{\frac{2\pi \i \ell  }{N_*} } - 1  }\delta_{ab}\equiv w_{a}(N)\delta_{ab}.
\end{equation}
This leads to equation (\ref{wa}) in Section \ref{sec:genefollow}.

\section{Monotonous operator algebra}\label{app:malg}

There is an interesting algebraic structure underlying the monotonous states if we treat the monotonous operator \( V_{\alpha \beta} = (\psi_{\alpha i} \chi_{\beta i} + \psi_{\beta i} \chi_{\alpha i}) \) and their Hermitian conjugates as a generator of a Lie algebra. To close the algebra, we also introduce a linear combination of fermionic bilinear operators and the identity: \( N_{\alpha \beta} = (n_{\psi})_{\alpha \beta} + (n_{\chi})_{\alpha \beta} - \delta_{\alpha \beta} N \), where \((n_{\psi})_{\alpha \beta} = \psi_{\alpha j} \bar{\psi}_{\beta j}\) and \((n_{\chi})_{\alpha \beta} = \chi_{\alpha i} \bar{\chi}_{\beta i}\). Here, Greek indices range from \(1\) to \(K\), while Latin indices range from \(1\) to \(N\) (and are summed over).

The commutation relations between \(V_{\alpha \beta}\), \(N_{\alpha \beta}\), and their adjoints are as follows:
\begin{equation} \label{sp(2M) algebra}
\begin{aligned}
 [V_{\alpha \beta}, V^{\dagger}_{\gamma \delta}] &= - \delta_{\alpha \gamma} N_{\alpha \beta} - \delta_{\beta \delta} N_{\gamma \alpha} - \delta_{\beta \gamma} N_{\delta \alpha} - \delta_{\alpha \delta} N_{\gamma \beta},\\
[V_{\alpha \beta}, N_{\gamma \delta}] &= \delta_{\alpha \gamma} V_{\delta \beta} + \delta_{\beta \gamma} V_{\alpha \delta}, \quad [V^{\dagger}_{\alpha \beta}, N_{\gamma \delta}] = - \delta_{\alpha \delta} V^{\dagger}_{\gamma \beta} - \delta_{\beta \delta} V^{\dagger}_{\alpha \gamma},\\
 [N_{\alpha \beta}, N_{\gamma \delta}] &= \delta_{\beta \gamma} N_{\alpha \delta} - \delta_{\alpha \delta} N_{\gamma \beta}.
\end{aligned}
\end{equation}
All other commutators vanish. One can recognize that the structure agrees with the commutation relations of the symplectic Lie algebra \(\mathfrak{sp}(2K)\); see equation (2.5) of \cite{rangarajan1992canonical} which can be matched to (\ref{sp(2M) algebra}) by identifying \(C = -N\), \(L = -V\), and \(R = -V^{\dagger}\).

The Cartan subalgebra, the maximal set of simultaneously diagonalizable elements, is given by \(\{ N_{\alpha \alpha} \}\), where \(\alpha\) ranges from \(1\) to \(K\). States are labeled by \(\ket{N_{1}, N_{2}, \dots, N_{K}}\), which represent the eigenvalues of \(N_{11}, N_{22}, \dots, N_{KK}\). Each irreducible representation is labeled by the highest weight \((N_{1, \text{max}}, N_{2, \text{max}}, \dots, N_{K, \text{max}})\).

In the special case \(K = 2\), we are dealing with \(\mathfrak{sp}(4)\). 
One can verify that \(V_{\alpha \beta}\), \(V^{\dagger}_{\alpha \beta}\), \(N_{12}\), and \(N_{21}\) act as raising or lowering operators. We can denote \(N_{11}\) and \(N_{22}\) as \(H_{1}\) and \(H_{2}\), respectively, and the remaining generators as \(E_{1}, \dots, E_{8}\). In this setup, the generators \(V\), \(V^{\dagger}\), and \(N\) form a Cartan-Weyl basis \((H_{1}, H_{2}, E_{1}, \dots, E_{8})\).
The actions of these generators on the states \(\ket{N_{1}, N_{2}}\) reflect the root system of type \(C_2\):
\begin{align*}
   N_{12} \ket{N_{1}, N_{2}} &= \ket{N_{1} + 1, N_{2} - 1}, & N_{21} \ket{N_{1}, N_{2}} &= \ket{N_{1} - 1, N_{2} + 1}, \\
 V_{11} \ket{N_{1}, N_{2}} &= \ket{N_{1} - 2, N_{2}}, & V^{\dagger}_{11} \ket{N_{1}, N_{2}} &= \ket{N_{1} + 2, N_{2}}, \\
   V_{22} \ket{N_{1}, N_{2}} &= \ket{N_{1}, N_{2} + 2}, & V^{\dagger}_{22} \ket{N_{1}, N_{2}} &= \ket{N_{1}, N_{2} - 2}, \\
    V_{12} \ket{N_{1}, N_{2}} &= \ket{N_{1} - 1, N_{2} - 1}, & V^{\dagger}_{12} \ket{N_{1}, N_{2}} &= \ket{N_{1} + 1, N_{2} + 1}.
\end{align*}
The dimension for each irreducible representation \((N_{1, \text{max}}, N_{2, \text{max}})\), with the convention \(N_{1, \text{max}} \geq N_{2, \text{max}}\), is given by \cite{hall2013lie}:
\[
\dim (N_{1, \text{max}}, N_{2, \text{max}}) = \frac{1}{6} (N_{1, \text{max}} - N_{2, \text{max}} + 1)(N_{2, \text{max}} + 1)(N_{1, \text{max}} + 2)(N_{1, \text{max}} + N_{2, \text{max}} + 3).
\]
In our $K=2$ two flavor model, we have $N_{1,\textrm{max}} = N_{2,\textrm{max}} = N$. For $N=4$, we have $\textrm{dim} = 55$, which indeed agrees with results from exact diagonalization.

\bibliography{references}

\bibliographystyle{utphys}

\end{document}